\documentclass[final]{siamltex}
\usepackage{amssymb}
\usepackage{showkeys}
\usepackage{graphicx}
\usepackage{amsmath}
\usepackage{amssymb}
\usepackage{graphicx}
\usepackage{epsfig}
\usepackage{labelfig}
\usepackage{amsmath}
\usepackage{color}
\usepackage{afterpage}
\usepackage{verbatim}
\usepackage{umlaut}
\usepackage{afterpage}
\usepackage[small]{caption}

\makeatletter

\long\def\addtocontents#1#2{%
  \protected@write\@auxout
    {\let\label\@gobble \let\index\@gobble \let\glossary\@gobble}%
    {\string\@writefile{#1}{#2}}}

\newcommand\tableofcontents{%
    \@starttoc{toc}%
    }
\newcommand*\l@section[2]{%
  \ifnum \c@tocdepth >\z@
    \addpenalty\@secpenalty
    \addvspace{1.0em \@plus\p@}%
    \setlength\@tempdima{1.5em}%
    \begingroup
      \parindent \z@ \rightskip \@pnumwidth
      \parfillskip -\@pnumwidth
      \leavevmode \bfseries
      \advance\leftskip\@tempdima
      \hskip -\leftskip
      #1\nobreak\hfil \nobreak\hb@xt@\@pnumwidth{\hss #2}\par
    \endgroup
  \fi}
\newcommand*\l@subsection{\@dottedtocline{2}{1.5em}{2.3em}}
\newcommand*\l@subsubsection{\@dottedtocline{3}{3.8em}{3.2em}}
\newcommand*\l@paragraph{\@dottedtocline{4}{7.0em}{4.1em}}
\newcommand*\l@subparagraph{\@dottedtocline{5}{10em}{5em}}

\makeatother

\input{tcilatex}
\begin{document}

\title{Carleman estimates for global uniqueness, stability and numerical
methods for coefficient inverse problems }
\author{Michael V. Klibanov$^{\ast }$}
\thanks{Department of Mathematics and Statistics, University of North
Carolina at Charlotte, Charlotte, NC 28223, USA, email: mklibanv{@}uncc.edu}
\maketitle

\begin{abstract}
This is a review paper of the role of Carleman estimates in the theory of
Multidimensional Coefficient Inverse Problems since the first inception of
this idea in 1981.
\end{abstract}

\tableofcontents

\section{Introduction}

\label{sec:1}

This is a review paper of the Bukhgeim-Klibanov method (BK). The author
considers this paper as an introductory material for BK. Because of many
publications on BK, the author restricts himself to citations of those works
which he is best familiar with. An interested reader can find further
citations as well as analytical details in publications cited here. Three
topics are discussed in this paper: (1) Global uniqueness for
Multidimensional Coefficient Inverse Problems (MCIPs) with single
measurement data via BK, (2) Global stability both for MCIPs and some
ill-posed Cauchy problems via Carleman estimates, and (3) Related convergent
numerical methods both for MCIPs and some ill-posed Cauchy problems.

BK was introduced in three originating papers of Bukhgeim and Klibanov in
1981 \cite{BukhKlib,Bukh1,Klib1}. Until now BK remains the only technique
enabling to prove global uniqueness and stability theorems for MCIPs with
single measurement data. The state of the art in the field of Inverse
Problems in 1981 is well reflected in the following citation from the paper
\cite{BukhKlib}. \textquotedblleft \emph{Uniqueness theorems for
multidimensional inverse problems have at present been obtained mainly in
classes of piecewise analytic functions and similar classes or
locally...Moreover, the technique of investigating these problems has, as a
rule depended in an essential way on the type of the differential equation.
In this note a new method of investigating inverse problems is proposed that
is based on weighted a priori estimates. This method makes it possible to
consider in a unified way a broad class of inverse problems for those
equations }$Pu=f$\emph{\ \ for which the solution of the Cauchy problem
admits a Carleman estimate...The theorems of \S 1 were proved by M.V.
Klibanov and those of \S 2 by A.L. Buhgeim. They were obtained
simultaneously and independently.}"

While the paper \cite{BukhKlib} contains announcement of results and an
indication of the proof, papers \cite{Bukh1,Klib1} contain first complete
proofs. Also, see, e.g. \cite%
{Bukh2,EEK,Kalt,Klib2,Klib3,Klib4,KlibMaxw,Klib5,KlibPar1,KlibSurvey,KlibPar2,KT,KYam,Klib6}
and Sections 1.10 and 1.11 of the book \cite{BK} for some follow up
publications of these authors on BK. Prior publications \cite%
{BukhKlib,Bukh1,Klib1} only the so-called \textbf{local} uniqueness theorems
were known for MCIPs with single measurement data. The term \emph{local}
means here that unknown spatially dependent coefficients were assumed to be
either piecewise analytic functions, or functions represented via truncated
Fourier-like series, or sufficiently small perturbations of constants. At
that time the common desire of many mathematicians working on MCIPs was to
prove \textbf{global} uniqueness theorems. That is, to prove uniqueness for
the case when the unknown coefficient $a\left( x\right) ,x\in \mathbb{R}%
^{n},n\geq 2$ satisfies only some natural conditions, such as, e.g. $a\in
C^{k}$. However, it was unclear at that time how to do this. Indeed,
standard methods were basically based on integral equations and did not work
for this goal. The absence of global uniqueness theorems was the main
stumbling block in the field of Inverse Problems in 1970s. This is what has
originally motivated the author in 1979 to think about moving away from
traditional techniques. The author has spent two years to figure out the
solution.

This paper is focused only on MCIPs with single measurement data. BK is
based on a special use of Carleman estimates for MCIPs. Roughly speaking, as
soon as Carleman estimate is valid for a PDE operator, BK can be applied.
Therefore, the generality of BK is due to the fact that Carleman estimates
are valid for hyperbolic, parabolic, elliptic, the non-stationary Schr\"{o}%
dinger and some other operators. Schematically, BK consists of the following
two steps:

\textbf{Step 1}. Given an MCIP, figure out whether a proper Carleman
estimate is valid for the PDE operator of this problem. If not, derive a
proper Carleman estimate (if possible).

\textbf{Step 2}. Given a proper Carleman estimate, apply BK.

Carleman estimates were originated in 1939 in a remarkable work of a
distinguished Swedish mathematician Torsten Carleman \cite{Carl}. Since then
and up to now they have been traditionally applied by many authors to proofs
of uniqueness theorems for various ill-posed problems for PDEs with the
Cauchy data on non-characteristic hypersurfaces, see, e.g. the paper of
Calderon \cite{Cald} and the book of H\"{o}rmander \cite{Horm}. While in
these references Carleman estimates were used for functions with compact
support, the book of Lavrent'ev, Romanov and Shishatskii \cite{LRS} (Chapter
4) uses functions with non-compact support.\ As a result, the technique of
this book allowed to prove not only uniqueness but H\"{o}lder stability
results as well for some ill-posed Cauchy problems. Thus, it was briefly
noticed in earlier publications \cite{Klib3,Klib5} with the reference to
\cite{LRS} that the applicability of BK to an MCIP usually automatically
implies the H\"{o}lder stability estimate for this MCIP. Since the author is
concerned with MCIPs, multiple quite interesting works on applications of
Carleman estimates to various ill-posed Cauchy problems are not cited here.

BK applies Carleman estimates, in a specially designed way, to proofs of
global uniqueness and stability results for MCIPs. MCIPs are substantially
different from those Cauchy problems. Indeed, in such a Cauchy problem all
coefficients of the corresponding PDE operator are known, the initial
condition is unknown, and the Cauchy data at a non-characteristic
hypersurface are known. It is required then to reconstruct the solution $u$
of the corresponding PDE. On the other hand, in an MCIP at least one of
coefficients, $a\left( x\right) ,$ of the corresponding PDE operator is
unknown, the initial condition is known, and the lateral Cauchy data are
known as well. It is required then to reconstruct the pair of functions $%
\left( a,u\right) .$ Those Cauchy problem are linear and any MCIP is
nonlinear.

The term \textquotedblleft MCIP with single measurement data" means the
problem of the recovery of one of coefficients of a PDE from a boundary
measurement generated by a single set of initial conditions. In the case of
either the point source or a plane wave this means either a single position
of that source or a single direction of that incident plane wave. More
generally, this is a single pair of initial conditions for a hyperbolic PDE
and a single initial condition for a parabolic PDE. Sometimes a few initial
sets of initial conditions are allowed. In the case when $k$ coefficients
are unknown, $k$ sets of initial conditions are allowed, which means $k$
measurements. MCIPs with single measurement are non-overdetermined ones. The
non-overdetermination means that the number of free variables in the data
equals the number of free variables in the unknown coefficient. The single
measurement case is the one with the minimal amount of the available
information. Hence, this is the most economical way of data collection. In
particular, in military applications the single measurement case is far
preferable to the case of many measurements. This is because an installation
of each source carries a serious risk for life of soldiers on a battlefield.

There is a single condition of BK, which has been viewed as a drawback from
the applied standpoint for a long time since 1981, This condition is still
not lifted. Specifically, BK requires that at least one initial condition to
be non-zero in the entire domain of interest. However, after getting an
extensive recent numerical experience with the approximately globally
convergent numerical method for MCIPs (see the book of Beilina and Klibanov
\cite{BK} and Section 6 below), the author believes now that this drawback
is an absolutely insignificant one precisely\emph{\ }from the applied
standpoint (although the mathematical question remains open). Indeed, the
most interesting case in applications is the case when the initial condition
is the function $\delta \left( x-x_{0}\right) $ with a fixed position of the
source $x_{0}.$ However, replacement of this function by its approximation
via a narrow Gaussian $\delta _{\varepsilon }\left( x-x_{0}\right) ,$
\begin{equation*}
\delta _{\varepsilon }\left( x-x_{0}\right) =C_{\varepsilon }\exp \left( -%
\frac{\left\vert x-x_{0}\right\vert ^{2}}{\varepsilon ^{2}}\right)
,\int\limits_{\mathbb{R}^{n}}\delta _{\varepsilon }\left( x-x_{0}\right) dx=1
\end{equation*}%
immediately lets BK working ($\varepsilon >0$ is sufficiently small here).
The corresponding boundary data, which model the data resulting from a
measurement, have only an insignificant change.\ Therefore, if a numerical
method for a corresponding MCIP is stable, as it must be, then this change
should affect the solution only insignificantly. Furthermore, physicists and
engineers are indifferent to such a replacement because of the above
reasons. This is why functions $\delta \left( x-x_{0}\right) $\ and $\delta
_{\varepsilon }\left( x-x_{0}\right) $\ are equivalent precisely from the
applied standpoint.

Still, the author has proved \cite{Klib6} uniqueness theorem for an MCIP for
the equation $u_{tt}=\Delta u+a\left( x,y,z\right) u,\left( x,y,z\right) \in
\mathbb{R}^{3}$ for the case of an incident plane wave with $u\left(
x,y,z,0\right) =0,u_{t}\left( x,y,z,0\right) =\delta \left( z\right) $ and
under the assumption that this equation for the function $u$ is written in
finite differences with respect to $\left( x,y\right) .$

The idea of BK is described in Section 3. Five examples of this section show
how BK works. In a less general form examples of Sections 3.2, 3.3.2 and 3.4
were first published in originating works \cite{BukhKlib,Klib1}, also see
\cite{BK,Klib5,KT} for more general forms of these three examples as well as
for two other examples of Section 3. In principle, it is possible to
formulate BK in a general abstract form, see, e.g. the earlier paper of the
author \cite{Klib2} for this form. However, it is not necessary to do so for
the understanding of BK. Previously published relevant results are discussed
in Section 4 as well as in the end of each of Sections 2,3,5,6.

\section{Carleman Estimates, H\"{o}lder Stability and the
Quasi-Reversibility Method}

\label{sec:2}

\subsection{Definition of the Carleman estimate}

\label{sec:2.1}

We now introduce the notion of the pointwise Carleman estimate for a general
Partial Differential Operator of the second order. Let $G\subset \mathbb{R}%
^{n}$ be a bounded domain with a piecewise smooth boundary $\partial G.$ Let
the function $\xi \in C^{2}\left( \overline{G}\right) $ and $\left\vert
\nabla \xi \right\vert \neq 0$ in $\overline{G}.$ For a number $c\geq 0$
denote
\begin{equation*}
\xi _{c}=\left\{ x\in \overline{G}:\xi \left( x\right) =c\right\}
,G_{c}=\left\{ x\in G:\xi \left( x\right) >c\right\} .
\end{equation*}%
Assume that $G_{c}\neq \varnothing .$ Let $\Gamma _{c}\subseteq \partial G$
a part of the boundary $\partial G$ defined as
\begin{equation*}
\Gamma _{c}=\left\{ x\in \partial G:\xi \left( x\right) \geq c\right\} .
\end{equation*}%
Let $G_{c}\neq \varnothing .$ Then the boundary of the domain $G_{c}$
consists of two parts,%
\begin{equation}
\partial G_{c}=\partial _{1}G_{c}\cup \partial _{2}G_{c},\partial
_{1}G_{c}=\xi _{c},\partial _{2}G_{c}=\Gamma _{c}.  \label{2.1}
\end{equation}%
Let $\lambda >1$ be a parameter, which we will consider to be large.
Consider the function $\varphi _{\lambda }\left( x\right) ,$%
\begin{equation}
\varphi _{\lambda }\left( x\right) =\exp \left( \lambda \xi \left( x\right)
\right) .  \label{2.2}
\end{equation}%
It follows from (\ref{2.1}), (\ref{2.2}) that
\begin{equation}
\min_{\overline{G}_{c}}\varphi _{\lambda }\left( x\right) =\varphi _{\lambda
}\left( x\right) \mid _{\xi _{c}}\equiv \exp \left( \lambda c\right) .
\label{2.3}
\end{equation}

Let $A\left( x,D\right) $ be a linear Partial Differential Operator of the
second order with real valued coefficients in $G$ and with its principal
part $A_{0}\left( x,D\right) ,$
\begin{eqnarray}
A\left( x,D\right) u &=&\sum\limits_{\left\vert \alpha \right\vert \leq
2}a_{\alpha }\left( x\right) D^{\alpha }u,\text{ }A_{0}\left( x,D\right)
u=\sum\limits_{\left\vert \alpha \right\vert =2}a_{\alpha }\left( x\right)
D^{\alpha }u.  \label{2.4} \\
a_{\alpha } &\in &C^{1}\left( \overline{G}\right) \text{ for }\left\vert
\alpha \right\vert =2,K:=\max_{\left\vert \alpha \right\vert =2}\left(
\left\Vert a_{\alpha }\right\Vert _{C^{1}\left( \overline{G}_{c}\right)
}\right) \text{; }a_{\alpha }\in C\left( \overline{G}\right) \text{ for }%
\left\vert \alpha \right\vert =0,1.  \label{2.5}
\end{eqnarray}

\textbf{Definition 2.1}. \emph{Let }$G_{c}\neq \varnothing .$ \emph{We say
that the operator }$A_{0}\left( x,D\right) $\emph{\ admits pointwise\
Carleman estimate in the domain }$G_{c}$\emph{\ with the Carleman Weight
Function (CWF) }$\varphi _{\lambda }\left( x\right) $\emph{\ if there exist
constants }$\lambda _{0}\left( \Omega ,K\right) >1,C_{0}=C_{0}\left( \Omega
,K\right) \geq 0,C\left( \Omega ,K\right) >0$\emph{\ depending only on the
domain }$G_{c}$\emph{\ and the number }$K$\emph{, such that the following a
priori estimate holds}%
\begin{eqnarray}
\left( A_{0}u\right) ^{2}\varphi _{\lambda }^{2}\left( x\right) &\geq &\frac{%
C_{0}}{\lambda }\sum\limits_{\left\vert \alpha \right\vert =2}\left(
D^{\alpha }u\right) ^{2}\varphi _{\lambda }^{2}\left( x\right) +C\lambda
\left( \nabla u\right) ^{2}\varphi _{\lambda }^{2}\left( x\right) +C\lambda
^{3}u^{2}\varphi _{\lambda }^{2}\left( x\right) +\func{div}U,  \label{2.6} \\
\forall \lambda &\geq &\lambda _{0},\forall u\in C^{2}\left( \overline{G}%
\right) ,\forall x\in G_{c}.  \label{2.7}
\end{eqnarray}%
\emph{In (\ref{2.6}) the term under the divergence sign satisfies the
following estimate}
\begin{equation}
\left\vert U\right\vert \leq C\lambda ^{3}\left[ \left( \nabla u\right)
^{2}+u^{2}\right] \varphi _{\lambda }^{2}\left( x\right) +\frac{C_{0}}{%
\lambda }\sum\limits_{\left\vert \alpha \right\vert =2}\left( D^{\alpha
}u\right) ^{2}\varphi _{\lambda }^{2}\left( x\right) .  \label{2.8}
\end{equation}

In the case of parabolic and elliptic operators $C_{0}>0$ and $C_{0}=0$ in
the case of a hyperbolic operator. Lemma 2.1 is elementary.

\textbf{Lemma 2.1}.\emph{\ Let conditions (\ref{2.5}) imposed on
coefficients of the operator }$A$ \emph{be valid. Suppose that the Carleman
estimate (\ref{2.6})-(\ref{2.8}) is valid for the principal part }$%
A_{0}\left( x,D\right) $\emph{\ of the operator }$A\left( x,D\right) .$\emph{%
\ Then it is also valid \ for the operator }$A\left( x,D\right) $\emph{,
although with a different constant }$\lambda _{0}.$ \emph{In other words,
the Carleman estimate depends only on the principal part of the operator. }

\textbf{Proof.} We have
\begin{equation}
\left( Au\right) ^{2}\varphi _{\lambda }^{2}\left( x\right) \geq \left(
A_{0}u\right) ^{2}\varphi _{\lambda }^{2}\left( x\right) -M\left[ \left(
\nabla u\right) ^{2}+u^{2}\right] \varphi _{\lambda }^{2}\left( x\right) ,
\label{2.9}
\end{equation}%
where $M>0$ is a constant depending only on the maximum of norms $\left\Vert
a_{\alpha }\right\Vert _{C\left( \overline{G}\right) },\left\vert \alpha
\right\vert =0,1.$ Substituting (\ref{2.9}) in (\ref{2.6}) and taking $%
\lambda $ sufficiently large, we again obtain (\ref{2.6}). $\square $

\subsection{H\"{o}lder stability}

\label{sec:2.2}

\bigskip Consider the following Cauchy problem for the differential
inequality%
\begin{eqnarray}
\left\vert A_{0}u\right\vert &\leq &B\left( \left\vert \nabla u\right\vert
+\left\vert u\right\vert +\left\vert f\right\vert \right) ,\forall x\in
G_{c},  \label{2.10} \\
u &\mid &_{\Gamma _{c}}=g_{0}\left( x\right) ,\partial _{n}u\mid _{\Gamma
_{c}}=g_{1}\left( x\right) ,  \label{2.11}
\end{eqnarray}%
where $B=const.>0$ and $f\in L_{2}\left( G_{c}\right) $ is a function.
Clearly, functions $g_{0},g_{1}$ in (\ref{2.11}) are the Cauchy data for the
function $u$. In particular, equation $Au=f$ with the boundary data (\ref%
{2.11}) can be reduced to the problem (\ref{2.10}), (\ref{2.11}). We want to
estimate the function $u$ via functions $f,g_{0},g_{1}.$ Such estimates were
derived in Chapter 4 of the book of Lavrent'ev, Romanov and Shishatskii \cite%
{LRS} for parabolic, elliptic and hyperbolic operators.

\textbf{Theorem 2.1} (H\"{o}lder stability estimate). \emph{Assume that
conditions (\ref{2.5}) hold and that the Carleman estimate of Definition 2.1
is valid.\ Suppose that there exists a sufficiently small number }$%
\varepsilon >0$\emph{\ such that the domain }$G_{c+3\varepsilon }\neq
\varnothing .$\ \emph{Denote }$m=\max_{\overline{G}_{c}}\xi \left( x\right)
. $\emph{\ Define the number }$\beta =2\varepsilon /\left( 3m+2\varepsilon
\right) \in \left( 0,1\right) .$ \emph{Let functions }$g_{0},g_{1},f$ \
\emph{be such that} $g_{0}\in H^{1}\left( \Gamma _{c}\right) ,g_{1}\in
L_{2}\left( \Gamma _{c}\right) ,f\in L_{2}\left( G_{c}\right) .$ \emph{Let
the function }$u\in C^{2}\left( \overline{G}_{c}\right) $\emph{\ satisfies
conditions (\ref{2.10}), (\ref{2.11}). Then there exists a sufficiently
small number }$\delta _{0}=\delta _{0}\left( \varepsilon ,m,B,K,G_{c}\right)
\in \left( 0,1\right) $\emph{\ and a constant }$C_{1}=C_{1}\left(
\varepsilon ,m,B,K,G_{c}\right) >0$\emph{\ such that if }$\delta \in \left(
0,\delta _{0}\right) $\emph{, }%
\begin{equation}
\left\Vert f\right\Vert _{L_{2}\left( G_{c}\right) }+\left\Vert
g_{0}\right\Vert _{H^{1}\left( \Gamma _{c}\right) }+\left\Vert
g_{1}\right\Vert _{L_{2}\left( \Gamma _{c}\right) }\leq \delta
\label{2.11_1}
\end{equation}%
\emph{then the following H\"{o}lder stability estimate holds}%
\begin{equation}
\left\Vert u\right\Vert _{H^{1}\left( G_{c+3\varepsilon }\right) }\leq
C_{1}\left( 1+\left\Vert u\right\Vert _{H^{1}\left( G_{c}\right) }\right)
\delta ^{\beta },\forall \delta \in \left( 0,\delta _{0}\right) ,
\label{2.11_2}
\end{equation}

\textbf{Remark 2.1.} Estimate (\ref{2.11_2}) is H\"{o}lder stability
estimate because $\beta \in \left( 0,1\right) .$ If we would have $\beta =1,$
then (\ref{2.11_2}) would become the Lipschitz stability estimate. Because
of the presence of the term $\left\Vert u\right\Vert _{H^{1}\left(
G_{c}\right) },$ this is the so-called \textquotedblleft conditional
stability estimate", which are common in the theory of ill-posed problems
\cite{BKok,BK,EHN,Kab,T}. Indeed, the presence of this term actually means
that we assume an \emph{a priori }given upper bound for the norm $\left\Vert
u\right\Vert _{H^{1}\left( G_{c}\right) }.$

\textbf{Proof of Theorem 2.1}. In this proof $C=C\left( \varepsilon
,K,G_{c}\right) $ and $C_{1}=C_{1}\left( \varepsilon ,m,B,K,G_{c}\right) $
denote different positive constants depending on listed parameters but
independent on the function $u$ and the parameter $\lambda .$ Obviously $%
G_{c+3\varepsilon }\subset G_{c+2\varepsilon }\subset G_{c+\varepsilon
}\subset G_{c}.$ Since $G_{c+3\varepsilon }\neq \varnothing ,$ then $%
G_{c+2\varepsilon },G_{c+\varepsilon },G_{c}\neq \varnothing .$ Let $\chi
\left( x\right) $ be a function such that%
\begin{equation}
\chi \in C^{2}\left( \overline{G}_{c}\right) ,\chi \left( x\right) =\left\{
\begin{array}{c}
1,x\in G_{c+2\varepsilon }, \\
0,x\in G_{c}\diagdown G_{c+\varepsilon }, \\
\in \left[ 0,1\right] ,x\in G_{c+\varepsilon }\diagdown G_{c+2\varepsilon }.%
\end{array}%
\right.  \label{2.11_3}
\end{equation}%
Consider the function $v$,
\begin{equation}
v=\chi u.  \label{2.11_30}
\end{equation}%
Then (\ref{2.10}), (\ref{2.11}) and (\ref{2.11_3}) imply%
\begin{eqnarray}
\left\vert A_{0}v\right\vert &\leq &C_{1}\left[ \left\vert \nabla
v\right\vert +\left\vert v\right\vert +\left\vert \nabla \chi \right\vert
\left\vert \nabla u\right\vert +\left( \sum\limits_{\left\vert \alpha
\right\vert =2}\left\vert D^{\alpha }\chi \right\vert \right) \left\vert
u\right\vert +\left\vert f\right\vert \right] ,\forall x\in G_{c},
\label{2.11_4} \\
v &\mid &_{\Gamma _{c}}=\chi g_{0},\partial _{n}v\mid _{\Gamma
_{c}}=g_{0}\partial _{n}\chi +\chi g_{1},  \label{2.11_5}
\end{eqnarray}%
\begin{equation}
v\left( x\right) =0,x\in G_{c}\diagdown G_{c+\varepsilon }.  \label{2.11_6}
\end{equation}%
Square both sides of (\ref{2.11_4}), multiply by $\varphi _{\lambda
}^{2}\left( x\right) $ and apply (\ref{2.6}) ignoring $C_{0}$. We obtain
\begin{eqnarray*}
C_{1}f^{2}\varphi _{\lambda }^{2}\left( x\right) +C_{1}\left\vert \nabla
\chi \right\vert ^{2}\left\vert \nabla u\right\vert ^{2}+C_{1}\left(
\sum\limits_{\left\vert \alpha \right\vert =2}\left\vert D^{\alpha }\chi
\right\vert ^{2}\right) \left\vert u\right\vert ^{2}-\func{div}U &\geq & \\
C\lambda \left( 1-\frac{C_{1}}{\lambda }\right) \left( \nabla v\right)
^{2}\varphi _{\lambda }^{2}\left( x\right) +C\lambda ^{3}\left( 1-\frac{C_{1}%
}{\lambda ^{3}}\right) v^{2}\varphi _{\lambda }^{2}\left( x\right) ,\forall
\lambda &>&\lambda _{0},\forall x\in G_{c}.
\end{eqnarray*}%
Choose $\lambda >\lambda _{1}:=\max \left( \lambda _{0},2C_{1}\right) $ so
large that $C_{1}/\lambda <1/2.$Then with a different constant $C$%
\begin{eqnarray*}
C_{1}f^{2}\varphi _{\lambda }^{2}\left( x\right) +C_{1}\left\vert \nabla
\chi \right\vert ^{2}\left\vert \nabla u\right\vert ^{2}+C_{1}\left(
\sum\limits_{\left\vert \alpha \right\vert =2}\left\vert D^{\alpha }\chi
\right\vert ^{2}\right) \left\vert u\right\vert ^{2}-\func{div}U &\geq & \\
C\lambda \left( \nabla u\right) ^{2}\varphi _{\lambda }^{2}\left( x\right)
+C\lambda ^{3}u^{2}\varphi _{\lambda }^{2}\left( x\right) ,\forall \lambda
&>&\lambda _{1},\forall x\in G_{c}.
\end{eqnarray*}%
Integrate this inequality over $G_{c}$ using Gauss-Ostrogradsky formula as
well as (\ref{2.1}), (\ref{2.3}), (\ref{2.8}), (\ref{2.11_3}), (\ref{2.11_5}%
) and (\ref{2.11_6}). We obtain%
\begin{eqnarray}
&&C_{1}e^{2\lambda m}\int\limits_{G_{c}}f^{2}dx+C_{1}\lambda ^{3}e^{2\lambda
m}\int\limits_{\Gamma _{c}}\left[ \left( \nabla g_{0}\right) ^{2}+g_{1}^{2}%
\right] dS_{x}  \notag \\
&&+C_{1}\exp \left[ 2\lambda \left( c+2\varepsilon \right) \right]
\int\limits_{G_{c+\varepsilon }\diagdown G_{c+2\varepsilon }}\left(
\left\vert \nabla u\right\vert ^{2}+u^{2}\right) dx  \label{2.12} \\
&\geq &\lambda \int\limits_{G_{c}}\left( \nabla v\right) ^{2}\varphi
_{\lambda }^{2}dx+\lambda ^{3}\int\limits_{G_{c}}v^{2}\varphi _{\lambda
}^{2}dx.  \notag
\end{eqnarray}%
Since $G_{c+3\varepsilon }\subset G_{c+2\varepsilon }\subset G_{c}$, then
strengthening inequality (\ref{2.12}) and using (\ref{2.11_3}), (\ref%
{2.11_30}), we obtain%
\begin{eqnarray*}
&&C_{1}e^{2\lambda m}\int\limits_{G_{c}}f^{2}dx+C_{1}\lambda ^{3}e^{2\lambda
m}\int\limits_{\Gamma _{c}}\left[ \left( \nabla g_{0}\right) ^{2}+g_{1}^{2}%
\right] dS_{x}+C_{1}\exp \left[ 2\lambda \left( c+2\varepsilon \right) %
\right] \int\limits_{G_{c+\varepsilon }\diagdown G_{c+2\varepsilon }}\left(
\left\vert \nabla u\right\vert ^{2}+u^{2}\right) dx \\
&\geq &\lambda \int\limits_{G_{c+3\varepsilon }}\left( \nabla u\right)
^{2}\varphi _{\lambda }^{2}dx+\lambda ^{3}\int\limits_{G_{c+3\varepsilon
}}u^{2}\varphi _{\lambda }^{2}dx\geq \lambda \exp \left[ 2\lambda \left(
c+3\varepsilon \right) \right] \int\limits_{G_{c+3\varepsilon }}\left[
\left( \nabla u\right) ^{2}+u^{2}\right] dx.
\end{eqnarray*}%
Hence, we have established that
\begin{eqnarray*}
&&C_{1}e^{2\lambda m}\int\limits_{G_{c}}f^{2}dx+C_{1}\lambda ^{3}e^{2\lambda
m}\int\limits_{\Gamma _{c}}\left[ \left( \nabla g_{0}\right) ^{2}+g_{1}^{2}%
\right] dS_{x}+C_{1}\exp \left[ 2\lambda \left( c+2\varepsilon \right) %
\right] \left\Vert u\right\Vert _{H^{1}\left( G_{c}\right) }^{2} \\
&\geq &\lambda \exp \left[ 2\lambda \left( c+3\varepsilon \right) \right]
\left\Vert u\right\Vert _{H^{1}\left( G_{c+3\varepsilon }\right) }^{2}.
\end{eqnarray*}%
Divide both sides of this inequality by $\lambda \exp \left[ 2\lambda \left(
c+3\varepsilon \right) \right] $.\ Hence, there exists a number

$\lambda _{2}=\lambda _{2}\left( \varepsilon ,m,B,K,G_{c}\right) >\lambda
_{1}$ such that%
\begin{equation}
\left[ \int\limits_{G_{c}}f^{2}dx+\int\limits_{\Gamma _{c}}\left[ \left(
\nabla g_{0}\right) ^{2}+g_{1}^{2}\right] dS_{x}\right] C_{1}e^{3\lambda
m}+C_{1}\exp \left[ -2\lambda \varepsilon \right] \left\Vert u\right\Vert
_{H^{1}\left( G_{c}\right) }^{2}\geq \left\Vert u\right\Vert _{H^{1}\left(
G_{c+3\varepsilon }\right) }^{2},\forall \lambda >\lambda _{2}.  \label{2.13}
\end{equation}%
Using (\ref{2.11_1}), we obtain
\begin{equation}
\left\Vert u\right\Vert _{H^{1}\left( G_{c+2\varepsilon }\right) }^{2}\leq
C_{1}\left( \delta ^{2}e^{3\lambda m}+e^{-2\lambda \varepsilon }\left\Vert
u\right\Vert _{H^{1}\left( G_{c}\right) }^{2}\right) .  \label{2.14}
\end{equation}%
We now balance two terms in the right hand side of (\ref{2.14}) via choosing
$\lambda =\lambda \left( \delta \right) $ such that
\begin{equation*}
\delta ^{2}e^{3\lambda m}=e^{-2\lambda \varepsilon }.
\end{equation*}%
Hence,
\begin{equation}
\lambda =\ln \left( \delta ^{-2\left( 3m+2\varepsilon \right) ^{-1}}\right) .
\label{2.15}
\end{equation}%
Hence we should have $\delta \in \left( 0,\delta _{0}\right) ,$ where the
number $\delta _{0}=\delta _{0}\left( \varepsilon ,m,B,K,G_{c}\right) $ is
so small that $\ln \left( \delta _{0}^{-2\left( 3m+2\varepsilon \right)
^{-1}}\right) >\lambda _{2}.$ The target estimate (\ref{2.11_2}) follows
from (\ref{2.14}) and (\ref{2.15}). $\square $

\textbf{Theorem 2.2} (uniqueness). \emph{Let conditions of Theorem 2.1 hold,
in (\ref{2.11}) }$g_{0}\left( x\right) \equiv g_{1}\left( x\right) \equiv
0,x\in \Gamma _{c}$\emph{\ and also }$f\left( x,t\right) \equiv 0.$\emph{\
Then }$u\left( x\right) \equiv 0$ \emph{\ for }$x\in G_{c}.$

This theorem immediately follows from Theorem 2.1. To prove convergence of
the Quasi-Reversibility Method (Section 2.5), we need to replace the
pointwise inequality (\ref{2.10}) with the following integral inequality%
\begin{equation}
\int\limits_{G_{c}}\left( Au\right) ^{2}dx\leq S^{2}.  \label{2.150}
\end{equation}

\textbf{Theorem 2.3}. \emph{Let the function }$u\in H^{2}\left( G_{c}\right)
$\emph{\ satisfies inequality (\ref{2.150}), }$u\mid _{\Gamma _{c}}=\partial
_{n}u\mid _{\Gamma _{c}}=0$\emph{\ and the number }$S\in \left( 0,\delta
\right) .$\emph{\ Assume that conditions (\ref{2.5}) hold and that the
Carleman estimate of Definition 2.1 is valid.\ Suppose that there exists a
sufficiently small number }$\varepsilon >0$\emph{\ such that the domain }$%
G_{c+3\varepsilon }\neq \varnothing .$\ \emph{Denote }$m=\max_{\overline{G}%
_{c}}\xi \left( x\right) .$\emph{\ Define the number }$\beta =2\varepsilon
/\left( 3m+2\varepsilon \right) \in \left( 0,1\right) .$ \emph{Then there
exists a sufficiently small number }$\delta _{0}=\delta _{0}\left(
\varepsilon ,m,A,G_{c}\right) \in \left( 0,1\right) $\emph{\ and a constant }%
$C_{1}=C_{1}\left( \varepsilon ,m,A,G_{c}\right) >0$\emph{\ such that if }$%
\delta \in \left( 0,\delta _{0}\right) $\emph{, then the following H\"{o}%
lder stability estimate holds}%
\begin{equation*}
\left\Vert u\right\Vert _{H^{1}\left( G_{c+3\varepsilon }\right) }\leq
C_{1}\left( 1+\left\Vert u\right\Vert _{H^{1}\left( G_{c}\right) }\right)
\delta ^{\beta },\forall \delta \in \left( 0,\delta _{0}\right) .
\end{equation*}

\textbf{Proof}. Assume first that the function $u\in C^{2}\left( \overline{G}%
_{c}\right) .$ We have
\begin{equation*}
S^{2}e^{2\lambda m}\geq \int\limits_{G_{c}}\left( Au\right) ^{2}\varphi
_{\lambda }^{2}\left( x\right) dx\geq \int\limits_{G_{c}}\left(
A_{0}u\right) ^{2}\varphi _{\lambda }^{2}\left( x\right)
dx-C_{1}\int\limits_{G_{c}}\left( \left( \nabla u\right) ^{2}+u^{2}\right)
\varphi _{\lambda }^{2}\left( x\right) dx.
\end{equation*}%
This is equivalent with%
\begin{equation*}
S^{2}e^{2\lambda m}+C_{1}\int\limits_{G_{c}}\left( \left( \nabla u\right)
^{2}+u^{2}\right) \varphi _{\lambda }^{2}\left( x\right) dx\geq
\int\limits_{G_{c}}\left( A_{0}u\right) ^{2}\varphi _{\lambda }^{2}\left(
x\right) dx.
\end{equation*}%
The rest of the proof is similar with the proof of Theorem 2.1. The
replacement of $u\in C^{2}\left( \overline{G}_{c}\right) $ with $u\in
H^{2}\left( G_{c}\right) $ can be done via density arguments. $\square $

\subsection{Derivation of the Carleman estimate for a parabolic operator}

\label{sec:2.3}

The goal of this Section is to present an example of the derivation of the
Carleman estimate. To choose the case of a simplified parabolic operator.\
Our derivation method is similar with the one of \S 1 of Chapter 4 of the
book of Lavrent'ev, Romanov and Shishatskii \cite{LRS}.

For any $x\in \mathbb{R}^{n}$ denote $y=\left( x_{2},..,x_{n}\right) .$ Let
numbers $\alpha ,\eta \in \left( 0,1\right) $ and $\alpha <\eta .$ Let $%
Y,T>0 $ be two arbitrary numbers. Consider the function $\psi \left(
x,t\right) ,$%
\begin{equation}
\psi \left( x,t\right) =x_{1}+\frac{\left\vert y\right\vert ^{2}}{2Y^{2}}+%
\frac{t^{2}}{2T^{2}}+\alpha .  \label{2.16_0}
\end{equation}%
Define the domain $G_{\eta }$ as%
\begin{equation*}
G_{\eta }=\left\{ \left( x,t\right) :\psi \left( x,t\right) <\eta
,x_{1}>0\right\} =\left\{ x_{1}+\frac{\left\vert y\right\vert ^{2}}{2Y^{2}}+%
\frac{t^{2}}{2T^{2}}+\alpha <\eta ,x_{1}>0\right\} .
\end{equation*}%
Let $\lambda ,\nu >1$ be two large parameters which we will choose later.
Consider the function $\varphi _{\lambda ,\nu }\left( x,t\right) ,$%
\begin{equation}
\varphi \left( x,t\right) =\exp \left( \lambda \psi ^{-\nu }\right) .
\label{2.16}
\end{equation}%
$\varphi _{\lambda ,\nu }\left( x,t\right) $ is the CWF for our parabolic
operator $L$ (below). To simplify notations, we use the notation $\varphi
\left( x,t\right) $ instead of $\varphi _{\lambda ,\nu }\left( x,t\right) .$
Hence, the boundary of the domain $G_{\eta }$ consists of a piece of the
hyperplane $\left\{ x_{1}=0\right\} $ and a piece of the paraboloid $\left\{
\psi \left( x,t\right) =\eta ,x_{1}>0\right\} ,$%
\begin{eqnarray}
\partial G_{\eta } &=&\partial _{1}G_{\eta }\cup \partial _{2}G_{\eta },
\label{2.17} \\
\partial _{1}G_{\eta } &=&\left\{ x_{1}=0,\frac{\left\vert y\right\vert ^{2}%
}{2Y^{2}}+\frac{t^{2}}{2T^{2}}+\alpha <\eta \right\} ,\partial _{2}G_{\eta
}=\left\{ x_{1}>0,x_{1}+\frac{\left\vert y\right\vert ^{2}}{2Y^{2}}+\frac{%
t^{2}}{2T^{2}}+\alpha <\eta \right\} .  \label{2.18}
\end{eqnarray}

Consider a function $a\left( x,t\right) $ for $\left( x,t\right) \in G_{\eta
}$ such that
\begin{equation}
a\in C^{1}\left( \overline{G}_{\eta }\right) ,K=\left\Vert a\right\Vert
_{C^{1}\left( \overline{G}_{\eta }\right) },a\left( x,t\right) \geq
a_{0}=const.>0\text{ for }\left( x,t\right) \in G_{\eta }.  \label{2.19}
\end{equation}%
Consider the parabolic operator $L,$%
\begin{equation}
Lu=u_{t}-a\left( x,t\right) \Delta u,\left( x,t\right) \in G_{\eta }.
\label{2.20}
\end{equation}%
By Definition 2.1 we want to estimate now $\left( Lu\right) ^{2}\varphi ^{2}$
from the below. Introduce the new function $v=u\varphi $ and express
derivatives of the function $u$ via derivatives of the function $v$, using (%
\ref{2.16_0}) and (\ref{2.16}).\emph{\ }Below $O\left( 1/\lambda \right)
,O\left( 1/\nu \right) $ denote different $C^{1}\left( \overline{G}_{\eta
}\right) -$functions, which are independent on the function $u,$ and such
that $\left\vert O\left( 1/\lambda \right) \right\vert \leq C/\lambda
,\left\vert O\left( 1/\nu \right) \right\vert \leq C/\nu ,\forall \lambda
,\nu \geq 1.$ Here and below in this section $C=C\left( a_{0},K,G_{\eta
}\right) $ denotes different positive constants depending only on listed
parameters.

\textbf{Lemma 2.1}. \emph{Suppose that the function }$a\left( x,t\right) $%
\emph{\ satisfies conditions (\ref{2.19}). Then} \emph{there exist
sufficiently large numbers }$\lambda _{0}=\lambda _{0}\left( a_{0},K,G_{\eta
}\right) >1,\nu _{0}=\nu _{0}\left( a_{0},K,G_{\eta }\right) >2$\emph{\ such
that for any function }$u\in C^{2,1}\left( \overline{G}_{\eta }\right) $%
\emph{\ the following estimate holds for all }$\lambda \geq \lambda _{0},\nu
\geq \nu _{0},\left( x,t\right) \in G_{\eta }$
\begin{eqnarray}
\left( Lu\right) ^{2}\psi ^{\nu +2}\varphi ^{2} &\geq &-C\lambda \nu \left(
\nabla u\right) ^{2}\varphi ^{2}+C\lambda ^{3}\nu ^{4}\psi ^{-2\nu
-2}\varphi ^{2}+\func{div}U_{1}+\partial _{t}V_{1},  \label{2.21} \\
\left\vert U_{1}\right\vert +\left\vert V_{1}\right\vert &\leq &C\lambda
^{3}\nu ^{3}\psi ^{-2\nu -2}\left( \left( \nabla u\right) ^{2}+u^{2}\right)
\varphi ^{2}.  \label{2.22}
\end{eqnarray}

\textbf{Proof}. We have $u=v\exp \left( -\lambda \psi ^{-\nu }\right) .$
Hence,%
\begin{eqnarray*}
u_{t} &=&\left( v_{t}+\frac{t}{T^{2}}\lambda \nu \psi ^{-\nu -1}v\right)
\exp \left( -\lambda \psi ^{-\nu }\right) , \\
u_{x_{1}} &=&\left( v_{x_{1}}+\lambda \nu \psi ^{-\nu -1}v\right) \exp
\left( -\lambda \psi ^{-\nu }\right) , \\
u_{x_{1}x_{1}} &=&\left[ v_{x_{1}x_{1}}+2\lambda \nu \psi ^{-\nu
-1}v_{x_{1}}+\lambda ^{2}\nu ^{2}\psi ^{-2\nu -2}\left( 1+O\left( \frac{1}{%
\lambda }\right) \right) v\right] \exp \left( -\lambda \psi ^{-\nu }\right) ,
\\
u_{x_{i}} &=&\left( v_{x_{i}}+\frac{x_{i}}{Y^{2}}\lambda \nu \psi ^{-\nu
-1}v\right) \exp \left( -\lambda \psi ^{-\nu }\right) ,i\in \left[ 2,n\right]
, \\
u_{x_{i}x_{i}} &=&\left[ v_{x_{i}x_{i}}+2\frac{x_{i}}{Y^{2}}\lambda \nu \psi
^{-\nu -1}v_{x_{i}}+\lambda ^{2}\nu ^{2}\psi ^{-2\nu -2}\left( \frac{%
x_{i}^{2}}{Y^{4}}+O\left( \frac{1}{\lambda }\right) \right) v\right] \exp
\left( -\lambda \psi ^{-\nu }\right) ,i\in \left[ 2,n\right] .
\end{eqnarray*}%
These equalities imply that
\begin{equation*}
\left( Lu\right) ^{2}\psi ^{\nu +2}\varphi ^{2}=
\end{equation*}%
\begin{equation*}
\left\{ v_{t}-a\Delta v-\left[ 2a\lambda \nu \psi ^{-\nu
-1}v_{x_{1}}+2a\lambda \nu \psi ^{-\nu -1}\sum\limits_{i=2}^{n}\frac{x_{i}}{%
Y^{2}}v_{x_{i}}\right] -a\lambda ^{2}\nu ^{2}\psi ^{-2\nu -2}\left( \left[
1+\sum\limits_{i=2}^{n}\frac{x_{i}^{2}}{Y^{4}}+O\left( \frac{1}{\lambda }%
\right) \right] \right) v\right\} \psi ^{\nu +2}.
\end{equation*}%
Denote%
\begin{eqnarray}
z_{1} &=&v_{t},z_{2}=-a\Delta v,  \label{2.23} \\
z_{3} &=&-\left[ 2a\lambda \nu \psi ^{-\nu -1}v_{x_{1}}+2a\lambda \nu \psi
^{-\nu -1}\sum\limits_{i=2}^{n}\frac{x_{i}}{Y^{2}}v_{x_{i}}\right] ,
\label{2.24} \\
z_{4} &=&-a\lambda ^{2}\nu ^{2}\psi ^{-2\nu -2}\left[ 1+\sum\limits_{i=2}^{n}%
\frac{x_{i}^{2}}{Y^{4}}+O\left( \frac{1}{\lambda }\right) \right] v.
\label{2.25}
\end{eqnarray}%
Hence,
\begin{equation}
\left( Lu\right) ^{2}\psi ^{\nu +2}\varphi ^{2}\geq \left(
z_{1}^{2}+2z_{1}z_{2}+2z_{1}z_{3}+z_{3}^{2}+2z_{2}z_{3}+2z_{3}z_{4}+2z_{1}z_{4}\right) \psi ^{\nu +2}.
\label{2.26}
\end{equation}

\textbf{Step 1}. Estimate $2z_{1}z_{2}\psi ^{\nu +2}$ from the below. By (%
\ref{2.23})
\begin{eqnarray*}
2z_{1}z_{2}\psi ^{\nu +2} &=&-2\sum\limits_{i=1}^{n}v_{t}v_{x_{i}x_{i}}a\psi
^{\nu +2}=\sum\limits_{i=1}^{n}\left( -2v_{t}v_{x_{i}}a\psi ^{\nu +2}\right)
_{x_{i}} \\
&&+2\sum\limits_{i=1}^{n}v_{tx_{i}}v_{x_{i}}a\psi ^{\nu
+2}+2v_{t}\sum\limits_{i=1}^{n}v_{x_{i}}\left( a\psi ^{\nu +2}\right)
_{x_{i}} \\
&=&\sum\limits_{i=1}^{n}\left( -2v_{t}v_{x_{i}}a\psi ^{\nu +2}\right)
_{x_{i}}+\left[ \left( \nabla v\right) ^{2}a\psi ^{\nu +2}\right]
_{t}-\left( \nabla v\right) ^{2}\left( a\psi ^{\nu +2}\right)
_{t}+2v_{t}\sum\limits_{i=1}^{n}v_{x_{i}}\left( a\psi ^{\nu +2}\right)
_{x_{i}}.
\end{eqnarray*}%
Thus, since $\psi <1,$ then
\begin{eqnarray}
2z_{1}z_{2}\psi ^{\nu +2} &\geq &2v_{t}\sum\limits_{i=1}^{n}v_{x_{i}}\left(
a\psi ^{\nu +2}\right) _{x_{i}}-C\nu \left( \nabla v\right) ^{2}+\func{div}%
U_{1,1}+\left( V_{1,1}\right) _{t},  \label{2.27} \\
\left\vert U_{1,1}\right\vert +\left\vert V_{1,1}\right\vert &\leq &C\nu
\left( v_{t}^{2}+\left( \nabla v\right) ^{2}\right) .  \label{2.28}
\end{eqnarray}

\textbf{Step 2}. Estimate $\left(
z_{1}^{2}+2z_{1}z_{2}+2z_{1}z_{3}+z_{3}^{2}\right) \psi ^{\nu +2}.$ Using (%
\ref{2.23}), (\ref{2.27}) and (\ref{2.28}), we obtain%
\begin{equation*}
\left( z_{1}^{2}+2z_{1}z_{2}+2z_{1}z_{3}+z_{3}^{2}\right) \psi ^{\nu +2}\geq
\end{equation*}%
\begin{eqnarray*}
&&z_{1}^{2}+z_{3}^{2}+2z_{1}\left(
z_{3}+\sum\limits_{i=1}^{n}v_{x_{i}}\left( a\psi ^{\nu +2}\right)
_{x_{i}}\right) -C\nu \left( \nabla v\right) ^{2}+\func{div}U_{1,1}+\left(
V_{1,1}\right) _{t} \\
&\geq &z_{1}^{2}+z_{3}^{2}-z_{1}^{2}-\left(
z_{3}+\sum\limits_{i=1}^{n}v_{x_{i}}\left( a\psi ^{\nu +2}\right)
_{x_{i}}\right) ^{2}-C\nu \left( \nabla v\right) ^{2}+\func{div}%
U_{1,1}+\left( V_{1,1}\right) _{t} \\
&=&z_{3}^{2}-z_{3}^{2}-2z_{3}\sum\limits_{i=1}^{n}v_{x_{i}}\left( a\psi
^{\nu +2}\right) _{x_{i}}-\left( \sum\limits_{i=1}^{n}v_{x_{i}}\left( a\psi
^{\nu +2}\right) _{x_{i}}\right) ^{2}-C\nu \left( \nabla v\right) ^{2}+\func{%
div}U_{1,1}+\left( V_{1,1}\right) _{t}.
\end{eqnarray*}%
Using (\ref{2.24}), we obtain%
\begin{equation*}
-2z_{3}\sum\limits_{i=1}^{n}v_{x_{i}}\left( a\psi ^{\nu +2}\right)
_{x_{i}}=4a^{2}\lambda \nu \left( \nu +2\right) \left[ v_{x_{1}}+\sum%
\limits_{i=2}^{n}\frac{x_{i}}{Y^{2}}v_{x_{i}}\right] ^{2}-C\lambda \nu
\left( \nabla v\right) ^{2}\geq -C\lambda \nu \left( \nabla v\right) ^{2}.
\end{equation*}%
Thus,
\begin{equation}
\left( z_{1}^{2}+2z_{1}z_{2}+2z_{1}z_{3}+z_{3}^{2}\right) \psi ^{\nu +2}\geq
-C\lambda \nu \left( \nabla v\right) ^{2}+\func{div}U_{1,1}+\left(
V_{1,1}\right) _{t}.  \label{2.29}
\end{equation}

\textbf{Step 3}. Estimate $2z_{2}z_{3}\psi ^{\nu +2},$%
\begin{equation}
2z_{2}z_{3}\psi ^{\nu +2}=4a^{2}\lambda \nu \psi \left(
v_{x_{1}}+\sum\limits_{i=2}^{n}\frac{x_{i}}{Y^{2}}v_{x_{i}}\right) \Delta v.
\label{2.30}
\end{equation}%
Consider for example the term $4a^{2}\lambda \nu \psi v_{x_{1}}\Delta v,$%
\begin{eqnarray*}
4a^{2}\lambda \nu \psi v_{x_{1}}\Delta v &=&4\lambda \nu
\sum\limits_{i=1}^{n}v_{x_{i}x_{i}}v_{x_{1}}a^{2}\psi
=\sum\limits_{i=1}^{n}\left( 4\lambda \nu a^{2}\nu _{x_{i}}v_{x_{1}}\psi
\right) _{x_{i}} \\
&&-\sum\limits_{i=1}^{n}4\lambda \nu a^{2}\psi \nu
_{x_{i}}v_{x_{i}x_{1}}-\sum\limits_{i=1}^{n}4\lambda \nu \left( a^{2}\psi
\right) _{x_{i}}\nu _{x_{i}}v_{x_{1}} \\
&\geq &\sum\limits_{i=1}^{n}\left( 4\lambda \nu a^{2}\nu
_{x_{i}}v_{x_{1}}\psi \right) _{x_{i}}+\left( \sum\limits_{i=1}^{n}2\lambda
\nu a^{2}\psi \nu _{x_{i}}^{2}\right) _{x_{1}}-2\lambda \nu \left( a^{2}\psi
\right) _{x_{1}}\left( \nabla v\right) ^{2}-C\lambda \nu \left( \nabla
v\right) ^{2} \\
&\geq &-C\lambda \nu \left( \nabla v\right) ^{2}+\func{div}U_{1,2}.
\end{eqnarray*}%
Thus, $4a^{2}\lambda \nu \psi v_{x_{1}}\Delta v\geq -C\lambda \nu \left(
\nabla v\right) ^{2}+\func{div}U_{1,2}.$ Similarly we obtain using (\ref%
{2.30})%
\begin{equation}
2z_{2}z_{3}\psi ^{\nu +2}\geq -C\lambda \nu \left( \nabla v\right) ^{2}+%
\func{div}U_{1,3}.  \label{2.31}
\end{equation}

\textbf{Step 4}. Estimate $2z_{3}z_{4}\psi ^{\nu +2},$%
\begin{eqnarray*}
2z_{3}z_{4}\psi ^{\nu +2} &=&4a^{2}\lambda ^{3}\nu ^{3}\psi ^{-2\nu -1}\left[
1+\sum\limits_{i=2}^{n}\frac{x_{i}^{2}}{Y^{4}}+O\left( \frac{1}{\lambda }%
\right) \right] \left[ v_{x_{1}}+\sum\limits_{i=2}^{n}\frac{x_{i}}{Y^{2}}%
v_{x_{i}}\right] v \\
&=&\left[ 2a^{2}\lambda ^{3}\nu ^{3}\psi ^{-2\nu -1}\left(
1+\sum\limits_{i=2}^{n}\frac{x_{i}^{2}}{Y^{4}}+O\left( \frac{1}{\lambda }%
\right) \right) v^{2}\right] _{x_{1}} \\
&&+\left[ 2a^{2}\lambda ^{3}\nu ^{3}\psi ^{-2\nu -1}\left(
1+\sum\limits_{j=2}^{n}\frac{x_{j}^{2}}{Y^{4}}+O\left( \frac{1}{\lambda }%
\right) \right) \left( \sum\limits_{i=2}^{n}\frac{x_{i}}{Y^{2}}\right) v^{2}%
\right] _{x_{i}} \\
&&+2a^{2}\lambda ^{3}\nu ^{3}\left( 2\nu +1\right) \psi ^{-2\nu -2}\left[
1+\left( \sum\limits_{i=2}^{n}\frac{x_{i}}{Y^{2}}\right) ^{2}+O\left( \frac{1%
}{\lambda }\right) +O\left( \frac{1}{\nu }\right) \right] v^{2}.
\end{eqnarray*}%
Thus,%
\begin{equation}
2z_{3}z_{4}\psi ^{\nu +2}\geq C\lambda ^{3}\nu ^{4}\psi ^{-2\nu -2}v^{2}+%
\func{div}U_{1,4}.  \label{2.32}
\end{equation}%
Similarly,
\begin{equation}
2z_{1}z_{4}\psi ^{\nu +2}\geq -C\lambda ^{2}\nu ^{3}\psi ^{-\nu -1}v^{2}+%
\func{div}U_{1,5}+\left( V_{1,2}\right) _{t}.  \label{2.33}
\end{equation}%
Summing up (\ref{2.29})-(\ref{2.33}) and replacing $v$ with $u=v\varphi
^{-1},$ we obtain (\ref{2.21}), (\ref{2.22}). $\square $

As one can see that we have one positive and one negative term in the right
hand side in estimate (\ref{2.21}). Therefore, we need to balance them
somehow to obtain only positive terms. To do this, we prove Lemma 2.2 first.

\textbf{Lemma 2.2}. \emph{Suppose that the function }$a\left( x,t\right) $%
\emph{\ satisfies conditions (\ref{2.19}). Then there exist sufficiently
large numbers }$\lambda _{0}=\lambda _{0}\left( a_{0},K,G_{\eta }\right)
>1,\nu _{0}=\nu _{0}\left( a_{0},K,G_{\eta }\right) >2$\emph{\ such that for
any function }$u\in C^{2,1}\left( \overline{G}_{\eta }\right) $\emph{\ the
following estimate holds for all }$\lambda \geq \lambda _{0},\nu \geq \nu
_{0},\left( x,t\right) \in G_{\eta }$%
\begin{eqnarray}
\left( u_{t}-a\left( x,t\right) \Delta u\right) u\varphi ^{2} &\geq &C\left(
\nabla u\right) ^{2}\varphi ^{2}-C\lambda ^{2}\nu ^{2}\psi ^{-2\nu
-2}\varphi ^{2}u^{2}+\func{div}U_{2}+\partial _{t}V_{2},  \label{2.34} \\
\left\vert U_{2}\right\vert +\left\vert V_{2}\right\vert &\leq &C\left(
\left\vert \nabla u\right\vert ^{2}+u^{2}\right) \varphi ^{2}.  \label{2.35}
\end{eqnarray}

\textbf{Proof}. We have%
\begin{eqnarray*}
\left( u_{t}-a\left( x,t\right) \Delta u\right) u\varphi ^{2} &=&\left(
\frac{1}{2}u^{2}\varphi ^{2}\right) _{t}+\frac{t}{2T^{2}}\lambda \nu \psi
^{-\nu -1}\varphi ^{2}u^{2}+\sum\limits_{i=1}^{n}\left( -a\left( x,t\right)
u_{x_{i}}u\varphi ^{2}\right) _{x_{i}}+a\left( \nabla u\right) ^{2}\varphi
^{2} \\
&&-2a\lambda \nu \sum\limits_{i=1}^{n}\psi _{x_{i}}\psi ^{-\nu
-1}u_{x_{i}}u\varphi ^{2}+\sum\limits_{i=1}^{n}a_{x_{i}}u_{x_{i}}u\varphi
^{2} \\
&\geq &C\left( \nabla u\right) ^{2}\varphi ^{2}-C\lambda ^{2}\nu ^{2}\psi
^{-2\nu -2}\varphi ^{2}u^{2}+\func{div}U_{2}+V_{2t}.\text{ }\square
\end{eqnarray*}

\textbf{Theorem 2.4}. \emph{Suppose that the function }$a\left( x,t\right) $%
\emph{\ satisfies conditions (\ref{2.19}). Then there exist sufficiently
large numbers }$\lambda _{0}=\lambda _{0}\left( a_{0},K,G_{\eta }\right)
>1,\nu _{0}=\nu _{0}\left( a_{0},K,G_{\eta }\right) >2$\emph{\ such that for
any function }$u\in C^{2,1}\left( \overline{G}_{\eta }\right) $\emph{\ the
following Carleman estimate holds for all }$\lambda \geq \lambda _{0},\nu
\geq \nu _{0},\left( x,t\right) \in G_{\eta }$%
\begin{eqnarray}
\left( Lu\right) ^{2}\varphi ^{2} &\geq &C\lambda \nu \left( \nabla u\right)
^{2}\varphi ^{2}+C\lambda ^{3}\nu ^{4}\psi ^{-2\nu -2}\varphi ^{2}u^{2}+%
\func{div}U+V_{t},  \label{2.36} \\
\left\vert U\right\vert +\left\vert V\right\vert &\leq &C\lambda ^{3}\nu
^{3}\psi ^{-2\nu -2}\left( \left( \nabla u\right) ^{2}+u^{2}\right) \varphi
^{2}.  \label{2.37}
\end{eqnarray}

\textbf{Proof}. Multiply (\ref{2.34}) and (\ref{2.35}) by $2\lambda \nu $
and sum up with (\ref{2.21}), (\ref{2.22}). We obtain%
\begin{equation}
\left( Lu\right) ^{2}\psi ^{\nu +2}\varphi ^{2}+2C\lambda \nu \left(
Lu\right) u\varphi ^{2}\geq C\lambda \nu \left( \nabla u\right) ^{2}\varphi
^{2}+C\lambda ^{3}\nu ^{4}\psi ^{-2\nu -2}\left( 1-\frac{1}{\nu }\right)
\varphi ^{2}u^{2}+\func{div}U+V_{t},  \label{2.38}
\end{equation}%
where the vector function $\left( U,V\right) $ satisfies (\ref{2.37}).
Choose $\nu _{0}=\nu _{0}\left( a_{0},K,G_{\eta }\right) >2.$ Also, since $%
\psi ^{\nu +2}<1,$ we have
\begin{equation*}
\left( Lu\right) ^{2}\psi ^{\nu +2}\varphi ^{2}+2\lambda \nu \left(
Lu\right) u\varphi ^{2}\leq 2\left( Lu\right) ^{2}\varphi ^{2}+\lambda
^{2}\nu ^{2}\varphi ^{2}u^{2}.
\end{equation*}%
Combining this with (\ref{2.38}), we obtain (\ref{2.36}). $\square $

We now want to incorporate higher order derivatives $u_{t},u_{x_{i}x_{j}}$
in the Carleman estimate of Theorem 2.3. To do this, we prove Lemma 2.4 and
Theorem 2.4.

\textbf{Lemma 2.4}. \emph{Suppose that the function }$a\left( x,t\right) $%
\emph{\ satisfies conditions (\ref{2.19}). Fix the number }$\nu :=\nu
_{0}\left( a_{0},K,G_{\eta }\right) >2$\emph{\ of Theorem 2.3. There exists
sufficiently large number }$\lambda _{0}=\lambda _{0}\left( a_{0},K,G_{\eta
}\right) >1$\emph{\ such that for any function }$u\in C^{3}\left( \overline{G%
}_{\eta }\right) $\emph{\ the following estimate holds for all }$\lambda
\geq \lambda _{0},\left( x,t\right) \in G_{\eta }$%
\begin{eqnarray}
\left( Lu\right) ^{2}\varphi ^{2} &\geq &\frac{1}{2}\left(
u_{t}^{2}+\sum\limits_{i,j=1}^{n}u_{x_{i}x_{j}}^{2}\right) \varphi
^{2}-C\lambda ^{2}\left( \nabla u\right) ^{2}\varphi ^{2}+\func{div}%
U_{3}+\partial _{t}V_{3},  \label{2.39} \\
\left\vert U_{3}\right\vert +\left\vert V_{3}\right\vert &\leq &C\left(
u_{t}^{2}+\sum\limits_{i,j=1}^{n}u_{x_{i}x_{j}}^{2}+\left( \nabla u\right)
^{2}\right) \varphi ^{2}.  \label{2.40}
\end{eqnarray}

\textbf{Proof}. We have%
\begin{equation}
\left( Lu\right) ^{2}\varphi ^{2}\geq \left( u_{t}^{2}+2au_{t}\Delta
u+a_{0}^{2}\left( \Delta u\right) ^{2}\right) \varphi ^{2}.  \label{2.41}
\end{equation}

\textbf{Step 1}. Estimate $2au_{t}\Delta u\varphi ^{2},$%
\begin{eqnarray*}
2au_{t}\Delta u\varphi ^{2}
&=&\sum\limits_{i=1}^{n}2au_{t}u_{x_{i}x_{i}}\varphi
^{2}=\sum\limits_{i=1}^{n}\left( 2au_{t}u_{x_{i}}\varphi ^{2}\right)
_{x_{i}}-2\sum\limits_{i=1}^{n}au_{tx_{i}}u_{x_{i}}\varphi ^{2} \\
&&-2\sum\limits_{i=1}^{n}a_{x_{i}}u_{t}u_{x_{i}}\varphi ^{2}-4\lambda \nu
\psi ^{-\nu -1}\left( \psi \right)
_{x_{i}}\sum\limits_{i=1}^{n}au_{t}u_{x_{i}}\varphi ^{2}.
\end{eqnarray*}%
We have
\begin{equation*}
-2\sum\limits_{i=1}^{n}au_{tx_{i}}u_{x_{i}}\varphi ^{2}=\left( -a\left(
\nabla u\right) ^{2}\varphi ^{2}\right) _{t}-4\lambda \nu \frac{t}{T^{2}}%
\psi ^{-\nu -1}a\left( \nabla u\right) ^{2}\varphi ^{2}+a_{t}\left( \nabla
u\right) ^{2}\varphi ^{2}.
\end{equation*}%
Since the number $\nu :=\nu _{0}\left( a_{0},K,G_{\eta }\right) $ depends on
the same parameters as the constant $C,$ we can incorporate $\nu $ in $C$.
Hence, we obtain%
\begin{equation}
2au_{t}\Delta u\varphi ^{2}\geq -u_{t}^{2}\varphi ^{2}-C\lambda ^{2}\left(
\nabla u\right) ^{2}\varphi ^{2}+\func{div}U_{3,1}+\left( V_{3,1}\right)
_{t}.  \label{2.42}
\end{equation}

\textbf{Step 2}. Estimate $\left( \Delta u\right) ^{2}\varphi ^{2},$%
\begin{eqnarray*}
\left( \Delta u\right) ^{2}\varphi ^{2}
&=&\sum\limits_{i,j=1}^{n}u_{x_{i}x_{i}}u_{x_{j}x_{j}}\varphi
^{2}=\sum\limits_{j=1}^{n}\left(
\sum\limits_{i=1}^{n}u_{x_{i}x_{i}}u_{x_{j}}\varphi ^{2}\right) _{x_{j}} \\
&&-\sum\limits_{i=1}^{n}\sum\limits_{j=1}^{n}u_{x_{i}x_{i}x_{j}}u_{x_{j}}%
\varphi ^{2}+2\lambda \nu \psi ^{-\nu
-1}\sum\limits_{i,j=1}^{n}u_{x_{i}x_{i}}u_{x_{j}}\psi _{x_{j}}\varphi ^{2} \\
&\geq &\sum\limits_{i=1}^{n}\left(
-\sum\limits_{j=1}^{n}u_{x_{i}x_{j}}u_{x_{j}}\varphi ^{2}\right) _{x_{i}}+%
\frac{1}{2}\sum\limits_{i,j=1}^{n}u_{x_{i}x_{j}}^{2}\varphi ^{2}-C\lambda
^{2}\left( \nabla u\right) ^{2}\varphi ^{2} \\
+\sum\limits_{j=1}^{n}\left(
\sum\limits_{i=1}^{n}u_{x_{i}x_{i}}u_{x_{j}}\varphi ^{2}\right) _{x_{j}} &=&%
\frac{1}{2}\sum\limits_{i,j=1}^{n}u_{x_{i}x_{j}}^{2}\varphi ^{2}-C\lambda
^{2}\left( \nabla u\right) ^{2}\varphi ^{2}+\func{div}U_{3,2},
\end{eqnarray*}%
Hence, we have obtained that
\begin{eqnarray}
\left( \Delta u\right) ^{2}\varphi ^{2} &\geq &\frac{1}{2}%
\sum\limits_{i,j=1}^{n}u_{x_{i}x_{j}}^{2}\varphi ^{2}-C\lambda ^{2}\left(
\nabla u\right) ^{2}\varphi ^{2}+\func{div}U_{3,2},  \label{2.43} \\
\left\vert U_{3,2}\right\vert &\leq &C\left(
\sum\limits_{i,j=1}^{n}u_{x_{i}x_{j}}^{2}+\left( \nabla u\right) ^{2}\right)
\varphi ^{2}.  \label{2.44}
\end{eqnarray}%
Comparing (\ref{2.41}), (\ref{2.42}), (\ref{2.43}) and (\ref{2.44}), we
obtain (\ref{2.39}) and (\ref{2.40}). $\square $

\textbf{Theorem 2.5.} \emph{Suppose that the function }$a\left( x,t\right) $%
\emph{\ satisfies conditions (\ref{2.19}). Fix the number }$\nu :=\nu
_{0}\left( a_{0},K,G_{\eta }\right) >2$\emph{\ of Theorem 2.3. Then there
exists a sufficiently large number }$\lambda _{0}=\lambda _{0}\left(
a_{0},K,G_{\eta }\right) >1$\emph{\ such that for any function }$u\in
C^{3}\left( \overline{G}_{\eta }\right) $\emph{\ the following estimate
holds for all }$\lambda \geq \lambda _{0},\left( x,t\right) \in G_{\eta }$%
\begin{eqnarray}
\left( Lu\right) ^{2}\varphi ^{2} &\geq &\frac{C_{0}}{\lambda }\left(
u_{t}^{2}+\sum\limits_{i,j=1}^{n}u_{x_{i}x_{j}}^{2}\right) \varphi
^{2}+C\lambda \left( \nabla u\right) ^{2}\varphi ^{2}+C\lambda
^{3}u^{2}\varphi ^{2}+\func{div}U+V_{t},  \label{2.45} \\
\left\vert U\right\vert +\left\vert V\right\vert &\leq &\frac{C_{0}}{\lambda
}\left( u_{t}^{2}+\sum\limits_{i,j=1}^{n}u_{x_{i}x_{j}}^{2}\right) \varphi
^{2}+C\lambda ^{3}\left( \left( \nabla u\right) ^{2}+u^{2}\right) \varphi
^{2},  \label{2.46}
\end{eqnarray}%
\emph{where the constant }$C_{0}=C_{0}\left( a_{0},K,G_{\eta }\right) >0$%
\emph{\ depends on the same parameters as the constant }$C$\emph{\ above. }

\textbf{Proof}. Divide inequality (\ref{2.39}) by $b\lambda $ with an
appropriate constant $b=b\left( a_{0},K,G_{\eta }\right) >1.$ Next, add the
resulting inequality to (\ref{2.36}), where set $\nu :=\nu _{0}.$ Then we
obtain (\ref{2.45}), (\ref{2.46}). $\square $

\textbf{Remark 2.2}. If one would set above $u_{t}\equiv 0$ and would ignore
the term $t^{2}/T^{2}$ in the function $\psi ,$ then one would obtain
analogs of these results for the general elliptic operator. A close analog
of Theorem 2.4 is valid for a general parabolic operator of the second
order, see \S 1 of Chapter 4 of the book of Lavrent'ev, Romanov and
Shishatskii \cite{LRS}. Therefore, Theorems 2.1-2.3 are valid for this
operator.

\subsection{Carleman estimate for a hyperbolic operator}

\label{sec:2.4}

Theorem 2.5 is proven in the book of Beilina and Klibanov \cite{BK}. Similar
theorems were established in earlier books of Isakov \cite{Is0,Is} and
Klibanov and Timonov \cite{KT}. Theorem 2.6 can be found in \S 4 of Chapter
4 of the book of Lavrent'ev, Romanov and Shishatskii \cite{LRS}. We do not
reproduce proofs here for brevity.

For brevity we consider a simple domain $\Omega =\left\{ \left\vert
x\right\vert <R\right\} \subset \mathbb{R}^{n}.$ Let $T=const.>0.$ Denote
\begin{equation*}
Q_{T}=\Omega \times \left( 0,T\right) ,Q_{T}^{\pm }=\Omega \times \left(
-T,T\right) ,S_{T}=\partial \Omega \times \left( 0,T\right) ,S_{T}^{\pm
}=\partial \Omega \times \left( -T,T\right) .
\end{equation*}%
Choose a point $x_{0}\in \mathbb{R}^{n}$. In particular, we can have $%
x_{0}\in \Omega .$ Let the number $\eta \in \left( 0,1\right) .$ Let $%
\lambda >1$ be a large parameter$.$ Define functions $\xi \left( x,t\right)
,\varphi \left( x,t\right) $ as%
\begin{equation}
\xi \left( x,t\right) =\left\vert x-x_{0}\right\vert ^{2}-\eta t^{2},\varphi
\left( x,t\right) =\exp \left[ \lambda \xi \left( x,t\right) \right] .
\label{2.47}
\end{equation}%
For a number $\gamma >0$ define the domain $G_{\gamma }$ as%
\begin{equation}
G_{\gamma }=\left\{ \xi \left( x,t\right) >\gamma \right\} .  \label{2.471}
\end{equation}%
One can choose $\gamma $ such that%
\begin{equation}
G_{\gamma }\subset Q_{T}^{\pm }.  \label{2.472}
\end{equation}

\textbf{Theorem 2.5.} \emph{Let }$\Omega =\left\{ \left\vert x\right\vert
<R\right\} \subset \mathbb{R}^{n},n\geq 2.$\emph{\ Let conditions (\ref{2.47}%
), (\ref{2.48}) and (\ref{2.49}) be in place. Let }$d=const.\geq 1.$ \emph{%
Let} \emph{the function }$c\left( x\right) $\emph{\ satisfies the following
conditions}%
\begin{eqnarray}
c^{-2}\left( x\right) &\in &\left[ 1,d\right] ,\forall x\in \overline{\Omega
},c\in C^{1}\left( \overline{\Omega }\right) ,  \label{2.48} \\
\left( x-x_{0},\nabla c^{-2}\left( x\right) \right) &\geq &0,\text{ }\forall
x\in \overline{\Omega },  \label{2.49}
\end{eqnarray}%
\emph{for a certain point }$x_{0}\in \Omega ,$ \emph{where }$\left( \cdot
,\cdot \right) $ \emph{denotes the scalar product in }$\mathbb{R}^{n}$\emph{%
. Let }$Lu=u_{tt}-c^{2}\left( x\right) \Delta u$\emph{\ be a hyperbolic
operator. Let}
\begin{equation}
P=P\left( x_{0},\Omega \right) =\max_{x\in \overline{\Omega }}\left\vert
x-x_{0}\right\vert .  \label{2.49_1}
\end{equation}%
\emph{Then there exists a number }$\eta _{0}=\eta _{0}\left( \Omega
,d,P,\left\Vert \nabla c\right\Vert _{C\left( \overline{\Omega }\right)
}\right) \in \left( 0,1\right) $ \emph{such that for any }$\eta \in \left(
0,\eta _{0}\right) $\emph{\ one can choose a sufficiently large number }$%
\lambda _{0}=\lambda _{0}\left( d,P,\left\Vert \nabla c\right\Vert _{C\left(
\overline{\Omega }\right) },\eta ,\gamma \right) >1$\emph{\ and the number }$%
C=C\left( d,P,\left\Vert \nabla c\right\Vert _{C\left( \overline{\Omega }%
\right) },\eta ,\gamma \right) >0$\emph{, such that for all }$u\in
C^{2}\left( \overline{G}_{\gamma }\right) $\emph{\ and for all }$\lambda
\geq \lambda _{0}$\emph{\ the following pointwise Carleman estimate holds}
\begin{equation*}
\left( Lu\right) ^{2}\varphi ^{2}\geq C\lambda \left( \left\vert \nabla
u\right\vert ^{2}+u_{t}^{2}\right) \varphi ^{2}+\lambda ^{3}u^{2}\varphi
^{2}+\func{div}U+V_{t}\text{, \emph{in} }G_{\gamma },
\end{equation*}%
\emph{where }
\begin{eqnarray}
\left\vert U\right\vert &\leq &C\lambda ^{3}\left( \left\vert \nabla
u\right\vert ^{2}+u_{t}^{2}+u^{2}\right) \varphi ^{2},  \label{2.50} \\
\left\vert V\right\vert &\leq &C\lambda ^{3}\left[ \left\vert t\right\vert
\left( u_{t}^{2}+\left\vert \nabla u\right\vert ^{2}+u^{2}\right) +\left(
\left\vert \nabla u\right\vert +\left\vert u\right\vert \right) \left\vert
u_{t}\right\vert \right] \varphi ^{2}.  \label{2.51}
\end{eqnarray}%
\emph{\ } \emph{In particular, (\ref{2.51}) implies that if either }$u\left(
x,0\right) =0$\emph{\ or }$u_{t}\left( x,0\right) =0,$ \emph{then }
\begin{equation}
V\left( x,0\right) =0.  \label{2.52}
\end{equation}

\textbf{Theorem 2.6}. \emph{Let }$c\left( x\right) \equiv 1$\emph{. Then
Theorem 2.5 is valid for any }$\eta \in \left( 0,1\right) .$

\subsection{The Quasi-Reversibility Method (QRM)}

\label{sec:2.5}

In this section we use notations of Sections 2.1, 2.2. Let $A$ be the
operator of Section 2.1. QRM delivers an approximate solution of the
following Cauchy problem%
\begin{eqnarray}
Au &=&f,x\in G_{c},  \label{2.500} \\
f &\in &L_{2}\left( G_{c}\right) ,u\in H_{0}^{2}\left( G_{c}\right) =\left\{
u\in H^{2}\left( G_{c}\right) :u\mid _{\Gamma _{c}}=\partial _{n}u\mid
_{\Gamma _{c}}=0\right\} .  \label{2.501}
\end{eqnarray}%
To find that approximate solution, QRM minimizes the following Tikhonov
functional with the regularization parameter $\gamma $%
\begin{equation}
J_{\gamma }\left( u\right) =\left\Vert Au-f\right\Vert _{L_{2}\left(
G_{c}\right) }^{2}+\gamma \left\Vert u\right\Vert _{H^{2}\left( G_{c}\right)
}^{2},u\in H_{0}^{2}\left( G_{c}\right) .  \label{2.502}
\end{equation}%
The variational principle implies that any minimizer $u_{\gamma }\in
H_{0}^{2}\left( G_{c}\right) $ satisfies the following integral identity%
\begin{equation}
\left( Au_{\gamma },Av\right) +\gamma \left[ u,v\right] =\left( f,Av\right)
,\forall v\in H_{0}^{2}\left( G_{c}\right) ,  \label{2.503}
\end{equation}%
where $\left( ,\right) $ and $\left[ ,\right] $ are scalar products in $%
L_{2}\left( G_{c}\right) $ and $H^{2}\left( G_{c}\right) $ respectively.
Riesz theorem and (\ref{2.503}) imply Lemma 2.5.

\textbf{Lemma 2.5}. \emph{Let }$A$\emph{\ be the operator defined in (\ref%
{2.4}), (\ref{2.5}) and the function }$f\in L_{2}\left( G_{c}\right) $\emph{%
. Then for any }$\gamma >0$\emph{\ there exists unique minimizer }$u\in
H_{0}^{2}\left( G_{c}\right) $\emph{\ of the functional (\ref{2.502}).\
Furthermore, with a constant }$\overline{C}=\overline{C}\left( A\right) $%
\emph{\ the following estimate holds }$\left\Vert u_{\gamma }\right\Vert
_{H^{2}\left( G_{c}\right) }\leq \overline{C}/\sqrt{\gamma }.$

\textbf{Theorem 2.7} (convergence). \emph{Assume that conditions (\ref{2.5})
hold and that the Carleman estimate of Definition 2.1 is valid. Suppose that
there exists a sufficiently small number }$\varepsilon >0$\emph{\ such that
the domain }$G_{c+3\varepsilon }\neq \varnothing .$\emph{\ Let the function }%
$u^{\ast }$\emph{\ be the exact solution of the problem (\ref{2.500}), (\ref%
{2.501}) with the exact function }$f^{\ast }\in L_{2}\left( G_{c}\right) .$%
\emph{\ Let }$\left\Vert f-f^{\ast }\right\Vert _{L_{2}\left( G_{c}\right)
}\leq \delta $\emph{\ and }$\gamma =\delta ^{2}.$\emph{\ Denote }$m=\max_{%
\overline{G}_{c}}\xi \left( x\right) .$\emph{\ Define the number }$\beta
=2\varepsilon /\left( 3m+2\varepsilon \right) \in \left( 0,1\right) .$ \emph{%
There exists a sufficiently small number }$\delta _{0}=\delta _{0}\left(
\varepsilon ,m,A,G_{c},\left\Vert u^{\ast }\right\Vert _{H^{2}\left(
G_{c}\right) }\right) \in \left( 0,1\right) $\emph{\ and a constant }$%
C_{1}=C_{1}\left( \varepsilon ,m,A,G_{c}\right) >0$\emph{\ such that if }$%
\delta \in \left( 0,\delta _{0}\right) $\emph{, then the following
convergence rate is valid }%
\begin{equation*}
\left\Vert u_{\gamma }-u^{\ast }\right\Vert _{H^{1}\left( G_{c+3\varepsilon
}\right) }\leq C_{1}\left( 1+\left\Vert u^{\ast }\right\Vert _{H^{2}\left(
G_{c}\right) }\right) \delta ^{\beta /2},\forall \delta \in \left( 0,\delta
_{0}\right) .
\end{equation*}

\textbf{Proof}. We have
\begin{equation*}
\left( Au^{\ast },Av\right) +\gamma \left[ u^{\ast },v\right] =\left(
f^{\ast },Av\right) +\gamma \left[ u^{\ast },v\right] ,\forall v\in
H_{0}^{2}\left( G_{c}\right) .
\end{equation*}%
Subtract this identity from (\ref{2.503}) and denote $w_{\gamma }=u_{\gamma
}-u^{\ast },g=f-f^{\ast }.$ We obtain%
\begin{equation}
\left( Aw_{\gamma },Av\right) +\gamma \left[ w_{\gamma },v\right] =\left(
g,Av\right) -\gamma \left[ u^{\ast },v\right] ,\forall v\in H_{0}^{2}\left(
G_{c}\right) .  \label{2.504}
\end{equation}%
Setting in (\ref{2.504}) $v:=w_{\gamma }$ and using Cauchy-Bunyakovsky
inequality, we obtain%
\begin{eqnarray*}
\left\Vert Aw_{\gamma }\right\Vert _{L_{2}\left( G_{c}\right) }^{2} &\leq
&\left( 1+\left\Vert u^{\ast }\right\Vert _{H^{2}\left( G_{c}\right)
}^{2}\right) \delta ^{2}\leq \delta , \\
\left\Vert w_{\gamma }\right\Vert _{H^{2}\left( G_{c}\right) }^{2} &\leq
&1+\left\Vert u^{\ast }\right\Vert _{H^{2}\left( G_{c}\right) }^{2}.
\end{eqnarray*}%
The rest of the proof follows from Theorem 2.3. $\square $

\subsection{Published results about QRM}

\label{sec:2.6}

As it is clear from (\ref{2.502}), QRM is a special form of the Tikhonov
regularization functional, see \cite{BKok,BK,EHN,T} for the theory of this
functional. In \textquotedblleft conventional" Tikhonov functional the
originating operator is continuous. On the other hand in QRM the originating
operator is a PDE operator, which is continuous only if its domain is $H^{2}$%
. QRM is well suitable for providing approximate solutions for ill-posed
Cauchy problems for PDEs, including boundary value problems with
over-determined boundary conditions. QRM was first introduced by Lattes and
Lions in their book \cite{LL}. This book shows how to apply QRM to ill-posed
Cauchy problems for all three main types of PDE operators of the second
order: elliptic, parabolic and hyperbolic. Although convergence theorems
were proven in \cite{LL}, convergence rates for QRM were not established
there.

The first work where a Carleman estimate was applied to get convergence rate
of QRM was one of Klibanov and Santosa \cite{KlibSant}. In \cite{KlibSant}
QRM was applied to the Cauchy problem for the Laplace equation. In Chapter 2
of the book of Klibanov and Timonov \cite{KT} Carleman estimates were also
applied to establish convergence rate of QRM for ill-posed Cauchy problems
for elliptic, parabolic and hyperbolic PDEs. Next, Carleman estimates were
used to prove convergence rate of QRM for the Cauchy problem for the Laplace
equation by Bourgeois \cite{Bourg1}, Bourgeois and Darde \cite{Bourg2} and
Cao, Klibanov and Pereverzev \cite{Cao}. The QRM for the problem of
determining of the initial condition in the parabolic PDE from boundary
measurements was considered by the author in \cite{Kl1,Kltherm}. Papers \cite%
{Bourg2,Cao,KlibSant} contain numerical results. As to the application of
QRM to the problem with the lateral Cauchy data for the hyperbolic PDE, see
Sections 5.4 and 5.5.

While above citations of QRM are concerned only with linear problems, it was
recently applied by the author with coauthors to solve MCIPs with
backscattering data via the approximately globally convergent method \cite%
{KuzhKl,KPK,KBK,KBKSNF,IEEE}, also see chapter 6 of \cite{BK}. The main
difference between the latter application of the QRM and the conventional
one is that MCIPs are nonlinear.

\section{The Bukhgeim-Klibanov Method}

\label{sec:3}

\subsection{Estimating an integral}

\label{sec:3.1}

First, we estimate a Volterra-like integral with a weight function. For the
first time an analog of Lemma 3.1 was proven by the author in \cite{Klib1}.
Next, that proof was published in some of above cited follow up papers of
the author about BK. The estimate of this lemma with the parameter $%
1/\lambda $ in it was first published in the book of Klibanov and Timonov
\cite{KT}, also see Section 1.10.3 in the book \cite{BK}.

\textbf{Lemma 3.1.} \emph{Let the function }$\varphi \in C^{1}\left[ 0,a%
\right] $\emph{\ and }$\varphi ^{\prime }\left( t\right) \leq -b$\emph{\ in }%
$\left[ 0,a\right] $\emph{, where }$b=const>0$\emph{. For a function }$g\in
L_{2}\left( -a,a\right) $\emph{\ consider the integral}
\begin{equation*}
I\left( g,\lambda \right) =\int\nolimits_{-a}^{a}\left(
\int\nolimits_{0}^{t}g\left( \tau \right) d\tau \right) ^{2}\exp \left[
2\lambda \varphi \left( t^{2}\right) \right] dt,\lambda =const.>0.
\end{equation*}%
\emph{Then, }
\begin{equation*}
I\left( g,\lambda \right) \leq \frac{1}{4\lambda b}\int%
\nolimits_{-a}^{a}g^{2}\left( t\right) \exp \left[ 2\lambda \varphi \left(
t^{2}\right) \right] dt.
\end{equation*}

\textbf{Proof.} We have for $t>0$%
\begin{equation*}
t\exp \left[ 2\lambda \varphi \left( t^{2}\right) \right] =t\frac{4\lambda
\varphi ^{\prime }\left( t^{2}\right) }{4\lambda \varphi ^{\prime }\left(
t^{2}\right) }\exp \left[ 2\lambda \varphi \left( t^{2}\right) \right]
\end{equation*}%
\begin{equation*}
=\frac{1}{4\lambda \varphi ^{\prime }\left( t^{2}\right) }\frac{d}{dt}%
\left\{ \exp \left[ 2\lambda \varphi \left( t^{2}\right) \right] \right\} =-%
\frac{1}{4\lambda \varphi ^{\prime }\left( t^{2}\right) }\frac{d}{dt}\left\{
-\exp \left[ 2\lambda \varphi \left( t^{2}\right) \right] \right\}
\end{equation*}%
\begin{equation*}
\leq \frac{1}{4\lambda b}\frac{d}{dt}\left\{ -\exp \left[ 2\lambda \varphi
\left( t^{2}\right) \right] \right\} .
\end{equation*}%
Hence,
\begin{equation*}
\int_{0}^{a}\left( \int_{0}^{t}g\left( \tau \right) d\tau \right) ^{2}\exp
\left( 2\lambda \varphi \left( t^{2}\right) \right) dt\leq \int_{0}^{a}\exp
\left( 2\lambda \varphi \left( t^{2}\right) \right) t\left(
\int_{0}^{t}g^{2}\left( \tau \right) d\tau \right) dt
\end{equation*}%
\begin{equation*}
\leq \frac{1}{4\lambda b}\int_{0}^{a}\frac{d}{dt}\left[ -\exp \left(
2\lambda \varphi \left( t^{2}\right) \right) \right] \left(
\int_{0}^{t}g^{2}\left( \tau \right) d\tau \right) dt
\end{equation*}%
\begin{equation*}
=-\frac{1}{4\lambda b}\exp \left( 2\lambda \varphi \left( a^{2}\right)
\right) \int_{0}^{a}g^{2}\left( \tau \right) d\tau +\frac{1}{4\lambda b}%
\int_{0}^{a}g^{2}\left( \tau \right) \exp \left( 2\lambda \varphi \left(
t^{2}\right) \right) dt
\end{equation*}%
\begin{equation*}
\leq \frac{1}{4\lambda b}\int_{0}^{a}g^{2}\left( \tau \right) \exp \left(
2\lambda \varphi \left( t^{2}\right) \right) dt.
\end{equation*}%
Thus, we have proved that
\begin{equation*}
\int_{0}^{a}\exp \left( 2\lambda \varphi \left( t^{2}\right) \right) \left(
\int_{0}^{t}g\left( \tau \right) d\tau \right) ^{2}dt\leq \frac{1}{4\lambda b%
}\int_{0}^{a}g^{2}\left( \tau \right) \exp \left( 2\lambda \varphi \left(
t^{2}\right) \right) dt.
\end{equation*}%
Similarly,
\begin{equation*}
\int_{-a}^{0}\exp \left( 2\lambda \varphi \left( t^{2}\right) \right) \left(
\int_{0}^{t}g\left( \tau \right) d\tau \right) ^{2}dt\leq \frac{1}{4\lambda b%
}\int_{-a}^{0}g^{2}\left( \tau \right) \exp \left( 2\lambda \varphi \left(
t^{2}\right) \right) dt.\text{\quad }\square
\end{equation*}

\subsection{An MCIP for a hyperbolic equation}

\label{sec:3.2}

In this section we use notations of Section 2.4. Let functions $a_{\alpha
}\left( x,t\right) \in C\left( \overline{Q}_{T}\right) ,\left\vert \alpha
\right\vert \leq 1$ and the function $c\left( x\right) \in C^{1}\left(
\overline{\Omega }\right) ,c\left( x\right) \geq const>0.$ Let the function $%
u\in C^{2}\left( \overline{Q}_{T}\right) $ be the solution of the following
initial boundary value problem
\begin{eqnarray}
c\left( x\right) u_{tt} &=&\Delta u+\sum\limits_{\left\vert \alpha
\right\vert \leq 1}a_{\alpha }\left( x\right) D_{x}^{\alpha }u\text{, \ in }%
Q_{T},  \label{3.1} \\
u\left( x,0\right) &=&f_{0}\left( x\right) ,\text{ \ }u_{t}\left( x,0\right)
=f_{1}\left( x\right) ,  \label{3.2} \\
u|_{S_{T}} &=&p\left( x,t\right) ,S_{T}=\partial \Omega \times \left(
0,T\right) .  \label{3.3}
\end{eqnarray}

\textbf{Coefficient Inverse Problem for the Hyperbolic Equation (\ref{3.1}).}
Let the Neumann boundary condition be known,
\begin{equation}
\frac{\partial u}{\partial n}|_{S_{T}}=q\left( x,t\right) .  \label{3.4}
\end{equation}%
Determine one of $x-$dependent coefficients of equation (\ref{3.1}),
assuming that other coefficients are known, so as and functions $%
f_{0},f_{1},p,q$ in (\ref{3.2})-(\ref{3.4}).

\textbf{Theorem 3.1.} \emph{Let the domain }$\Omega =\left\{ \left\vert
x\right\vert <R\right\} \subset \mathbb{R}^{n},n\geq 2.$ \emph{Denote} $%
b\left( x\right) =1/\sqrt{c\left( x\right) }.$ \emph{Let the function }$%
b\left( x\right) $\emph{\ satisfies conditions (\ref{2.48}), (\ref{2.49}).\
In addition, let coefficients }$a_{\alpha }\in C\left( \overline{\Omega }%
\right) .$\emph{\ \ Consider two cases:}

\textbf{Case 1.}\emph{\ The coefficient }$c\left( x\right) $\emph{\ is
unknown and all other coefficients }$a_{\alpha }\left( x,t\right) $\emph{\
are known.\ In this case we assume that}
\begin{equation}
\Delta f_{0}\left( x\right) +\sum\limits_{\left\vert \alpha \right\vert \leq
1}a_{\alpha }\left( x\right) D_{x}^{\alpha }f_{0}\left( x\right) \neq 0\text{
\emph{for} }x\in \overline{\Omega }.  \label{3.5}
\end{equation}%
\emph{Then for a sufficiently large }$T>0$\emph{\ there exists at most one
pair of functions }$\left( u,c\right) $\emph{\ satisfying (\ref{3.1})-(\ref%
{3.4}) and such that }$u\in C^{4}\left( \overline{Q}_{T}\right) .$

\textbf{Case 2.}\emph{\ Let }$\alpha _{0}$ \emph{be one of multi-indices in (%
\ref{3.1}), }$\left\vert \alpha _{0}\right\vert \leq 1$\emph{.\ Let the
coefficient }$a_{\alpha _{0}}\left( x\right) $\emph{\ be unknown and all
other coefficients are known.\ In this case we assume that }
\begin{equation}
D_{x}^{\alpha _{0}}f_{0}\left( x\right) \neq 0\text{ }\emph{for}\text{ }x\in
\overline{\Omega }.  \label{3.6}
\end{equation}%
\emph{Then for a sufficiently large }$T>0$\emph{\ there exists at most one
pair of functions }$\left( u,a_{\alpha _{0}}\right) $\emph{\ satisfying (\ref%
{3.1})-(\ref{3.4}) and such that }$u\in C^{3+\left\vert \alpha
_{0}\right\vert }\left( \overline{Q}_{T}\right) $\emph{. }

\emph{If in (\ref{3.2}) }$f_{0}\left( x\right) \equiv 0,$\emph{\ then
conditions of these two cases should be imposed on the function }$%
f_{1}\left( x\right) $\emph{, the required smoothness of the function }$u$%
\emph{\ should be }$u\in C^{5}\left( \overline{Q}_{T}\right) $ \emph{in Case
1 and }$u\in C^{4+\left\vert \alpha _{0}\right\vert }\left( \overline{Q}%
_{T}\right) $\emph{\ in Case 2, and the above statements about uniqueness
would still hold.}

\textbf{Proof.} First, we note that if $f_{0}\left( x\right) \equiv 0,$\emph{%
\ }then one should consider in this proof $u_{t}$ instead of $u,$ and the
rest of the proof is the same as the one below. We prove this theorem only
for Case 1, since Case 2 is similar. Assume that there exist two solutions $%
\left( u_{1},c_{1}\right) $ and $\left( u_{2},c_{2}\right) $. Denote $%
\widetilde{u}=u_{1}-u_{2},\widetilde{c}=c_{1}-c_{2}$. Since
\begin{equation*}
c_{1}u_{1tt}-c_{2}u_{2tt}=c_{1}u_{1tt}-c_{1}u_{2tt}+\left(
c_{1}-c_{2}\right) u_{2tt}=c_{1}\widetilde{u}_{tt}+\widetilde{c}u_{2tt},
\end{equation*}%
then (\ref{3.1})-(\ref{3.4}) lead to
\begin{eqnarray}
L\widetilde{u} &=&c_{1}\left( x\right) \widetilde{u}_{tt}-\Delta \widetilde{u%
}-\sum\limits_{j=1}^{n}a_{\alpha }\left( x\right) D_{x}^{\alpha }\widetilde{u%
}=\widetilde{c}\left( x\right) B\left( x,t\right) ,\text{ in }Q_{T},
\label{3.7} \\
\widetilde{u}\left( x,0\right) &=&0,\widetilde{u}_{t}\left( x,0\right) =0,
\label{3.9} \\
\widetilde{u}|_{S_{T}} &=&\frac{\partial \widetilde{u}}{\partial n}%
|_{S_{T}}=0,  \label{3.101} \\
B\left( x,t\right) &:&=-u_{2tt}\left( x,t\right) .  \label{3.7_1}
\end{eqnarray}%
Setting in (\ref{3.1}) $c:=c_{2},u:=u_{2},t:=0$ and using (\ref{3.5}) and (%
\ref{3.101}), we obtain
\begin{equation*}
B\left( x,0\right) =-c_{2}^{-1}\left( x\right) \left( \Delta f_{0}\left(
x\right) +\sum\limits_{\left\vert \alpha \right\vert \leq 1}a_{\alpha
}\left( x\right) D_{x}^{\alpha }f_{0}\left( x\right) \right) \neq 0\text{
for }x\in \overline{\Omega }.
\end{equation*}%
Hence, there exists a sufficiently small positive number $\varepsilon $,
such that
\begin{equation}
B\left( x,t\right) \neq 0\text{ in }\overline{Q}_{\varepsilon }=\overline{%
\Omega }\times \text{ }\left[ 0,\varepsilon \right] .  \label{3.11}
\end{equation}%
By (\ref{3.7})
\begin{equation*}
\widetilde{c}\left( x\right) =\frac{L\widetilde{u}}{B\left( x,t\right) }%
\text{ in }\overline{Q}_{\varepsilon }.
\end{equation*}%
Hence,
\begin{equation*}
\frac{\partial }{\partial t}\left[ \widetilde{c}\left( x\right) \right] =%
\frac{\partial }{\partial t}\left[ \frac{L\widetilde{u}}{B\left( x,t\right) }%
\right] =0\text{ in }\overline{Q}_{\varepsilon }.
\end{equation*}%
Or
\begin{equation}
L\widetilde{u}_{t}=\frac{B_{t}}{B}\left( L\widetilde{u}\right) \text{ in }%
\overline{Q}_{\varepsilon }.  \label{3.12}
\end{equation}%
Denote
\begin{equation}
h\left( x,t\right) =\frac{B_{t}}{B}\left( x,t\right) .  \label{3.13}
\end{equation}%
By (\ref{3.11}) and (\ref{3.13})
\begin{equation}
h\in C^{2}\left( \overline{Q}_{\varepsilon }\right) .  \label{3.14}
\end{equation}%
Denote
\begin{equation}
v\left( x,t\right) =\widetilde{u}_{t}\left( x,t\right) -h\widetilde{u}\left(
x,t\right) .  \label{3.15}
\end{equation}%
We can consider (\ref{3.15}) as an ordinary differential equation with
respect to $\widetilde{u}\left( x,t\right) $ with the initial condition from
(\ref{3.8}), i.e. $\widetilde{u}\left( x,0\right) =0.$ Hence, using (\ref%
{3.13}), (\ref{3.14}) and (\ref{3.15}), we obtain
\begin{eqnarray}
\widetilde{u}\left( x,t\right) &=&\int\limits_{0}^{t}K\left( x,t,\tau
\right) v\left( x,\tau \right) d\tau ,  \label{3.16} \\
K\left( x,t,\tau \right) &=&\frac{B\left( x,t\right) }{B\left( x,\tau
\right) }\in C^{2}\left( \overline{\Omega }\times \left[ 0,\varepsilon %
\right] \times \left[ 0,\varepsilon \right] \right) ,  \label{3.17} \\
v\left( x,0\right) &=&0.  \label{3.18}
\end{eqnarray}%
Using (\ref{3.13}), (\ref{3.14}), (\ref{3.15}), (\ref{3.16}) and (\ref{3.17}%
), we obtain in $\overline{Q}_{\varepsilon }$
\begin{eqnarray*}
c_{1}\left( \widetilde{u}_{t}\right) _{tt}-hc_{1}\widetilde{u}_{tt}
&=&c_{1}v_{tt}+2c_{1}h_{t}\widetilde{u}_{t}+c_{1}h_{tt}\widetilde{u} \\
&=&c_{1}v_{tt}+2c_{1}h_{t}v+2c_{1}h_{t}\int\limits_{0}^{t}K_{t}\left(
x,t,\tau \right) v\left( x,\tau \right) d\tau
+c_{1}h_{tt}\int\limits_{0}^{t}K\left( x,t,\tau \right) v\left( x,\tau
\right) d\tau ,
\end{eqnarray*}%
\begin{eqnarray*}
\Delta \widetilde{u}_{t}-h\Delta \widetilde{u} &=&\Delta v+2\nabla h\nabla
\widetilde{u}+\Delta h\widetilde{u} \\
&=&\Delta v+2\nabla h\nabla \left( \int\limits_{0}^{t}K\left( x,t,\tau
\right) v\left( x,\tau \right) d\tau \right) +\Delta
h\int\limits_{0}^{t}K\left( x,t,\tau \right) v\left( x,\tau \right) d\tau .
\end{eqnarray*}%
Since by (\ref{3.12}) and (\ref{3.13}) $L\widetilde{u}_{t}-h\cdot L%
\widetilde{u}=0$ in $Q_{\varepsilon },$ then two recent formulas, boundary
conditions (\ref{3.9}), the initial condition (\ref{3.18}) as well as (\ref%
{3.14}) and (\ref{3.17}) lead to
\begin{eqnarray}
\left\vert c_{1}\left( x\right) v_{tt}-\Delta v\right\vert &\leq &M\left[
\left\vert \nabla v\right\vert \left( x,t\right) +\left\vert v\right\vert
\left( x,t\right) +\int\limits_{0}^{t}\left( \left\vert \nabla v\right\vert
+\left\vert v\right\vert \right) \left( x,\tau \right) d\tau \right] \text{
in }\overline{Q}_{\varepsilon },  \label{3.19} \\
v &\mid &_{S_{\varepsilon }}=\frac{\partial v}{\partial n}\mid
_{S_{\varepsilon }}=0,  \label{3.20} \\
v\left( x,0\right) &=&0,  \label{3.21}
\end{eqnarray}%
where $M>0$ is a constant independent on $v,x,t.$ The idea now is to apply
the Carleman estimate of Theorem 2.5 to the problem (\ref{3.19})-(\ref{3.21}%
) and estimate integrals using Lemma 3.1.

Let the point $x_{0}\in \Omega ,$ the number $P=P\left( x_{0},\Omega \right)
$ be the one defined in (\ref{2.49_1}) and $\eta _{0}=\eta _{0}\left(
d,P,\left\Vert \nabla c\right\Vert _{C\left( \overline{\Omega }\right)
}\right) \in \left( 0,1\right) $ be the number considered in Theorem 2.5.
Choose an arbitrary number $\eta \in \left( 0,\eta _{0}\right) .$ Assuming
that $\varepsilon $ is so small that $\eta _{0}\varepsilon ^{2}<R^{2},$
consider the domain $G_{\eta \varepsilon ^{2}}^{+}$,
\begin{equation*}
G_{\eta \varepsilon ^{2}}^{+}=\left\{ \left( x,t\right) :\left\vert
x\right\vert ^{2}-\eta t^{2}>R^{2}-\eta \varepsilon ^{2},t>0,\left\vert
x\right\vert <R\right\} .
\end{equation*}%
Then $G_{\eta \varepsilon ^{2}}^{+}\subset \overline{Q}_{\varepsilon }.$ Let
$S_{\varepsilon }=\partial \Omega \times \left( 0,\varepsilon \right) .$ The
boundary of $G_{\eta \varepsilon ^{2}}^{+}$ consists of three parts,%
\begin{eqnarray*}
\partial G_{\eta \varepsilon ^{2}}^{+} &=&\bigcup\limits_{i=1}^{3}\partial
_{i}G_{\eta \varepsilon ^{2}}^{+}, \\
\partial _{1}G_{\eta \varepsilon ^{2}}^{+} &=&S_{\varepsilon }\cap \overline{%
G}_{\eta \varepsilon ^{2}}^{+}, \\
\partial _{2}G_{\eta \varepsilon ^{2}}^{+} &=&\left\{ \left\vert
x\right\vert \in \left( \sqrt{R^{2}-\eta \varepsilon ^{2}},R\right)
,t=0\right\} , \\
\partial _{3}G_{\eta \varepsilon ^{2}}^{+} &=&\left\{ \left\vert
x\right\vert ^{2}-\eta t^{2}=R^{2}-\eta \varepsilon ^{2},t>0,\left\vert
x\right\vert <R\right\} .
\end{eqnarray*}
Square both sides of inequality (\ref{3.19}), multiply by the function $%
\varphi ^{2}\left( x,t\right) $ defined in subsection 2.4, apply Theorem 2.5
and Gauss-Ostrogradsky formula. By (\ref{2.50}) and (\ref{3.20}) integral
over $\partial _{1}G_{\eta \varepsilon ^{2}}^{+}$ equals zero. By (\ref{2.52}%
) and (\ref{3.21}) integral over $\partial _{2}G_{\eta \varepsilon ^{2}}^{+}$
also equals zero. Hence, we obtain with a different constant $M$%
\begin{eqnarray}
C\lambda \int\limits_{G_{\eta \varepsilon ^{2}}^{+}}\left( \nabla v\right)
^{2}\varphi ^{2}dxdt+C\lambda ^{3}\int\limits_{G_{\eta \varepsilon
^{2}}^{+}}v^{2}\varphi ^{2}dxdt &\leq &M\int\limits_{G_{\eta \varepsilon
^{2}}^{+}}\left[ \left( \nabla v\right) ^{2}+v^{2}\right] \varphi ^{2}dxdt
\notag \\
&&M\int\limits_{G_{\eta \varepsilon ^{2}}^{+}}\left(
\int\limits_{0}^{t}\left( \left\vert \nabla v\right\vert +\left\vert
v\right\vert \right) \left( x,\tau \right) d\tau \right) ^{2}\varphi ^{2}dxdt
\label{3.21_1} \\
+C\lambda ^{3}\exp \left[ 2\lambda \left( R^{2}-\eta \varepsilon ^{2}\right) %
\right] \int\limits_{\partial _{3}G_{\eta \varepsilon ^{2}}^{+}}\left[
\left( \nabla v\right) ^{2}+v^{2}\right] dS_{x,t},\forall \lambda &>&\lambda
_{0}.  \notag
\end{eqnarray}

For an arbitrary point $\left( x,0\right) \in \partial _{2}G_{\eta
\varepsilon ^{2}}^{+}$ consider the straight line, which is parallel to the $%
t-$axis and passes through the point $\left( x,0\right) .$ Then this line
intersects the hypersurface $\partial _{3}G_{\eta \varepsilon ^{2}}^{+}$ at
the point $\left( x,t\left( x\right) \right) .$ Hence, applying Lemma 3.1 to
any function $g\in C\left( \overline{G}_{\eta \varepsilon ^{2}}^{+}\right) ,$
we obtain%
\begin{equation}
\int\limits_{G_{\eta \varepsilon ^{2}}^{+}}\left(
\int\limits_{0}^{t}\left\vert g\right\vert \left( x,\tau \right) d\tau
\right) ^{2}\varphi ^{2}dxdt=\int\limits_{\partial _{3}G_{\eta \varepsilon
^{2}}^{+}}dx\left[ \int\limits_{0}^{t\left( x\right) }\left(
\int\limits_{0}^{t}\left\vert g\right\vert \left( x,\tau \right) d\tau
\right) ^{2}\varphi ^{2}dt\right] \leq \frac{C}{\lambda }\int\limits_{G_{%
\eta \varepsilon ^{2}}^{+}}g^{2}\varphi ^{2}dxdt.  \label{3.21_2}
\end{equation}

Using (\ref{3.21_1}) and (\ref{3.21_2}) and choosing a sufficiently large
number $\lambda _{1}=\lambda _{1}\left( M,\lambda _{0}\right) >\lambda _{0}$%
, we obtain with a different constant $C$%
\begin{equation*}
C\lambda \int\limits_{G_{\eta \varepsilon ^{2}}^{+}}\left( \nabla v\right)
^{2}\varphi ^{2}dxdt+C\lambda ^{3}\int\limits_{G_{\eta \varepsilon
^{2}}^{+}}v^{2}\varphi ^{2}dxdt\leq C\lambda ^{3}\exp \left[ 2\lambda \left(
R^{2}-\eta \varepsilon ^{2}\right) \right] \int\limits_{\partial _{3}G_{\eta
\varepsilon ^{2}}^{+}}\left[ \left( \nabla v\right) ^{2}+v^{2}\right]
dS_{x,t},\forall \lambda >\lambda _{1}.
\end{equation*}%
Let $\delta \in \left( 0,\eta \varepsilon ^{2}\right) $ be any number. Then $%
G_{\delta }^{+}\subset G_{\eta \varepsilon ^{2}}^{+}$ and $\varphi
^{2}\left( x,t\right) >\exp \left[ 2\lambda \left( R^{2}-\delta \right) %
\right] $ in $G_{\delta }^{+}.$ Hence, strengthening the last inequality, we
obtain with a different constant $C$%
\begin{equation*}
\lambda ^{3}\exp \left[ 2\lambda \left( R^{2}-\delta \right) \right]
\int\limits_{G_{\delta }^{+}}v^{2}dxdt\leq C\lambda ^{3}\exp \left[ 2\lambda
\left( R^{2}-\eta \varepsilon ^{2}\right) \right] \int\limits_{\partial
_{3}G_{\eta \varepsilon ^{2}}^{+}}\left[ \left( \nabla v\right) ^{2}+v^{2}%
\right] dS_{x,t},\forall \lambda >\lambda _{1}.
\end{equation*}%
Dividing by $\lambda ^{3}\exp \left[ 2\lambda \left( R^{2}-\delta \right) %
\right] ,$ we obtain%
\begin{equation*}
\int\limits_{G_{\delta }^{+}}v^{2}dxdt\leq C\exp \left[ -2\lambda \left(
\eta \varepsilon ^{2}-\delta \right) \right] \int\limits_{\partial
_{3}G_{\eta \varepsilon ^{2}}^{+}}\left[ \left( \nabla v\right) ^{2}+v^{2}%
\right] dS_{x,t}.
\end{equation*}%
Setting here $\lambda \rightarrow \infty ,$ we obtain $v\left( x,t\right) =0$
in $G_{\delta }^{+}.$ Since numbers $\eta \in \left( 0,\eta _{0}\right) $
and $\delta \in \left( 0,\eta \varepsilon ^{2}\right) $ are arbitrary ones,
then $v\left( x,t\right) =0$ in $G_{\eta _{0}\varepsilon ^{2}}^{+}.$

Hence, by (\ref{3.16}) $\widetilde{u}\left( x,t\right) =0$ in $G_{\eta
_{0}\varepsilon ^{2}}^{+}.$ Hence, setting $t=0$ in (\ref{3.7}) and using (%
\ref{3.5}), we obtain
\begin{equation}
\widetilde{c}\left( x\right) =0\text{ for }x\in \left\{ \left\vert
x\right\vert \in \left( \sqrt{R^{2}-\eta _{0}\varepsilon ^{2}},R\right)
\right\} .  \label{3.22}
\end{equation}

Substitute (\ref{3.22}) in (\ref{3.7}) and use (\ref{3.8}) and (\ref{3.9}).
We obtain for $\left( x,t\right) \in \left\{ \left\vert x\right\vert \in
\left( \sqrt{R^{2}-\eta _{0}\varepsilon ^{2}},R\right) \right\} \times
\left( 0,T\right) $
\begin{eqnarray}
L\widetilde{u} &=&c_{1}\left( x\right) \widetilde{u}_{tt}-\Delta \widetilde{u%
}-\sum\limits_{j=1}^{n}a_{\alpha }\left( x\right) D_{x}^{\alpha }\widetilde{u%
}=0,  \label{3.23} \\
\widetilde{u}\left( x,0\right) &=&0,\widetilde{u}_{t}\left( x,0\right) =0,
\label{3.25} \\
\widetilde{u}|_{S_{T}} &=&\frac{\partial \widetilde{u}}{\partial n}%
|_{S_{T}}=0.  \label{3.26}
\end{eqnarray}%
Consider an arbitrary number $t_{0}\in \left( 0,T-\varepsilon \right) .$ And
consider the domain $G_{\eta _{0}\varepsilon ^{2}}\left( t_{0}\right) ,$
\begin{equation*}
G_{\eta _{0}\varepsilon ^{2}}\left( t_{0}\right) =\left\{ \left( x,t\right)
:\left\vert x\right\vert ^{2}-\eta _{0}\left( t-t_{0}\right) ^{2}>R^{2}-\eta
_{0}\varepsilon ^{2},t>0,\left\vert x\right\vert <R\right\} .
\end{equation*}%
Since $t_{0}\in \left( 0,T-\varepsilon \right) ,$ then $t\in \left(
0,T\right) $ in this domain. Hence,
\begin{equation*}
G_{\eta _{0}\varepsilon ^{2}}\left( t_{0}\right) \subset \left\{ \left\vert
x\right\vert \in \left( \sqrt{R^{2}-\eta _{0}\varepsilon ^{2}},R\right)
\right\} \times \left( 0,T\right) .
\end{equation*}%
Hence, using conditions (\ref{3.25}), (\ref{3.26}), we can apply Theorems
2.3, 2.5 to the domain $G_{\eta _{0}\varepsilon ^{2}}\left( t_{0}\right) $.
Therefore, $\widetilde{u}\left( x,t\right) =0$ for $\left( x,t\right) \in
G_{\eta _{0}\varepsilon ^{2}}\left( t_{0}\right) .$ Since $t_{0}$ is an
arbitrary number of the interval $\left( 0,T-\varepsilon \right) ,$ then,
varying this number, we obtain that
\begin{equation*}
\widetilde{u}\left( x,t\right) =0\text{ for }\left( x,t\right) \in \left\{
\left\vert x\right\vert \in \left( \sqrt{R^{2}-\eta _{0}\varepsilon ^{2}}%
,R\right) \right\} \times \left( 0,T-\varepsilon \right) .
\end{equation*}%
Therefore, we now can replace in (\ref{3.7})-(\ref{3.10}) sets $Q_{T}$ and $%
S_{T}$ with sets $Q_{T}^{\varepsilon }$ and $S_{T}^{\varepsilon }$
respectively, where
\begin{equation*}
Q_{T}^{\varepsilon }=\left\{ \left\vert x\right\vert <\sqrt{R^{2}-\eta
_{0}\varepsilon ^{2}}\right\} \times \left( 0,T-\varepsilon \right)
,S_{T}^{\varepsilon }=\left\{ \left\vert x\right\vert =\sqrt{R^{2}-\eta
_{0}\varepsilon ^{2}}\right\} \times \left( 0,T-\varepsilon \right) .
\end{equation*}%
Next, we repeat the above proof. Hence, we obtain instead of (\ref{3.22})
\begin{equation*}
\widetilde{c}\left( x\right) =0\text{ for }x\in \left\{ \left\vert
x\right\vert \in \left( \sqrt{R^{2}-2\eta _{0}\varepsilon ^{2}},R\right)
\right\} .
\end{equation*}%
Since $\varepsilon >0$ is sufficiently small, we can choose $\varepsilon $
such that $R^{2}=k\eta _{0}\varepsilon ^{2}$ where $k=k\left( R,\varepsilon
\right) \geq 1$ is an integer. Suppose that
\begin{equation*}
T>k\varepsilon =\frac{R^{2}}{\eta _{0}\varepsilon }.
\end{equation*}%
Then we can repeat this process $k$ times until the entire domain $\Omega
=\left\{ \left\vert x\right\vert <R\right\} $ will be covered. Thus, we
obtain that $\widetilde{c}\left( x\right) =0$ in $\Omega .$ Hence, the right
hand side of equation (\ref{3.7}) is identical zero. This and the standard
energy estimate imply that $\widetilde{u}\left( x,t\right) =0$ in $Q_{T}.$ $%
\square $

An inconvenient point of Theorem 3.1 is that the observation time $T$ is
assumed to be sufficiently large. An experience of the author of working
with experimental data \cite{BK,BK4,KFBPS,KBKSNF,IEEE} indicates that this
is not a severe restriction in applications.\ Indeed, usually the
pre-processing procedure of the measured signal leaves only a small portion
of the time dependent curve to work with. Still, it is possible to restrict
the value of $T$ via imposing the condition $f_{1}\left( x\right) \equiv 0.$
The proof of Theorem 3.2 partially uses arguments of works of Imanuvilov and
Yamamoto \cite{Y2,Y3}.

\textbf{Theorem 3.2.} \emph{Assume that all conditions of Theorem 3.1 are
satisfied. In addition, assume that the function }$f_{1}\left( x\right)
\equiv 0.$\emph{\ Then Theorem 3.1 holds if }
\begin{equation}
T>\frac{R}{\sqrt{\eta _{0}}}.  \label{3.27}
\end{equation}%
\emph{In particular if }$c\left( x\right) \equiv 1,$\emph{\ then it is
sufficient to have }$T>R.$

\textbf{Proof.} Just as in the proof of Theorem 3.1, we consider now only
Case 1, keep notations of Theorem 3.1. Introduce the function $w\left(
x,t\right) =\widetilde{u}_{tt}\left( x,t\right) .$ Then by (\ref{3.7})-(\ref%
{3.10})
\begin{eqnarray}
c_{1}\left( x\right) w_{tt}-\Delta w-\sum\limits_{j=1}^{n}a_{\alpha }\left(
x\right) D_{x}^{\alpha }w &=&-\widetilde{c}\left( x\right) \partial
_{t}^{4}u_{2},\text{ in }Q_{T},  \label{3.28} \\
w_{t}\left( x,0\right) &=&0,  \notag \\
w|_{S_{T}} &=&\frac{\partial w}{\partial n}|_{S_{T}}=0,  \notag \\
w\left( x,0\right) &=&-\widetilde{c}\left( x\right) p\left( x\right) ,
\label{3.29} \\
p\left( x\right) &=&c_{1}^{-1}\left( x\right) \left( \Delta f_{0}\left(
x\right) +\sum\limits_{\left\vert \alpha \right\vert \leq 1}a_{\alpha
}\left( x\right) D_{x}^{\alpha }f_{0}\left( x\right) \right) \neq 0,x\in
\overline{\Omega }  \label{3.30}
\end{eqnarray}%
By (\ref{3.29}) and (\ref{3.30})
\begin{equation*}
-\widetilde{c}\left( x\right) =\frac{w\left( x,0\right) }{p\left( x\right) }=%
\frac{1}{p\left( x\right) }\left[ w\left( x,t\right)
-\int\limits_{0}^{t}w_{t}\left( x,\tau \right) d\tau \right] .
\end{equation*}%
Substituting this formula in (\ref{3.28}) and proceeding similarly with the
proof of Theorem 3.1, we obtain that $\widetilde{c}\left( x\right) =0$ in $%
\Omega $ and $w\left( x,t\right) =\widetilde{u}\left( x,t\right) =0$ in $%
Q_{T}.$ $\square $

\subsection{MCIPs for parabolic equations}

\label{sec:3.3}

In this subsection we prove uniqueness theorems for three MCIPs for
parabolic PDEs. Unlike the hyperbolic case, conditions imposed on the
principal part of the corresponding elliptic operator are not restrictive in
two out of three of these problems. The reason is that Carleman estimate can
be proven for a quite general parabolic operator, see \S 1 of Chapter 4 of
the book \cite{LRS}. Below $C^{k+\beta },C^{2k+\beta ,k+\beta /2}$ are H\"{o}%
lder spaces, where $k\geq 0$ is an integer and $\beta \in \left( 0,1\right) $%
.

\subsubsection{The first MCIP for a parabolic equation}

\label{sec:3.3.1}

Denote $D_{T}^{n+1}=\mathbb{R}^{n}\times \left( 0,T\right) .$ Consider the
Cauchy problem for the following parabolic equation
\begin{eqnarray}
c\left( x\right) u_{t} &=&\Delta u+\sum\limits_{\left\vert \alpha
\right\vert \leq 1}a_{\alpha }\left( x\right) D_{x}^{\alpha }u\text{ \ in }%
D_{T}^{n+1},  \label{3.33} \\
u\left( x,0\right) &=&f_{0}\left( x\right) ,  \label{3.34} \\
c,a_{\alpha } &\in &C^{\beta }\left( \mathbb{R}^{n}\right) ,\text{ }c\left(
x\right) \in \left[ 1,d\right] ,\text{ }f_{0}\in C^{2+\beta }\left( \mathbb{R%
}^{n}\right) .  \label{3.35}
\end{eqnarray}%
The problem (\ref{3.33})-(\ref{3.35}) has unique solution $u\in C^{2+\beta
,1+\beta /2}\left( \overline{D}_{T}^{n+1}\right) ,$ see the book of
Ladyzhenskaya, Solonnikov and Uralceva \cite{LSU}. Just as in Section 3.2,
we assume that $\Omega =\left\{ \left\vert x\right\vert <R\right\} \subset
\mathbb{R}^{n},n\geq 2.$ Let $\Gamma \subseteq \partial \Omega $ be a part
of the boundary of the domain $\Omega $, $T=const.>0$ and $\Gamma
_{T}=\Gamma \times \left( 0,T\right) $.

\textbf{The First Parabolic Coefficient Inverse Problem. }Suppose that one
of coefficients in equation (\ref{3.33}) is unknown inside of the domain $%
\Omega $ and is known outside of it. Also, assume that all other
coefficients in (\ref{3.33}) are known and conditions (\ref{3.34}), (\ref%
{3.35}) are satisfied. Determine that unknown coefficient inside of $\Omega
, $ assuming that the following functions $p\left( x,t\right) $ and $q\left(
x,t\right) $ are known
\begin{equation}
u\mid _{\Gamma _{T}}=p\left( x,t\right) ,\quad \frac{\partial u}{\partial n}%
\mid _{\Gamma _{T}}=q\left( x,t\right) .  \label{3.36}
\end{equation}

It is yet unclear how to prove a uniqueness theorem for this CIP
\textquotedblleft straightforwardly\textquotedblright . The reason is that
one cannot extend properly the solution of the problem (\ref{3.33}), (\ref%
{3.34}) in $\left\{ t<0\right\} .$ Thus, the idea here is to consider an
associated MCIP for the hyperbolic PDE (\ref{3.1}) using a connection
between these two CIPs via an analog of the Laplace transform. Next, since
the Laplace transform is one-to-one, then Theorem 3.1 will provide the
desired uniqueness result.

That associated hyperbolic Cauchy problem is
\begin{eqnarray}
c\left( x\right) v_{tt} &=&\Delta v+\sum\limits_{\left\vert \alpha
\right\vert \leq 1}a_{\alpha }\left( x\right) D_{x}^{\alpha }v\text{ \ in }%
D_{\infty }^{n+1}=\mathbb{R}^{n}\times \left( 0,\infty \right) ,
\label{3.37} \\
v|_{t=0} &=&0,\text{ }v_{t}|_{t=0}=f_{0}\left( x\right) .  \label{3.38}
\end{eqnarray}%
In addition to (\ref{3.35}), we assume that the coefficients $c\left(
x\right) ,a_{\alpha }\left( x\right) $ and the initial condition $%
f_{0}\left( x\right) $ are so smooth that the solution $v$ of the problem (%
\ref{3.37}), (\ref{3.38}) is such that (a) $v\in C^{5}\left( \overline{D}%
_{\infty }^{n+1}\right) $ if the function $c\left( x\right) $ is unknown,
and (b) $v\in C^{4+\left\vert \alpha \right\vert }\left( \overline{D}%
_{\infty }^{n+1}\right) $ if the function $c\left( x\right) $ is \ known and
any of functions $a_{\alpha }\left( x\right) $ is unknown.

Consider the Laplace-like transform $\mathcal{R}$ of Reznickaya \cite{Rezn}.
Since the publication \cite{Rezn} in 1973 this transform is widely used \cite%
{BK,KT,LRS}. The following connection between solutions $u$ and $v$ of
parabolic and hyperbolic Cauchy problems (\ref{3.33}), (\ref{3.34}) and (\ref%
{3.37}), (\ref{3.38}) can be easily verified
\begin{equation}
u\left( x,t\right) =\frac{1}{2t\sqrt{\pi t}}\int_{0}^{\infty }\exp \left[ -%
\frac{\tau ^{2}}{4t}\right] \tau v\left( x,\tau \right) d\tau :=\mathcal{R}v.
\label{3.39}
\end{equation}%
Using an analogy with the Laplace transform, one can easily prove that the
operator $\mathcal{R}$ is one-to-one. Hence, given functions $p,q$ in (\ref%
{3.36}), the following two functions $\overline{p}\left( x,t\right) $ and $%
\overline{q}\left( x,t\right) $ can be uniquely determined
\begin{equation}
v|_{\Gamma _{\infty }}=\overline{p}\left( x,t\right) ,\text{\quad }\frac{%
\partial v}{\partial n}|_{\Gamma _{\infty }}=\overline{q}\left( x,t\right) .
\label{3.40}
\end{equation}%
Therefore, we have reduced the First Parabolic Coefficient Inverse Problem
to the hyperbolic CIP (\ref{3.37}), (\ref{3.38}), (\ref{3.40}).

To apply Theorem 3\textbf{.}1, we need to replace $\Gamma _{\infty }$ in (%
\ref{3.40}) with $S_{\infty }.$ To do this, we observe that, using (\ref%
{3.36}) and the fact that the unknown coefficient is given outside of the
domain $\Omega $, one can uniquely determine the function $u\left(
x,t\right) $ for $\left( x,t\right) \in \left( \mathbb{R}^{n}\diagdown
\Omega \right) \times \left( 0,T\right) .$ Indeed, this follows from Remark
2.2. Therefore we can uniquely determine functions $u,\partial _{n}u$ at $%
S_{T}.$ Hence, we can replace in (\ref{3.40}) $\Gamma _{\infty }$ with $%
S_{\infty }.$ Hence, Theorem 3.1 implies Theorem 3.3.

\textbf{Theorem 3.3. }\emph{Assume that conditions (\ref{3.35}) hold. Also,
assume that coefficients }$c\left( x\right) ,a_{\alpha }\left( x\right) $%
\emph{\ and the initial condition }$f_{0}\left( x\right) $\emph{\ are so
smooth that the solution }$v$\emph{\ of the problem (\ref{3.37}), (\ref{3.38}%
) is such that: }

\emph{(a) }$v\in C^{5}\left( \overline{D}_{\infty }^{n+1}\right) $\emph{\ if
the function }$c\left( x\right) $\emph{\ is unknown, and (b) }$v\in
C^{4+\left\vert \alpha \right\vert }\left( \overline{D}_{\infty
}^{n+1}\right) $\emph{\ if the function }$a_{\alpha }\left( x\right) $\
\emph{is unknown. Let the domain }$\Omega =\left\{ \left\vert x\right\vert
<R\right\} \subset \mathbb{R}^{n},n\geq 2.$ \emph{Denote} $b\left( x\right)
=1/\sqrt{c\left( x\right) }.$ \emph{Let the function }$b\left( x\right) $
\emph{satisfies conditions (\ref{2.48}), (\ref{2.49}). Suppose that
conditions of either of Cases 1 or 2 of Theorem 3.1 hold. Then conclusions
of Theorem 3.1 are true for the inverse problem (\ref{3.33})-(\ref{3.36}).}

\subsubsection{The second MCIP for the parabolic equation}

\label{sec:3.3.2}

Two points of Theorem 3.3 are inconvenient ones. First, one needs to reduce
the parabolic CIP to the hyperbolic CIP via inverting the transform (\ref%
{3.39}). Second, one needs to use a special form of the elliptic operator in
(\ref{3.37}) with the restrictive condition (\ref{2.49}) imposed on the
coefficient $c\left( x\right) $. Although (\ref{2.49}) holds for the case $%
c\left( x\right) \equiv 1$, still the question remains whether it is
possible to prove uniqueness of an MCIP for the case of a general parabolic
operator of the second order. We show in this Section that the latter is
possible, provided that one can guarantee the existence of the solution of
the parabolic PDE for $t\in \left( -T,T\right) $ and that the function $%
u\left( x,0\right) $ is known. On the other hand, it was noticed in the
paper of Yamamoto and Zou \cite{Y9} that the measurement of the temperature
at $t=\theta >0$ is often easier to achieve than the measurement at the
initial time moment $t=0$. Thus, that condition likely has a good applied
sense.

Let $\Omega \subset \mathbb{R}^{n}$ be either finite or infinite connected
domain with the piecewise smooth boundary $\partial \Omega $, $\Gamma
\subseteq \partial \Omega $ be a part of this boundary and $T=const>0$.
Denote $\Gamma _{T}^{\pm }=\Gamma \times \left( -T,T\right) .$ Let $L$ be
the elliptic operator in $Q_{T}^{\pm }$ of the form
\begin{eqnarray}
Lu &=&\sum\limits_{\left\vert \alpha \right\vert \leq 2}a_{\alpha }\left(
x\right) D_{x}^{\alpha }u,\left( x,t\right) \in Q_{T}^{\pm },  \label{3.41}
\\
a_{\alpha } &\in &C^{1}\left( \overline{\Omega }\right) ,\left\vert \alpha
\right\vert =2,a_{\alpha }\in C\left( \overline{\Omega }\right) ,\left\vert
\alpha \right\vert =0,1,  \label{3.42} \\
\mu _{1}\left\vert \xi \right\vert ^{2} &\leq &\sum\limits_{\left\vert
\alpha \right\vert =2}a_{\alpha }\left( x\right) \xi ^{\alpha }\leq \mu
_{2}\left\vert \xi \right\vert ^{2},\forall \xi \in \mathbb{R}^{n},\forall
x\in \overline{\Omega };\mu _{1},\mu _{2}=const.>0.  \label{3.43}
\end{eqnarray}

\textbf{The Second Parabolic Coefficient Inverse Problem. }Assume that one
of coefficients $a_{\alpha _{0}}\left( x\right) $\ of the operator $L$\ is
unknown in $\Omega $, and all other coefficients of $L$\ are known in $%
\Omega $. Let the function $u\in C^{4,2}\left( \overline{Q}_{T}^{\pm
}\right) $\ satisfies the parabolic equation
\begin{equation}
u_{t}=Lu+F\left( x,t\right) ,\text{ in }Q_{T}^{\pm }.  \label{3.44}
\end{equation}%
Determine the coefficient $a_{\alpha _{0}}\left( x\right) $ for $x\in \Omega
,$ assuming that the function $F\left( x,t\right) $ is known in $Q_{T}^{\pm
} $ and that the following functions $f_{0}\left( x\right) ,p\left(
x,t\right) $\ and $q\left( x,t\right) $\ are known as well
\begin{eqnarray}
u\left( x,0\right) &=&f_{0}\left( x\right) ,x\in \Omega ,  \label{3.45} \\
u|_{\Gamma _{T}^{\pm }} &=&p\left( x,t\right) ,\frac{\partial u}{\partial n}%
|_{\Gamma _{T}^{\pm }}=q\left( x,t\right) .  \label{3.46}
\end{eqnarray}

\textbf{Theorem 3.4}. \emph{Assume that conditions (\ref{3.42}) and (\ref%
{3.43}) are valid and that }%
\begin{equation}
D_{x}^{\alpha _{0}}f_{0}\left( x\right) \neq 0\emph{\ }\text{in}\emph{\ }%
\overline{\Omega }\emph{.}  \label{3.47}
\end{equation}%
\emph{\ Then there exists at most one solution }$\left( u,a_{\alpha
_{0}}\right) $ \emph{of the inverse problem (\ref{3.44})-(\ref{3.46}) such
that }$u\in C^{4,2}\left( \overline{Q}_{T}^{\pm }\right) $\emph{.}

\textbf{Proof.} Assume that there exist two solutions $\left(
u_{1},a_{\alpha _{0}}^{\left( 1\right) }\right) $ and $\left(
u_{2},a_{\alpha _{0}}^{\left( 2\right) }\right) $. Denote $\widetilde{u}%
=u_{1}-u_{2},\widetilde{a}=a_{\alpha _{0}}^{\left( 1\right) }-a_{\alpha
_{0}}^{\left( 2\right) }.$ Using (\ref{3.44})-(\ref{3.46}), we obtain%
\begin{eqnarray}
\widetilde{u}_{t}-L^{\left( 1\right) }\widetilde{u} &=&\widetilde{a}\left(
x\right) D^{\alpha _{0}}u_{2},  \label{3.48} \\
\widetilde{u}\left( x,0\right) &=&0,  \label{3.49} \\
\widetilde{u}|_{\Gamma _{T}^{\pm }} &=&\frac{\partial \widetilde{u}}{%
\partial n}\mid _{\Gamma _{T}^{\pm }}=0.  \label{3.50}
\end{eqnarray}%
Here $L^{\left( 1\right) }$ means that the coefficient $a_{\alpha
_{0}}\left( x\right) $ in the operator $L$ is replaced with $a_{\alpha
_{0}}^{\left( 1\right) }\left( x\right) .$ It follows from (\ref{3.47}) that
there exists a small number $\varepsilon >0$ such that $D^{\alpha
_{0}}u_{2}\neq 0$ in $\overline{Q}_{\varepsilon }^{\pm }.$ Denote for
brevity $g\left( x,t\right) =D^{\alpha _{0}}u_{2}\left( x,t\right) .$ By (%
\ref{3.48})%
\begin{equation*}
\widetilde{a}\left( x\right) =\frac{\widetilde{u}_{t}-L^{\left( 1\right) }%
\widetilde{u}}{g},\left( x,t\right) \in \text{ }\overline{Q}_{\varepsilon
}^{\pm }.
\end{equation*}%
Differentiating this equality with respect to $t$ and denoting
\begin{equation}
h\left( x,t\right) =\frac{g_{t}}{g},  \label{3.51}
\end{equation}%
we obtain%
\begin{equation}
\widetilde{u}_{tt}-L^{\left( 1\right) }\widetilde{u}_{t}=h\left( \widetilde{u%
}_{t}-L^{\left( 1\right) }\widetilde{u}\right) .  \label{3.52}
\end{equation}%
We now proceed similarly with the proof of Theorem 3.1. Denote
\begin{equation}
v=\widetilde{u}_{t}-h\widetilde{u}.  \label{3.53}
\end{equation}%
Solving the Ordinary Differential Equation (\ref{3.53}) with the zero
initial condition (\ref{3.49}) and taking into account (\ref{3.51}), we
obtain%
\begin{equation}
\widetilde{u}\left( x,t\right) =\int\limits_{0}^{t}K\left( x,t,\tau \right)
v\left( x,\tau \right) d\tau ,K\left( x,t,\tau \right) =\frac{g\left(
x,t\right) }{g\left( x,\tau \right) },\left( x,t\right) \in \overline{Q}%
_{\varepsilon }^{\pm }  \label{3.54}
\end{equation}%
Next, using (\ref{3.50}), (\ref{3.52}) and (\ref{3.53}), we obtain similarly
with (\ref{3.19}) and (\ref{3.20})
\begin{eqnarray}
\left\vert v_{t}-L_{0}^{\left( 1\right) }v\right\vert &\leq &M\left[
\left\vert \nabla v\right\vert \left( x,t\right) +\left\vert v\right\vert
\left( x,t\right) +sgn\left( t\right) \int\limits_{0}^{t}\left( \left\vert
\nabla v\right\vert +\left\vert v\right\vert \right) \left( x,\tau \right)
d\tau \right] \text{ in }\overline{Q}_{\varepsilon }^{\pm },  \label{3.55} \\
v &\mid &_{\Gamma _{\varepsilon }^{\pm }}=\frac{\partial v}{\partial n}\mid
_{\Gamma _{\varepsilon }^{\pm }}=0,  \label{3.56}
\end{eqnarray}%
where $L_{0}^{\left( 1\right) }$ is the principal part of the operator $%
L^{\left( 1\right) }$, $sgn\left( t\right) =1$ if $t>0$ and $sgn\left(
t\right) =-1$ if $t<0,$ and the positive constant $M$ is independent on $%
x,t,v.$ Without loss of generality we assume that
\begin{equation}
\Gamma =\left\{ x_{1}=0,\left\vert y\right\vert \leq Y\right\} .
\label{3.57}
\end{equation}%
Otherwise we can still obtain (\ref{3.57}) for at least a piece of $\Gamma $
via a change of variables. Consider such a point $x_{0}\in \Omega $ that the
straight line which is perpendicular to $\left\{ x_{1}=0\right\} $ and
passes through $x_{0}$, intersects $\Gamma ,$ and the segment of this
straight line which connects $x_{0}$ and $\Gamma ,$ lies inside of $\Omega .$
Without loss of generality we assume that
\begin{equation}
x_{0}\in \left\{ x_{1}>0,x_{1}+\frac{\left\vert y\right\vert ^{2}}{2Y^{2}}%
<\eta -\alpha \right\} ,  \label{3.58}
\end{equation}%
where numbers $\alpha ,\eta \in \left( 0,1\right) ,\alpha <\eta $ were
defined in subsection 2.3.

Let $\varphi \left( x,t\right) $ be the function defined in (\ref{2.16_0}), (%
\ref{2.16}). In (\ref{2.16_0}) choose the parameter $T$ so large that $%
\varepsilon ^{2}/\left( 2T^{2}\right) <\eta -\alpha .$ Hence, the domain $%
G_{\eta }\subset Q_{\varepsilon }^{\pm },$ where $G_{\eta }$ was defined in
subsection 2.3. Square both sides of inequality (\ref{3.55}), multiply by
the function $\varphi ^{2}\left( x,t\right) $, integrate over the domain $%
G_{\eta }$. Following Remark 2.2, we use the Carleman estimate of Theorem
2.3, assuming that it is valid for the operator $L_{0}^{\left( 1\right) }.$
In addition, we use (\ref{3.56}). Also, fix the parameter $\nu :=\nu
_{0}\left( K,G_{\eta },\mu _{1},\mu _{2}\right) .$ We obtain with a new
constant $M$%
\begin{eqnarray*}
C\lambda \int\limits_{G_{\eta }}\left( \nabla v\right) ^{2}\varphi
^{2}dxdt+C\lambda ^{3}\int\limits_{G_{\eta }}v^{2}\varphi ^{2}dxdt &\leq
&M\int\limits_{G_{\eta }}\left[ \left( \nabla v\right) ^{2}+v^{2}\right]
\varphi ^{2}dxdt+M\int\limits_{G_{\eta }}\left[ \int\limits_{0}^{t}\left(
\left\vert \nabla v\right\vert +\left\vert v\right\vert \right) d\tau \right]
^{2}\varphi ^{2}dxdt \\
&&+C\lambda ^{3}\exp \left[ 2\lambda \eta ^{-\nu }\right] \int\limits_{%
\partial _{2}G_{\eta }}\left[ \left( \nabla v\right) ^{2}+v^{2}\right] dS.
\end{eqnarray*}%
Using Lemma 3.1, we obtain similarly with the proof of inequality (\ref%
{3.21_2}) with a new constant $M$%
\begin{equation*}
M\int\limits_{G_{\eta }}\left[ \int\limits_{0}^{t}\left( \left\vert \nabla
v\right\vert +\left\vert v\right\vert \right) d\tau \right] ^{2}\varphi
^{2}dxdt\leq \frac{M}{\lambda }\int\limits_{G_{\eta }}\left[ \left( \nabla
v\right) ^{2}+v^{2}\right] \varphi ^{2}dxdt.
\end{equation*}%
Hence, choosing sufficiently large $\lambda _{1}=\lambda _{1}\left(
M,K,G_{\eta },\mu _{1},\mu _{2}\right) >\lambda _{0},$ we obtain%
\begin{equation}
\lambda ^{3}\int\limits_{G_{\eta }}v^{2}\varphi ^{2}dxdt\leq C\lambda
^{3}\exp \left[ 2\lambda \eta ^{-\nu }\right] \int\limits_{\partial
_{2}G_{\eta }}\left[ \left( \nabla v\right) ^{2}+v^{2}\right] dS,\forall
\lambda >\lambda _{1}.  \label{3.59}
\end{equation}%
Let $\delta \in \left( 0,\eta -\alpha \right) $ be so small that the point $%
\left( x_{0},0\right) \in G_{\eta -\alpha }.$ Then making (\ref{3.59})
stronger, we obtain%
\begin{equation*}
\lambda ^{3}\exp \left[ 2\lambda \left( \eta -\delta \right) ^{-\nu }\right]
\int\limits_{G_{\eta -\delta }}v^{2}dxdt\leq C\lambda ^{3}\exp \left[
2\lambda \eta ^{-\nu }\right] \int\limits_{\partial _{2}G_{\eta }}\left[
\left( \nabla v\right) ^{2}+v^{2}\right] dS,\forall \lambda >\lambda _{1}.
\end{equation*}%
Dividing this inequality by $\lambda ^{3}\exp \left[ 2\lambda \left( \eta
-\delta \right) ^{-\nu }\right] ,$ we obtain%
\begin{equation*}
\int\limits_{G_{\eta -\delta }}v^{2}dxdt\leq C\exp \left\{ -2\lambda \left[
\left( \eta -\delta \right) ^{-\nu }-\eta ^{-\nu }\right] \right\} ,\forall
\lambda >\lambda _{1}.
\end{equation*}%
Setting here $\lambda \rightarrow \infty ,$ we obtain%
\begin{equation*}
\int\limits_{G_{\eta -\delta }}v^{2}dxdt=0.
\end{equation*}%
Hence, $v\left( x,t\right) =0$ in $G_{\eta -\delta }.$ Hence, (\ref{3.54})
implies that $\widetilde{u}\left( x,t\right) =0$ in $G_{\eta -\delta }.$
Substituting this in (\ref{3.48}) and using $D^{\alpha _{0}}u_{2}\left(
x,t\right) \neq 0$ in $\overline{G}_{\eta -\delta },$ we obtain $\widetilde{a%
}\left( x\right) =0$ in $G_{\eta -\delta }\cap \left\{ t=0\right\} .$ In
particular $\widetilde{a}\left( x_{0}\right) =0.$ It is clear that rotating
and moving the coordinate system, one can cover the entire domain $\Omega $
by paraboloids like $G_{\eta -\delta }\cap \left\{ t=0\right\} .$ Thus, $%
\widetilde{a}\left( x\right) \equiv 0.$ This, (\ref{3.48}), (\ref{3.50}),
Theorems 2.3, 2.4 and Remark 2.2 imply that $u_{1}\left( x,t\right)
=u_{2}\left( x,t\right) $ in $Q_{T}^{\pm }.$ $\square $

\subsubsection{An MCIP for a parabolic equation with final over determination%
}

\label{sec:3.3.3}

Let $D_{T}^{n+1}=\mathbb{R}^{n}\times \left( 0,T\right) $ and $L$ be the
elliptic operator in $\mathbb{R}^{n}$, whose coefficients depend only on $x,$%
\begin{eqnarray}
Lu &=&\sum\limits_{\left\vert \alpha \right\vert \leq 2}a_{\alpha }\left(
x\right) D_{x}^{\alpha }u,  \label{3.60} \\
a_{\alpha } &\in &C^{2+\beta }\left( \mathbb{R}^{n}\right) ,\beta \in \left(
0,1\right) ,  \label{3.61} \\
\mu _{1}\left\vert \xi \right\vert ^{2} &\leq &\sum\limits_{\left\vert
\alpha \right\vert =2}a_{\alpha }\left( x\right) \xi ^{\alpha }\leq \mu
_{2}\left\vert \xi \right\vert ^{2},\forall x,\xi \in \mathbb{R}^{n},\mu
_{1},\mu _{2}=const.>0.  \label{3.62}
\end{eqnarray}%
Consider the following Cauchy problem
\begin{eqnarray}
u_{t} &=&Lu\text{ in }D_{T}^{n+1},\quad u\in C^{4+\beta ,2+\beta /2}\left(
\overline{D}_{T}^{n+1}\right) ,  \label{3.63} \\
u|_{t=0} &=&f\left( x\right) \in C^{4+\beta }\left( \mathbb{R}^{n}\right) .
\label{3.64}
\end{eqnarray}%
It is well known that the problem (\ref{3.63}), (\ref{3.64}) has unique
solution \cite{LSU}.

\textbf{The Parabolic Coefficient Inverse Problem with\ Final
Overdetermination.} Let $T_{0}\in \left( 0,T\right) $ and $\Omega \subset
\mathbb{R}^{n}$ be a bounded domain. Suppose that the coefficient $a_{\alpha
_{0}}\left( x\right) $ of the operator $L$ is known inside of $\Omega $ and
is unknown outside of $\Omega $. Assume that the initial condition $f\left(
x\right) $ is also unknown. Determine both the coefficient $a_{\alpha
_{0}}\left( x\right) $ for $x\in \mathbb{R}^{n}\diagdown \Omega $ and the
initial condition $f\left( x\right) $ for $x\in \mathbb{R}^{n},$ assuming
that the following function $F\left( x\right) $ is known
\begin{equation}
F\left( x\right) =u\left( x,T_{0}\right) ,x\in \mathbb{R}^{n}.  \label{3.65}
\end{equation}

\textbf{Theorem 3.5}. \emph{Assume that conditions (\ref{3.60})-(\ref{3.62})
hold, all coefficients of the operator }$L$\emph{\ belong to }$C^{\infty
}\left( \Omega \right) $\emph{, and }
\begin{equation*}
D^{\alpha _{0}}F\left( x\right) \neq 0\text{, in }\mathbb{R}^{n}\diagdown
\Omega .
\end{equation*}%
\emph{Then there exists at most one pair vector function }$\left( u\left(
x,t\right) ,a_{\alpha _{0}}\left( x\right) ,f\left( x\right) \right) $\emph{%
\ satisfying conditions (\ref{3.63})-(\ref{3.65}).}

\textbf{Proof. }Consider the solution of the following hyperbolic Cauchy
problem
\begin{eqnarray*}
v_{tt} &=&Lv\text{ in }D_{\infty }^{n+1}, \\
v\left( x,0\right) &=&0,v_{t}\left( x,0\right) =f\left( x\right) .
\end{eqnarray*}%
Then by (\ref{3.39}) $u=\mathcal{R}v$. Hence, for any $x\in \mathbb{R}^{n}$
the function $u\left( x,t\right) $ is analytic as a function of the real
variable $t>0$. We now show that the function $u\left( x,t\right) $ can be
uniquely determined for $\left( x,t\right) \in \Omega \times \left(
0,T\right) .$ Since all coefficients $a_{\alpha }\in C^{\infty }\left(
\Omega \right) $, then $u\in C^{\infty }\left( \Omega \times \left(
0,T\right) \right) ,$ see the book of Friedman \cite{F}. Hence, using (\ref%
{3.63}) and (\ref{3.65}), we obtain
\begin{equation*}
D_{t}^{k+1}u\left( x,T_{0}\right) =L^{k}\left[ F\left( x\right) \right]
,x\in \Omega ,k=0,1,\ldots .
\end{equation*}%
This means that one can uniquely determine all $t$ derivatives of the
function $u\left( x,t\right) $ at $t=T_{0}$ for all $x\in \Omega .$ Hence,
the analyticity of the function $u\left( x,t\right) $ with respect to $t$
implies that this function can be uniquely determined for $\left( x,t\right)
\in \Omega \times \left( 0,T\right) .$ Next, applying Theorem 3.4, we obtain
that the coefficient $a_{\alpha _{0}}\left( x\right) $ is uniquely
determined in the domain $\mathbb{R}^{n}\diagdown \Omega .$ The classical
theorem about the uniqueness of the solution of the parabolic equation with
the reversed time \cite{F,LRS} implies that the initial condition $f\left(
x\right) $ is also uniquely determined. $\square $

\subsection{An MCIP for an elliptic equation}

\label{sec:3.4}

We now consider an elliptic analog of the Second Parabolic Coefficient
Inverse Problem. Let $\Omega \subset \mathbb{R}^{n}$ be either finite or
infinite domain with the piecewise smooth boundary $\partial \Omega $ and
let $\Gamma \subset \partial \Omega $ be a part of this boundary. Let $%
T=const>0$. We keep notations of Section 3.3. Let $L$ be an elliptic
operator in $\Omega $,
\begin{eqnarray}
Lu &=&\sum\limits_{\left\vert \alpha \right\vert \leq 2}a_{\alpha }\left(
x\right) D_{x}^{\alpha }u,x\in \Omega ,  \label{3.66} \\
a_{\alpha } &\in &C^{1}\left( \overline{\Omega }\right) ,\left\vert \alpha
\right\vert =2;a_{\alpha }\in C\left( \overline{\Omega }\right) ,\left\vert
\alpha \right\vert =0,1,  \label{3.67} \\
\mu _{1}\left\vert \xi \right\vert ^{2} &\leq &\sum\limits_{\left\vert
\alpha \right\vert =2}a_{\alpha }\left( x\right) \xi ^{\alpha }\leq \mu
_{2}\left\vert \xi \right\vert ^{2},\forall \xi \in \mathbb{R}^{n},\forall
x\in \Omega ;\mu _{1},\mu _{2}=const.>0.  \label{3.68}
\end{eqnarray}

\textbf{Coefficient Inverse Problem for an Elliptic Equation.} Let the
function $u\in C^{2}\left( \overline{Q}_{T}^{\pm }\right) $\ satisfies the
following conditions
\begin{eqnarray}
u_{tt}+Lu &=&F\left( x,t\right) \text{ in }Q_{T}^{\pm },  \label{3.69} \\
u\left( x,0\right) &=&f_{0}\left( x\right) \text{ in }\Omega ,  \label{3.70}
\\
u|_{\Gamma _{T}^{\pm }} &=&p\left( x,t\right) ,\quad \frac{\partial u}{%
\partial n}|_{\Gamma _{T}^{\pm }}=q\left( x,t\right) .  \label{3.71}
\end{eqnarray}%
Assume that the coefficient $a_{\alpha _{0}}\left( x\right) $ of the
operator $L$\ is unknown in $\Omega $ and all other coefficients are known.
Determine the coefficient $a_{\alpha _{0}}\left( x\right) $ from conditions (%
\ref{3.69})-(\ref{3.71}).

\textbf{Theorem 3.6}. \emph{Assume that }$D_{x}^{\alpha _{0}}f_{0}\left(
x\right) \neq 0$ \emph{in }$\overline{\Omega }$ \emph{and conditions (\ref%
{3.66})-(\ref{3.68}) hold. Then there exists at most one pair of functions }$%
\left( a_{\alpha _{0}}\left( x\right) ,u\left( x,t\right) \right) $\emph{\
satisfying (\ref{3.69})-(\ref{3.71}) and such that }$u\in C^{3+\left\vert
\alpha _{0}\right\vert }\left( \overline{Q}_{T}^{\pm }\right) .$\emph{\ }

It follows from Theorem 2.4 and Remark 2.2 that the proof of this theorem is
completely similar with the proof Theorem 3.4. Therefore, we omit this proof
here.

\section{Published Results About BK}

\label{sec:4}

Given a significant number of publications about BK, it would be quite space
consuming to tell details about the topic of each one. Therefore, the author
provides short comments about cited papers. An interesting reader is
referred to the corresponding paper for detail. Many works cited in this
section are devoted to either Lipschitz or H\"{o}lder global stability
estimates for MCIPs. H\"{o}lder stability estimate means $\left\Vert
a_{1}-a_{2}\right\Vert _{X}\leq C\left\Vert F_{1}-F_{2}\right\Vert
_{Y}^{\alpha },\alpha =const.\in \left( 0,1\right] .$ Here $a_{1}$ and $%
a_{2} $ are two unknown coefficients, corresponding to the data $F_{1}$ and $%
F_{2}$ respectively and $X$ and $Y$ are corresponding Banah spaces. The case
$\alpha =1$ is called Lipschitz stability estimate, which is obviously
stronger than $\alpha <1.$ This is why the Lipschitz stability estimate is
usually much harder to prove than the H\"{o}lder stability. On the other
hand, it was briefly mentioned in earlier papers of the author \cite%
{Klib3,Klib5} that usually the H\"{o}lder stability follows almost
immediately from BK.\ To do this, one needs to combine BK either with
Theorem 2.1 or with similar theorems of Chapter 4 of the book \cite{LRS}.
All stability estimates for finding coefficients mentioned in this section
are conditional stability estimates, as it is usually the case in the theory
of Ill-Posed problems \cite{BKok,BK,EHN,Kab,T}. In other words, some \emph{a
priori} upper bounds are imposed either on certain norms of coefficients of
interest or on certain norms of solutions of corresponding PDEs. The
constant $C$ depends on these bounds. In some works the problem of finding
an unknown coefficient is replaced with the problem of finding the function $%
f\left( x\right) $ in the source term $f\left( x\right) P\left( x,t\right) ,$
where the function $P\left( x,t\right) $ is known. This problem is almost
equivalent to a corresponding MCIP, since the interpretation of $f\left(
x\right) $ in this case is $f\left( x\right) =a_{1}\left( x\right)
-a_{2}\left( x\right) $, see (\ref{3.7}) and (\ref{3.7_1}). Since the
problem of finding the source term is linear, unlike an MCIP, then \emph{a
priori} bounds depend on the function $P\left( x,t\right) $.

\subsection{MCIPs for hyperbolic PDEs}

\label{sec:4 1}

Lipschitz stability is established for some of these MCIPs. This became
possible because the hyperbolic PDE can be solved in both forward and
backward directions of time. The idea behind Lipschitz stability estimates
is to use various combinations of BK with the Lipschitz stability estimate
for the Cauchy problem for the hyperbolic equation with Dirichlet and
Neumann data given at the lateral boundary of the time cylinder. For the
latter see Theorem 5.1 in Section 5.1 as well as originating papers of
Klibanov and Malinsky \cite{KlibM}\ and Kazemi and Klibanov \cite{Kaz}.

For the first time, the Lipschitz stability estimate for an MCIP with single
measurement data was obtained by Puel and Yamamoto \cite{Y1}. The initial
boundary value problem in \cite{Y1} is
\begin{eqnarray}
u_{tt} &=&\Delta u-p\left( x\right) u-f\left( x\right) P\left( x,t\right)
,\left( x,t\right) \in Q_{T},  \notag \\
u\left( x,0\right) &=&u_{t}\left( x,0\right) =0,  \label{7.1} \\
u &\mid &_{S_{T}}=0.  \notag
\end{eqnarray}%
The MCIP of \cite{Y1} consists in finding the source function $f\left(
x\right) ,$ assuming that the normal derivative $g\left( x,t\right)
=\partial _{n}u\mid _{S_{T}}$is known and $P\left( x,0\right) \neq 0$ in $%
\overline{\Omega }$. Hence, this MCIP can be obtained from the MCIP of
finding the unknown coefficient $p\left( x\right) $ by assuming that $%
f\left( x\right) =p_{1}\left( x\right) -p_{2}\left( x\right) ,$ where $p_{1}$
and $p_{2}$ are two possible coefficients. In this case in (\ref{7.1}) $%
p=p_{2},u=u_{1}-u_{2},P=u_{1},$ where $u_{1}$ and $u_{2}$ are solutions of
the problem (\ref{7.1}) with $p:=p_{1}$ and $p:=p_{2}$ respectively with the
same initial conditions
\begin{equation*}
u_{1}\left( x,0\right) =u_{2}\left( x,0\right) =P\left( x,0\right)
,u_{1t}\left( x,0\right) =u_{2t}\left( x,0\right)
\end{equation*}%
and the same Dirichlet boundary condition $u_{1}\mid _{S_{T}}=$ $u_{2}\mid
_{S_{T}}=P\mid _{S_{T}}.$ It was shown in \cite{Y1} that if $\Omega =\left\{
\left\vert x\right\vert <R\right\} $ and $T>2R,$ then
\begin{equation*}
\left\Vert f\right\Vert _{H^{3}\left( \Omega \right) }\leq C\left(
\sum\limits_{j=0}^{4}\left\Vert \partial _{t}^{j}g\right\Vert _{L_{2}\left(
S_{T}\right) }^{2}\right) ^{1/2}.
\end{equation*}%
Isakov and Yamamoto \cite{Y4} and Imanuvilov and Yamamoto \cite{Y2,Y3} have
obtained various\ Lipschitz stability estimates for MCIPs for hyperbolic
PDEs with the principal part of hyperbolic operators $\partial
_{t}^{2}-\Delta .$ Bellassoued \cite{Bell1} has proved Lipschitz stability
for an MCIP for the equation $u_{tt}=c^{2}\left( x\right) Au,$ where $A$ is
a self-adjoint elliptic operator of the second order with $x-$dependent
coefficients and $c\left( x\right) $ is the unknown coefficient.

Imanuvilov and Yamamoto \cite{Y5} have considered the case of the
determination of the coefficient $p\left( x\right) $ in equation%
\begin{equation}
u_{tt}=\func{div}\left( p\left( x\right) \nabla u\right) .  \label{7.8}
\end{equation}%
A new element here, compared with Theorem 3.1, is that both the function $p$
and its first derivatives are involved in equation (\ref{7.8}). On the other
hand, the machinery of Theorem 3.1 would require to consider derivatives $%
p_{x_{i}}$ as \textquotedblleft independent" functions.\ In turn, this would
require to use $n+1$ independent initial conditions. However, only one
initial condition was used in \cite{Y5}. H\"{o}lder stability estimate was
obtained in \cite{Y5}. This result was extended by Klibanov and Yamamoto
\cite{KYam} to the Lipschitz stability. The method of \cite{KYam} is
different in some respects from the one of \cite{Y5}, because in \cite{KYam}
a combination of ideas of Theorem 3.1 and Theorem 5.1 was used.

Publications cited above in this section used the assumption that the
Dirichlet boundary condition is known on the entire boundary $S_{T}$ and the
Neumann boundary condition is known on a part of the boundary$\partial
\Omega $ satisfying an appropriate geometrical condition. To obtain
logarithmic stability estimates for the case when the Neumann boundary
condition is known on an arbitrary piece of the boundary, Bellassoued \cite%
{Bell1} has proposed to use the so-called \textquotedblleft
Fourier-Bros-Iagolnitzer integral transformation" (FBI) with respect to $t$
. FBI transforms the hyperbolic equation in the elliptic one, where the
operator $\partial _{t}^{2}-\Delta $ is replaced with $\partial
_{t}^{2}+\Delta .$ While in \cite{Bell1} this was done for equation $%
u_{tt}=\Delta u-p\left( x\right) u,$ Bellassoued and Yamamoto \cite{Y6} have
extended this result to the case of equation (\ref{7.8}). We refer to
Robbiano \cite{Rob1,Rob2} for the introduction of the FBI transformation.

Liu and Triggiani \cite{Trig3} have considered the hyperbolic equation with
the damping term $q\left( x\right) u_{t},$
\begin{eqnarray*}
u_{tt} &=&\Delta u+q\left( x\right) u_{t},\left( x,t\right) \in Q_{T}, \\
u\left( x,\frac{T}{2}\right) &=&u_{0}\left( x\right) ,u_{t}\left( x,\frac{T}{%
2}\right) =u_{1}\left( x\right) , \\
\partial _{n}u &\mid &_{S_{T}}=g\left( x,t\right) .
\end{eqnarray*}%
The MCIP of \cite{Trig3} consists in determining the coefficient $q\left(
x\right) $ from the Dirichlet data $u\mid _{\Gamma _{T}}=f\left( x,t\right)
, $ where $\Gamma \subset \partial \Omega $ is a part of the boundary $%
\partial \Omega $ satisfying an appropriate geometrical condition. The
Lipschitz stability estimate for this MCIP was obtained in \cite{Trig3}.

As to other publications about the use of BK for MCIPs for hyperbolic PDEs,
we refer to works of Kha\u{\i}darov \cite{Khai}, Isakov \cite{Is0,IsMilan,Is}%
, Doubova and Osses \cite{DO}, Baudouin, Cr\'{e}peau and Velein \cite{Baud2}%
, Baudouin, Mercado and Osses \cite{BMO}, Yuan and Yamamoto \cite{Y21} and
Liu and Triggiani \cite{Trig4,Trig5}. All above results for MCIPs for
hyperbolic PDEs are obtained under assumptions like the one in (\ref{2.49}),
which is imposed on the coefficient $c\left( x\right) $ in the principal
part of the hyperbolic operator. This is because conditions like (\ref{2.49}%
) are the only known ones which guarantee the existence of the Carleman
estimate for the hyperbolic case on the one hand, and can be directly
analytically verified for generic functions on the other hand. Note that (%
\ref{2.49}) is valid of course for the case $c\left( x\right) \equiv 1.$

\subsection{MCIPs for parabolic PDEs}

\label{sec:4.2}

A comprehensive survey about the topic of this section as well as about some
related topics can be found in the paper of Yamamoto \cite{Y7}. Thus, the
author refers to \cite{Y7} for further references. It is a long standing
well known open problem to prove uniqueness theorems for MCIPs for parabolic
PDEs with single measurement in the case when the regular initial condition
is given at $\left\{ t=0\right\} $ and the equation is valid only for $t\in
\left( 0,T\right) .$ This is why the only case when the uniqueness can be
currently proven for this type of data is the one of Section 3.3.1, where
the inverse Reznickaya's transform (\ref{3.39}) was used to obtain the MCIP
for a similar hyperbolic PDE. Hence, we discuss in this section only the
case when the equation is valid for $t\in \left( -T,T\right) $ and the data
are given at $\left\{ t=0\right\} $ as well as on at least a part of the
lateral boundary. The only exception is the case of nonlinear parabolic PDEs.

An inconvenience of the conventional Carleman estimate of Theorem 2.4
(Section 2.3) is that it is valid only in the paraboloid $G_{\eta },$ which
is a subdomain of the time cylinder $Q_{T}^{\pm }.$ On the other hand,
Fursikov and Imanuvilov \cite{Furs, Im} have proved a radically new Carleman
estimate for an arbitrary parabolic operator of the second order. This
estimate is valid in the entire time cylinder $Q_{T}^{\pm }$, although the
Carleman Weight Function exponentially decays to zero at $t\rightarrow \pm
T. $ Using this fact, Imanuvilov and Yamamoto \cite{Y8} have proved, for the
first time, the Lipschitz stability estimate for an MCIP for a general
parabolic equation with $x-$dependent coefficients. In \cite{Y8} the
Dirichlet boundary condition for the solution of the forward problem was
known at the entire boundary, whereas the Neumann boundary condition was
known on any piece of the boundary. Starting from \cite{Y8}, Carleman
estimates of the Fursikov-Imanuvilov type became popular in the inverse
problems community, see, e.g. the papers of Baudouin and Puel \cite{Baud1}
and Cristofol, Gaitan and Ramoul \cite{Cr1}.

Yamamoto \cite{Y7} has obtained the H\"{o}lder stability estimate for the
case of the equation $u_{t}=\func{div}\left( p\left( x\right) \nabla
u\right) $ with the unknown coefficient $p\left( x\right) $ in the case when
two boundary conditions are given at any part $\Gamma $ of the boundary $%
\partial \Omega .$ The main difference between this result and the one of
Theorem 3.4 is that the machinery of Theorem 3.4 would require to treat
first derivatives of the function $p\left( x\right) $ as independent
functions, which would lead, in turn to the necessity to use $\left(
n+1\right) $ conditions at $\left\{ t=0\right\} .$ Unlike the latter, only
one condition at $\left\{ t=0\right\} $ is used in \cite{Y7}. It was pointed
out in Remark on page 41 of \cite{Y7} that if one boundary condition would
be known at the entire boundary $\partial \Omega $ and the second one would
be known only at $\Gamma ,$ then the Lipschitz stability estimate would be
obtained. In the latter case the Carleman estimate of the
Fursikov-Imanuvilov type would be used.

Yamamoto and Zou \cite{Y9} have considered an MCIP for the equation
\begin{eqnarray*}
u_{t} &=&\Delta u+p\left( x\right) u,\left( x,t\right) \in Q_{T}, \\
u &\mid &_{S_{T}}=\eta \left( x,t\right)
\end{eqnarray*}%
Let $\omega \subset \Omega $ be subdomain of the domain $\Omega .$ An
interesting new feature of this work is that the inverse problem consists in
the simultaneous reconstruction of both the coefficient $p\left( x\right) $
and the initial condition $\mu \left( x\right) =u\left( x,0\right) ,$
assuming that the following functions $f$ and $g$ are given%
\begin{equation*}
u\left( x,\theta \right) =f\left( x\right) ,u\mid _{\partial \omega \times
\left( 0,T\right) }=g\left( x,t\right) ,
\end{equation*}%
where $\theta =const.\in \left( 0,T\right) .$ First, using the technique of
\cite{Y8}, they proved Lipschitz stability estimate for the function $%
p\left( x\right) .$ Next, they proved logarithmic stability for the initial
condition $\mu \left( x\right) $ using the method of logarithmic convexity
of Payne \cite{Payne}. This method works for the case when the corresponding
elliptic operator is self-adjoint. In addition, they have constructed a
numerical method, which is based on the minimization of the Tikhonov
functional. A careful convergence analysis was provided. That analysis was
confirmed by a number of numerical experiments.

Surprisingly, L\"{u} \cite{Lu}, has applied BK, for the first time, to an
MCIP for the stochastic parabolic PDE.

The assumption of Theorem 3.4 that coefficients of the operator $L$ are
independent on $t$ was imposed only for the sake of simplicity. In fact, one
can allow all coefficients, except of $a_{\alpha _{0}}\left( x\right) ,$ to
be dependent on both $x$ and $t$. A direct analog of Theorem 3.4 is valid in
this case. To prove it, one should use the Carleman estimate of Theorem 2.5
(for $L_{0}^{\left( 1\right) }),$ a certain change of variables and the
assumption that the target coefficient $a_{\alpha _{0}}\left( x\right) $ is
known for $x\in \Gamma ,$ see works of the author \cite{Klib2,Klib4,Klib5}
as well as Theorem 1.10.7 in the book \cite{BK}. The same is true for
Theorem 3.6. This idea was used in the works of the author discussed in the
next paragraph.

BK was also applied to MCIPs for nonlinear parabolic PDEs. In the 1d case
the author has considered the inverse problem for the equation
\begin{equation*}
u_{t}=F\left( u_{xx},u_{x},u,q_{1}\left( u\right) ,...,q_{n}\left( u\right)
\right) ,\left( x,t\right) \in \left( 0,1\right) \times \left( 0,T\right) ,
\end{equation*}%
where $\partial _{u_{xx}}F\left( y\right) \neq 0,\forall y\in \mathbb{R}%
^{n+3}$ \cite{KlibPar1}. Let $\left\{ x_{i}\right\} _{i=1}^{n+1}\subset
\left( 0,1\right) $ be a sequence of points, such that $x_{i}\neq x_{j}$ if $%
i\neq j.$ The inverse problem in \cite{KlibPar1} consists in determining the
vector function $q\left( u\right) =\left( q_{1},...,q_{n}\right) \left(
u\right) ,$ assuming that the functions $f_{i}\left( t\right) =u\left(
x_{i},t\right) ,i\in \left[ 1,n+1\right] $ are known. It was assumed that $%
u_{x}>0$ in $\left[ 0,1\right] \times \left[ 0,T\right] .$ This inequality
can often be established via the maximum principle. Uniqueness theorem was
proved in \cite{KlibPar1}. The first step was to introduce a new spatial
variable $z$ and a new function $v\left( z,t\right) $ via $u\left( v\left(
z,t\right) ,t\right) :=z.$

In \cite{KlibPar2} \ and Chapter 4 of \cite{KT} the author considered an
MCIP for the nonlinear parabolic equation
\begin{equation*}
u_{t}=F\left(
u_{x_{1}x_{1}},...,u_{x_{n}x_{n}},u_{x_{1}},...,u_{x_{n}},u,x,t,q\left(
u,x_{2},...,x_{n}\right) \right) ,x\in \left\{ x_{1}\in \left( 0,1\right)
,y\in \Omega ^{\prime }\right\} ,t\in \left( 0,T\right) ,
\end{equation*}%
where $y=\left( x_{2},...,x_{n}\right) ,$ and $\partial
_{u_{x_{i}x_{i}}}F\left( z\right) \in \left[ \mu _{1},\mu _{2}\right]
,\forall z\in \mathbb{R}^{3n+3},$ where $\mu _{1},\mu _{2}=const.>0.$ Here $%
\Omega ^{\prime }\subseteq \mathbb{R}^{n-1}$ is an arbitrary domain. The
following functions $\varphi _{1},\varphi _{2},\psi _{1},\psi _{2}$ were
given in \cite{KlibPar2}%
\begin{equation*}
\varphi _{1}\left( y,t\right) =u\left( 0,y,t\right) ,\psi _{1}\left(
y,t\right) =u_{x_{1}}\left( 0,y,t\right) ,\varphi _{2}\left( y,t\right)
=u\left( 1,y,t\right) ,\psi _{2}\left( y,t\right) =u_{x_{1}}\left(
1,y,t\right) ,\left( y,t\right) \in \Omega ^{\prime }\times \left(
0,T\right) .
\end{equation*}%
It was required to reconstruct the function $q\left(
u,x_{2},...,x_{n}\right) .$ The first step was again to introduce a new
spatial variable $z$ and a new function $v\left( z,y,t\right) $ via $u\left(
v\left( z,y,t\right) ,t\right) :=z.$ Next, uniqueness theorem was proved.
However, a stability estimate was not established in \cite{KlibPar2}.\
Furthermore, unlike the linear case, H\"{o}lder stability estimate does not
follow automatically from BK in this nonlinear case. Thus, the H\"{o}lder
stability estimate for a similar inverse problem was proved in the paper of
Egger, Engl and Klibanov \cite{EEK} for the case of the equation
\begin{equation*}
u_{t}=u_{xx}+q\left( u\right) +f\left( x,t\right)
\end{equation*}%
with the unknown function $q.$ In addition, a numerical reconstruction
procedure was developed in \cite{EEK} via minimizing the Tikhonov
functional. Numerical results were also presented in \cite{EEK}. Other MCIPs
for nonlinear parabolic PDEs were treated via BK in Boulakia, Grandmont and
Osses \cite{Boul} and Kaltenbacher and Klibanov \cite{Kalt}.

Some other MCIPs for parabolic PDEs were treated by various modifications of
BK in papers of Bellassoued and Yamamoto \cite{Y20}, Benabdallah, Dermenjian
and Le Rousseau \cite{Ben2}, Benabdallah, Gaitan and Le Rousseau \cite{Ben3}
and Poisson \cite{Poisson}.

Isakov \cite{Is0,Is} has proved uniqueness of the parabolic MCIP with the
final overdetermination without the assumption of Theorem 3.5 of the
knowledge of the target coefficient in the domain $\Omega $.\ The initial
condition is also known in \cite{Is0,Is}. BK was not used in \cite{Is0,Is}.

\subsection{MCIPs for the non-stationary Schr\"{o}dinger equation}

\label{sec:4.3}

Baudouin and Puel \cite{Baud1} were the first ones who has applied BK to the
MCIP for the non-stationary Schr\"{o}dinger equation%
\begin{eqnarray}
iu_{t}+\Delta u+q\left( x\right) u &=&0,\left( x,t\right) \in Q_{T},i^{2}=-1,
\notag \\
u\left( x,0\right) &=&u_{0}\left( x\right) ,  \label{7.9} \\
u &\mid &_{S_{T}}=h\left( x,t\right) .  \notag
\end{eqnarray}
The inverse problem in \cite{Baud1} consists in determining the coefficient $%
q\left( x\right) $ from the Neumann boundary condition
\begin{equation}
\partial _{n}u\mid _{\Gamma _{T}}=g\left( x,t\right) ,\Gamma _{T}=\Gamma
\times \left( 0,T\right) ,  \label{7.10}
\end{equation}%
where $\Gamma \subseteq \partial \Omega $ is \ a part of the boundary
satisfying an appropriate geometrical condition. First, following the idea
of \cite{Y8}, an analog of the Carleman estimate of Fursikov and Imanuvilov
\cite{Furs, Im} was proved. Next, the Lipschitz stability for the MCIP (\ref%
{7.9}), (\ref{7.10}) was established.

For three other results of the topic of this section we refer to works of
Baudouin and Mercado \cite{Baud3}, Mercado, Osses and Rosier \cite{Mercado}
and Yuan and Yamamoto \cite{Y22}.

\subsection{Non-standard PDEs}

\label{sec:4.4}

Baudouin, Cerpa, Cr\'{e}peau and Mercado \cite{Baud4} have considered the
Kuramoto-Sivashinsky equation (KS) in 1d with $Q_{T}=\left( 0,1\right)
\times \left( 0,T\right) $, $_{x}$
\begin{eqnarray}
u_{t}+\left( \sigma \left( x\right) u_{xx}\right) _{xx}+\gamma \left(
x\right) u_{xx}+u\cdot u_{x} &=&g,  \label{7.11} \\
u\left( x,0\right) &=&u_{0}\left( x\right) ,  \label{7.12}
\end{eqnarray}%
\begin{eqnarray}
u &\mid &_{x=0}=h_{1}\left( t\right) ,u\mid _{x=1}=h_{2}\left( t\right) ,
\label{7.13} \\
u_{x} &\mid &_{x=0}=h_{3}\left( t\right) ,u_{x}\mid _{x=1}=h_{4}\left(
t\right) .  \label{7.14}
\end{eqnarray}%
Note that KS is a nonlinear equation. Conditions (\ref{7.11})-(\ref{7.14})
define the forward problem for KS. First, an existence, uniqueness and
stability theorem for this problem was proved in \cite{Baud4}. In the
inverse problem the following functions were assumed to be known%
\begin{equation*}
u\left( x,T_{0}\right) ,u_{xx}\left( 0,t\right) ,u_{xxx}\left( 0,t\right) ,
\end{equation*}%
where $T_{0}=const.\in \left( 0,T\right) .$ As to the knowledge of the
function $u\left( x,T_{0}\right) ,$ see the arguments in the beginning of
Section 4.2. The Lipschitz stability for the inverse problem was proved in
\cite{Baud4}.

Cavaterra, Lorenzi and Yamamoto \cite{Lorenzi} considered an MCIP for a PDE
whose principal part was a hyperbolic operator $\partial
_{t}^{2}-c^{2}\left( x\right) \Delta $ and lower order terms included
Volterra integrals
\begin{equation}
\int\limits_{0}^{t}\left( \cdot \right) d\tau .  \label{7.15}
\end{equation}%
The Lipschitz stability estimate was obtained in \cite{Lorenzi}. Romanov and
Yamamoto \cite{Rom3} obtained the H\"{o}lder stability estimate for an MCIP
for a hyperbolic-like PDE with integrals (\ref{7.15}). We also refer to the
work of Buhan and Osses \cite{Buhan}, where logarithmic stability estimate
for a hyperbolic-like coupler system of PDEs with integrals like the one in (%
\ref{7.15}) was obtained, a numerical method, based on the minimization of
the Tikhonov functional, was developed, and numerical results were presented.

\subsection{Coupled systems of PDEs}

\label{sec:4.5}

In this section we do not present coupled systems under discussion because
they are well known. Still, we present two systems of PDEs, which are not
conventionally known. An MCIP for the Maxwell equations was considered by
the author in \cite{KlibMaxw}. Let $\varepsilon \left( x\right) $ and $%
\sigma \left( x\right) ,x\in \mathbb{R}^{3}$ be the dielectric permittivity
and the electric conductivity coefficients respectively. It was required to
find both of them simultaneously, given the magnetic vector field $\mathbf{H}%
\left( x,t\right) $ outside of the domain of interest.\ It was assumed that
the magnetic permeability coefficient $\mu \left( x\right) \equiv 1.$
Uniqueness theorem was proved. We also refer to the book of Romanov and
Kabanikhin \cite{RomKab} for inverse problems for the Maxwell's system with
impulsive sources.

Yamamoto \cite{Y40} studied an inverse source problem for the Maxwell's
system. In this case the unknown source vector function $\mathbf{f}\left(
x\right) $ is three dimensional. Both electric and magnetic vector fields
were known at the boundary. Uniqueness theorem was proved. We also refer to
the papers of Li \cite{LiSIAM} and Li and Yamamoto \cite{LiYam} for the
cases of bi-isotropic and anisotropic Maxwell's system respectively. In \cite%
{LiSIAM} the Lipschitz stability estimate was proved, and in the paper \cite%
{LiYam} the H\"{o}lder stability was establishes.\ Note that in \cite{LiYam}
the unknown coefficients are actually matrices of dielectric and magnetic
permeability coefficients which are independent on one spatial variable but
dependent on time $t$.

Bellassoued, Cristofol and Soccorsi \cite{Bell2} obtained H\"{o}lder
stability estimate for an MCIP for the Maxwell's system. In this case both
functions $\varepsilon \left( x\right) $ and $\mu \left( x\right) $ were
unknown and $\sigma \left( x\right) \equiv 0.$ Tangential components of both
magnetic and electric field were measured at the boundary for two sets of
initial conditions, i.e. measurements were conducted twice.

Imanuvilov, Isakov and Yamamoto \cite{Y10} considered the MCIP of the
reconstruction of all three elastic coefficients $\lambda \left( x\right)
,\mu \left( x\right) ,\rho \left( x\right) $ in the time dependent Lam\'{e}
system. Two sets of initial data were used and they have generated two sets
of boundary conditions, which were used as the data for MCIP. The H\"{o}lder
stability estimate was obtained.

In \cite{Y10} the data for the inverse problem were given at the entire
boundary. Unlike this, Bellassoued, Imanuvilov, and Yamamoto \cite{Y11}
considered the MCIP of recovering of elastic coefficients $\lambda \left(
x\right) ,\mu \left( x\right) ,\rho \left( x\right) $ of the time dependent
Lam\'{e} system in the case when the data for the MCIP are given at an
arbitrary part $\Gamma \subset \partial \Omega $ of the boundary for $t\in
\left( 0,T\right) .$ In other words, the previous idea of Bellassoued \cite%
{Bell1} and Bellassoued and Yamamoto \cite{Y6} (Section 4.1) was extended
from a single hyperbolic equation to the to the case of Lam\'{e} system.
Similarly with \cite{Bell1,Y6} the Fourier-Bros-Iagolnitzer transformation
was applied. Logarithmic stability estimate was obtained in \cite{Y11}.

Liu and Triggiani \cite{Trig2} considered the following $2\times 2$ coupled
system of Schr\"{o}dinger equations in $Q_{T}$%
\begin{eqnarray*}
iw_{t}+\Delta w &=&m\left( x\right) \cdot \nabla w+n\left( x\right)
w+\varsigma \left( x\right) \cdot \nabla z+p\left( x\right) z, \\
iz_{t}+\Delta z &=&\mu \left( x\right) \cdot \nabla z+\sigma \left( x\right)
z+\psi \left( x\right) \cdot \nabla w+q\left( x\right) w, \\
w\left( x,\frac{T}{2}\right) &=&w_{0}\left( x\right) ,z\left( x,\frac{T}{2}%
\right) =z_{0}\left( x\right) , \\
\partial _{n}w &\mid &_{S_{T}}=g_{1}\left( x\right) ,\partial _{n}z\mid
_{S_{T}}=g_{2}\left( x\right) .
\end{eqnarray*}%
The MCIP in \cite{Trig2} consists in the determination of the pair of
unknown coefficients $\left( p,q\right) \left( x\right) ,x\in \Omega $ from
the Dirichlet boundary data
\begin{equation*}
w\mid _{\Gamma _{T}}=f_{1}\left( x,t\right) ,z\mid _{\Gamma
_{T}}=f_{2}\left( x,t\right) ,
\end{equation*}%
where $\Gamma \subset \partial \Omega $ is a part of the boundary satisfying
an appropriate geometrical condition. Uniqueness theorem was proved in \cite%
{Trig2}.

Fan, Di Cristo, Jiang and Nakamura \cite{Fan} proved Lipschitz stability
estimate for an MCIP for Navier-Stokes equations. The MCIP consists in
recovering of the viscosity function from a single boundary measurement.

Wu and Liu \cite{Wu2} considered the MCIP for the thermoelastic system with
memory (also, see their preceding work \cite{Wu1}). That system is
\begin{eqnarray*}
\mathbf{u}_{tt}-\alpha \Delta \mathbf{u}-\beta \nabla \func{div}\mathbf{u+}%
\gamma \nabla v &=&\sigma \left( x,t\right) \mathbf{p}\left( x\right)
,\left( x,t\right) \in Q_{T}, \\
v_{t}-\int\limits_{0}^{t}k\left( t-\tau \right) \Delta v\left( x,\tau
\right) d\tau +\gamma \func{div}\mathbf{u}_{t} &=&0,\left( x,t\right) \in
Q_{T}, \\
\mathbf{u}\left( x,0\right) &=&\mathbf{u}_{0}\left( x\right) ,\mathbf{u}%
_{t}\left( x,0\right) =\mathbf{u}_{1}\left( x\right) ,v\left( x,0\right)
=v_{1}\left( x\right) , \\
\mathbf{u} &\mid &_{S_{T}}=0,v\mid _{S_{T}}=0.
\end{eqnarray*}%
Here $\mathbf{u}=\left( u_{1},u_{2},u_{3}\right) ^{T}$ and $v$ are
displacement and temperature respectively. Let $\omega \subset \Omega $ be a
subdomain. The MCIP in \cite{Wu2} consists in the recovery of the source
term $\mathbf{p}\left( x\right) $, given the vector function $\mathbf{f}%
\left( x,t\right) =\mathbf{u}\left( x,t\right) $ for $\left( x,t\right) \in
\omega \times \left( 0,T\right) .$ The function $\sigma \left( x,t\right) $
is assumed to be known. Lipschitz stability estimate for this MCIP was
obtained in \cite{Wu2}.

Cristofol, Gaitan and Ramoul \cite{Cr1} considered the following parabolic $%
2\times 2$ system%
\begin{eqnarray*}
u_{t} &=&\Delta u+a\left( x\right) u+b\left( x\right) v,\left( x,t\right)
\in Q_{T}, \\
v_{t} &=&\Delta v+c\left( x\right) u+d\left( x\right) v,\left( x,t\right)
\in Q_{T}, \\
u\left( x,0\right) &=&u_{0}\left( x\right) ,v\left( x,0\right) =v_{0}\left(
x\right) . \\
u &\mid &_{S_{T}}=g\left( x,t\right) ,v\mid _{S_{T}}=h\left( x,t\right) .
\end{eqnarray*}%
This is a reaction-diffusion system. Let $\omega \subset \Omega $ be a
subdomain, $t_{0}\in \left( 0,T\right) $ and $T^{\prime }=\left(
t_{0}+T\right) /2$. The data for the MCIP in \cite{Cr1} are the following
functions%
\begin{equation*}
\Delta u\left( x,T^{\prime }\right) ,u\left( x,T^{\prime }\right) ,v\left(
x,T^{\prime }\right) ,v_{t}\mid _{\omega \times \left( t_{0},T\right) }.
\end{equation*}%
The MCIP consists in determining the vector function $\left(
b,u_{0},v_{0}\right) .$The function $b$ can be replaced with the function $a$%
. A new point of \cite{Cr1} is that only one function $v_{t}$ is measured in
$\omega ,$ unlike the conventional way of measuring both functions $u$ and $%
v.$ The Lipschitz stability estimate for the coefficient $b\left( x\right) $
as well as the logarithmic stability estimate for initial conditions $%
u_{0}\left( x\right) ,v_{0}\left( x\right) $ were obtained in \cite{Cr1}.

\section{Stability Estimates for Hyperbolic Equations and Inequalities With
Lateral Cauchy Data and Thermoacoustic Tomography}

\label{sec:5}

It was shown in Section 2.2 how Carleman estimates lead to H\"{o}lder
stability estimates for ill-posed Cauchy problems for PDEs, including
inequalities. In this section we obtain Lipschitz and logarithmic stability
estimates for the case of hyperbolic PDEs. We also show how these estimates
help to specify QRM for the hyperbolic case. In Section 5.5 we cite related
published results.

\subsection{Lipschitz stability}

\label{sec:5.1}

As usual, we consider the simplest case when the domain of interest $\Omega $
is a ball, $\Omega =\left\{ \left\vert x\right\vert <R\right\} \subset
\mathbb{R}^{n}$, where $R=const.>0.$ Let functions $c\left( x\right)
,b_{j}\left( x,t\right) $ satisfy the following conditions%
\begin{eqnarray}
c\left( x\right) &\in &C^{1}\left( \overline{\Omega }\right) ,c\left(
x\right) \neq 0\text{ in }\overline{\Omega },  \label{4.1} \\
b_{j} &\in &C\left( \overline{Q}_{T}^{\pm }\right) ,j=0,...,n+1.  \label{4.2}
\end{eqnarray}%
Let the function $u\in C^{2}\left( \overline{Q}_{T}^{\pm }\right) $ and the
function $f\in C\left( \overline{Q}_{T}^{\pm }\right) .$ Suppose that the
function $u$ is a solution of the following hyperbolic equation%
\begin{equation}
u_{tt}=c^{2}\left( x\right) \Delta u+\sum\limits_{j=1}^{n+1}b_{j}\left(
x,t\right) u_{x_{j}}+b_{0}\left( x,t\right) u+f\left( x,t\right) ,\left(
x,t\right) \in Q_{T}^{\pm },  \label{4.3}
\end{equation}%
where $u_{n+1}:=u_{t}.$ Consider Dirichlet and Neumann boundary conditions
for the function $u$ at the lateral side $S_{T}^{\pm }$ of the time cylinder
$Q_{T}^{\pm },$%
\begin{equation}
u\mid _{S_{T}^{\pm }}=p\left( x,t\right) ,\partial _{n}u\mid _{S_{T}^{\pm
}}=q\left( x,t\right) .  \label{4.4}
\end{equation}

\textbf{Problem 5.1}. Given conditions (\ref{4.3}), (\ref{4.4}), estimate
the function $u\in C^{2}\left( \overline{Q}_{T}^{\pm }\right) $ in the time
cylinder $Q_{T}^{\pm }$ via functions $p$, $q$ and $f$.

The of this section works for a more general case of a hyperbolic
inequality. Specifically, we consider the following problem.

\textbf{Problem 5.2}. Let $A=const.>0.$ Let the function $u\in C^{2}\left(
\overline{Q}_{T}^{\pm }\right) $ satisfies the following pointwise
inequality in the cylinder $Q_{T}^{\pm }$%
\begin{equation}
\left\vert u_{tt}-c^{2}\left( x\right) \Delta u\right\vert \leq A\left(
\left\vert \nabla u\right\vert +\left\vert u_{t}\right\vert +\left\vert
u\right\vert +\left\vert f\right\vert \right) ,\forall \left( x,t\right) \in
Q_{T}^{\pm },  \label{4.5}
\end{equation}%
where the function $f\in L_{2}\left( Q_{T}^{\pm }\right) .$ Estimate the
function $u$ via functions $p$, $q$ and $f$.

Since Problem 5.2 is more general than Problem 5.1, we study only Problem
5.2 in this section. Theorem 4.1 provides the Lipschitz stability estimate
for Problem 5.2. We now reformulate condition (\ref{2.49}) for the case of
the operator $\partial _{t}^{2}-c^{2}\left( x\right) \Delta $ in a stronger
form as
\begin{equation}
\left( x,\nabla c^{-2}\left( x\right) \right) \geq \alpha =const.>0\text{ in
}\overline{\Omega },  \label{4.6}
\end{equation}%
where $\alpha >0$ is a certain number and $\left( ,\right) $ is the scalar
product in $\mathbb{R}^{n}$. Hence, there exists a sufficiently small number
$\varepsilon =\varepsilon \left( \alpha ,\left\Vert \nabla c^{-2}\right\Vert
_{C^{1}\left( \overline{\Omega }\right) }\right) \in \left( 0,R\right) $
such that%
\begin{equation}
\left( x-x_{0},\nabla c^{-2}\left( x\right) \right) \geq \frac{\alpha }{2}>0%
\text{ in }\overline{\Omega },\forall x_{0}\in \left\{ \left\vert
x_{0}\right\vert \leq \varepsilon \right\} .  \label{4.6_1}
\end{equation}%
Inequality (\ref{4.6_1}) guarantees the Carleman estimate for the operator $%
\partial _{t}^{2}-c^{2}\left( x\right) \Delta $ (Theorem 2.5). This estimate
is also guaranteed if $c\equiv const.\neq 0.$

\textbf{Theorem 5.1}. \emph{Let the domain }$\Omega =\left\{ \left\vert
x\right\vert <R\right\} .$\emph{\ Let }$\alpha >0$ \emph{and }$d>1$\ \ \emph{%
be certain numbers. Let }%
\begin{equation}
c\in C^{1}\left( \overline{\Omega }\right) \emph{,}c^{-2}\left( x\right) \in %
\left[ 1,d\right] ,\left\Vert \nabla c^{-2}\right\Vert _{C^{1}\left(
\overline{\Omega }\right) }\leq d.  \label{4.6_2}
\end{equation}
\emph{In the case }$c\neq const.$\emph{\ we assume that condition (\ref{4.6}%
) is fulfilled. Let the function }$u\in C^{2}\left( \overline{Q}_{T}^{\pm
}\right) $\emph{\ satisfies inequality (\ref{4.5}) with Dirichlet and
Neumann boundary conditions (\ref{4.4}). Then there exists a constant }$\eta
_{0}=\eta _{0}\left( R,d,\alpha \right) \in \left( 0,1\right] $\emph{\ such
that if }%
\begin{equation}
T>\frac{R}{\sqrt{\eta _{0}}},  \label{4.7}
\end{equation}%
\emph{then with a constant }$K=K\left( A,R,T,d,\alpha \right) =const.>0$%
\emph{\ the following Lipschitz stability estimate holds for the function }$%
u,$%
\begin{equation}
\left\Vert u\right\Vert _{H^{1}\left( Q_{T}^{\pm }\right) }\leq K\left[
\left\Vert p\right\Vert _{H^{1}\left( S_{T}^{\pm }\right) }+\left\Vert
q\right\Vert _{L_{2}\left( S_{T}^{\pm }\right) }+\left\Vert f\right\Vert
_{L_{2}\left( Q_{T}^{\pm }\right) }\right] .  \label{4.8}
\end{equation}%
\emph{In particular, if }$c\left( x\right) \equiv 1,$\emph{\ then one can
take }$\eta _{0}=1$\emph{\ and in (\ref{4.7}) }$T>R$\emph{.}

\textbf{Proof}. In this proof $K=K\left( A,R,T,d,\alpha \right) $ denotes
different positive constants depending on listed parameters. We note first
that the constant $P\left( x_{0},\Omega \right) $ in (\ref{2.49_1}) can be
estimated as $P\left( x_{0},\Omega \right) \leq 2R.$ Hence, we can set in
Theorem 2.5 $\eta _{0}=\eta _{0}\left( R,d,\left\Vert c^{-2}\right\Vert
_{C^{1}\left( \overline{\Omega }\right) }\right) \in \left( 0,1\right) .$ By
(\ref{4.7}) we can choose a sufficiently small $\varepsilon $ in (\ref{4.6_1}%
) and then choose $\eta $ such that%
\begin{equation}
\eta \in \left( \frac{\left( R+\varepsilon \right) ^{2}}{T^{2}},\eta
_{0}\right) \subset \left( 0,1\right) ,  \label{4.7_1}
\end{equation}%
Similarly with Section 2.4 let $\gamma =const\in \left( 0,\varepsilon
^{2}/9\right) ,$%
\begin{eqnarray*}
\xi \left( x,t\right) &=&\left\vert x\right\vert ^{2}-\eta t^{2},\varphi
\left( x,t\right) =\exp \left[ \lambda \xi \left( x,t\right) \right] ,\left(
x,t\right) \in Q_{T}^{\pm }, \\
G_{\gamma } &=&\left\{ \xi \left( x,t\right) >\gamma ,x\in \Omega \right\}
=\left\{ \left\vert x\right\vert ^{2}-\eta t^{2}>\gamma ,x\in \Omega
\right\} ,
\end{eqnarray*}%
where $\lambda >1$ is a large parameter which we define later. By (\ref%
{4.7_1})
\begin{equation}
\overline{G}_{\gamma }\subset \left\{ \left\vert t\right\vert <T\right\} .
\label{4.9}
\end{equation}%
Choose a sufficiently small number $\delta $ such that $\gamma +3\delta \in
\left( 0,\varepsilon ^{2}/9\right) .$ Hence,%
\begin{equation}
G_{\gamma +3\delta }\neq \varnothing \text{ and }G_{\gamma +3\delta }\subset
G_{\gamma +2\delta }\subset G_{\gamma +\delta }\subset G_{\gamma }.
\label{4.9_1}
\end{equation}

Introduce a function $\chi _{\delta }\in C^{2}\left( \overline{Q}_{T}^{\pm
}\right) $ satisfying%
\begin{equation}
\chi _{\delta }\left( x,t\right) =\left\{
\begin{array}{c}
1,\left( x,t\right) \in G_{\gamma +2\delta }, \\
0,\left( x,t\right) \in Q_{T}^{\pm }\diagdown G_{\gamma +\delta }, \\
\text{between 0 and 1 otherwise.}%
\end{array}%
\right.  \label{4.10}
\end{equation}%
The existence of such functions is well known from the Real Analysis course.
Let%
\begin{equation}
v\left( x,t\right) =u\left( x,t\right) \chi _{\delta }\left( x,t\right) .
\label{4.11}
\end{equation}%
Multiplying both sides of (\ref{4.5}) by $\chi _{\delta }$ and using (\ref%
{4.4}), (\ref{4.9}), (\ref{4.10}) and (\ref{4.11}), we obtain%
\begin{equation}
\left\vert v_{tt}-c^{2}\left( x\right) \Delta v\right\vert \leq K\left(
\left\vert \nabla v\right\vert +\left\vert v_{t}\right\vert +\left\vert
v\right\vert +\left\vert f\right\vert \right) +K\left( 1-\chi _{\delta
}\right) \left( \left\vert \nabla u\right\vert +\left\vert u_{t}\right\vert
+\left\vert u\right\vert \right) ,\left( x,t\right) \in G_{\gamma },
\label{4.12}
\end{equation}%
\begin{equation}
v\mid _{S_{T}^{\pm }}=\chi _{\delta }\varphi ,\partial _{n}v\mid
_{S_{T}^{\pm }}=\chi _{\delta }\psi +\varphi \partial _{n}\chi _{\delta }.
\label{4.13}
\end{equation}%
Squaring both sides of (\ref{4.12}) and using Theorem 2.5, we obtain%
\begin{equation*}
K\left( \left\vert \nabla v\right\vert ^{2}+v_{t}^{2}+v^{2}+f^{2}\right)
\varphi ^{2}+K\left( 1-\chi _{\delta }\right) \left( \left\vert \nabla
u\right\vert ^{2}+u_{t}^{2}+u^{2}\right) \varphi ^{2}
\end{equation*}%
\begin{equation*}
\geq C\lambda \left( \left\vert \nabla v\right\vert ^{2}+v_{t}^{2}\right)
\varphi ^{2}+C\lambda ^{3}v^{2}\varphi ^{2}+\func{div}U+V_{t}\text{, in }%
G_{\gamma },\forall \lambda \geq \lambda _{0}.
\end{equation*}%
where the vector function $\left( U,V\right) $ satisfies conditions (\ref%
{2.50}), (\ref{2.51}) with the replacement of $u$ by $v$. Hence, (\ref{4.10}%
) implies that $U=V=0$ on $\left\{ \left( x,t\right) :\xi \left( x,t\right)
=\gamma ,x\in \Omega \right\} .$ Hence, integrating the letter inequality
over $G_{\gamma }$ and using Gauss-Ostrogradsky formula and (\ref{4.13}), we
obtain%
\begin{equation*}
C\int\limits_{G_{\gamma }}\lambda \left( \left\vert \nabla v\right\vert
^{2}+v_{t}^{2}\right) \varphi ^{2}dxdt+C\lambda ^{3}\int\limits_{G_{\gamma
}}v^{2}\varphi ^{2}dxdt\leq K\int\limits_{G_{\gamma }}\left( \left\vert
\nabla v\right\vert ^{2}+v_{t}^{2}+v^{2}+g^{2}\right) \varphi ^{2}
\end{equation*}%
\begin{equation*}
+K\exp \left[ 2\lambda \left( \gamma +2\delta \right) \right] \left\Vert
u\right\Vert _{H^{1}\left( Q_{T}^{\pm }\right) }^{2}+Ke^{2\lambda
R^{2}}\left( \left\Vert p\right\Vert _{H^{1}\left( S_{T}^{\pm }\right)
}^{2}+\left\Vert q\right\Vert _{L_{2}\left( S_{T}^{\pm }\right) }^{2}\right)
.
\end{equation*}%
Let $\lambda _{0}>1$ be the number of Theorem 2.5. There exists a number $%
\lambda _{1}=\lambda _{1}\left( \lambda _{0},C,K\right) \geq \lambda _{0}$
such that $C\lambda /2>K,\forall \lambda \geq \lambda _{1}.$ Hence,%
\begin{equation*}
\lambda \int\limits_{G_{\gamma }}\left( \left\vert \nabla v\right\vert
^{2}+v_{t}^{2}+v^{2}\right) \varphi ^{2}dxdt\leq
\end{equation*}%
\begin{equation*}
\leq K\exp \left[ 2\lambda \left( \gamma +2\delta \right) \right] \left\Vert
u\right\Vert _{H^{1}\left( Q_{T}^{\pm }\right) }^{2}+Ke^{2\lambda R^{2}}%
\left[ \left\Vert p\right\Vert _{H^{1}\left( S_{T}^{\pm }\right)
}^{2}+\left\Vert q\right\Vert _{L_{2}\left( S_{T}^{\pm }\right)
}^{2}+\left\Vert g\right\Vert _{L_{2}\left( S_{T}^{\pm }\right) }^{2}\right]
.
\end{equation*}%
By (\ref{4.9_1}) and (\ref{4.10})
\begin{eqnarray*}
\lambda \int\limits_{G_{\gamma }}\left( \left\vert \nabla v\right\vert
^{2}+v_{t}^{2}+v^{2}\right) \varphi ^{2}dxdt &\geq &\lambda
\int\limits_{G_{\gamma +3\delta }}\left( \left\vert \nabla v\right\vert
^{2}+v_{t}^{2}+v^{2}\right) \varphi ^{2}dxdt \\
&\geq &\exp \left[ 2\lambda \left( \gamma +3\delta \right) \right]
\left\Vert u\right\Vert _{H^{1}\left( G_{\gamma +3\delta }\right) }^{2}.
\end{eqnarray*}%
Hence,%
\begin{equation}
\left\Vert u\right\Vert _{H^{1}\left( G_{\gamma +2\delta }\right) }^{2}\leq
K\exp \left( -2\lambda \delta \right) \left\Vert u\right\Vert _{H^{1}\left(
Q_{T}^{\pm }\right) }^{2}+Ke^{2\lambda R^{2}}\left[ \left\Vert p\right\Vert
_{H^{1}\left( S_{T}^{\pm }\right) }^{2}+\left\Vert q\right\Vert
_{L_{2}\left( S_{T}^{\pm }\right) }^{2}+\left\Vert f\right\Vert
_{L_{2}\left( Q_{T}^{\pm }\right) }^{2}\right] .  \label{4.14}
\end{equation}%
Note that
\begin{equation}
G_{\gamma +3\delta }\cap \left\{ t=0\right\} =\left\{ x:\left\vert
x\right\vert \in \left( \sqrt{\gamma +3\delta },R\right) \right\} \supset
\left\{ x:\left\vert x\right\vert \in \left( \frac{\varepsilon }{3},R\right)
\right\} .  \label{4.15}
\end{equation}

Choose now a point $x_{0}$ such that $\left\vert x_{0}\right\vert =3\sqrt{%
\gamma +3\delta }.$ Then $\left\vert x_{0}\right\vert <\varepsilon .$
Consider now an arbitrary point $y\in \left\{ \left\vert x\right\vert \leq
\sqrt{\gamma +3\delta }\right\} .$ Then%
\begin{equation*}
\left\vert y-x_{0}\right\vert \geq \left\vert x_{0}\right\vert -\left\vert
y\right\vert =3\sqrt{\gamma +3\delta }-\left\vert y\right\vert \geq 3\sqrt{%
\gamma +3\delta }-\sqrt{\gamma +3\delta }=2\sqrt{\gamma +3\delta }>\sqrt{%
\gamma +3\delta }.
\end{equation*}%
Hence,
\begin{equation}
\left\{ \left\vert x\right\vert \leq \sqrt{\gamma +3\delta }\right\} \subset
\left\{ \left\vert y-x_{0}\right\vert >\sqrt{\gamma +3\delta }\right\} .
\label{4.16}
\end{equation}%
Introduce now the domain $G_{\gamma }\left( x_{0}\right) $ as%
\begin{equation*}
G_{\gamma }\left( x_{0}\right) =\left\{ \left\vert x-x_{0}\right\vert
^{2}-\eta t^{2}>\gamma ,x\in \Omega \right\} .
\end{equation*}%
It follows from (\ref{4.15}) and (\ref{4.16}) that there exists a
sufficiently small number $\sigma =\sigma \left( \varepsilon \right) $ such
that%
\begin{equation}
\left\{ t\in \left( 0,\sigma \right) \right\} \subset \left[ G_{\gamma
+3\delta }\cup G_{\gamma +3\delta }\left( x_{0}\right) \right] .
\label{4.17}
\end{equation}%
Since $\gamma +3\delta \in \left( 0,\varepsilon ^{2}/9\right) $ and $%
\left\vert x_{0}\right\vert =3\sqrt{\gamma +3\delta },$ then (\ref{4.7_1})
implies that $\overline{G}_{\gamma }\left( x_{0}\right) \subset \left\{
\left\vert t\right\vert <T\right\} .$ Next, we use (\ref{4.6_1}) and Theorem
2.5 and obtain similarly with (\ref{4.14})
\begin{equation*}
\left\Vert u\right\Vert _{H^{1}\left( G_{\gamma +3\delta }\left(
x_{0}\right) \right) }^{2}\leq K\exp \left( -2\lambda \delta \right)
\left\Vert u\right\Vert _{H^{1}\left( Q_{T}^{\pm }\right) }^{2}+Ke^{2\lambda
R^{2}}\left[ \left\Vert p\right\Vert _{H^{1}\left( S_{T}^{\pm }\right)
}^{2}+\left\Vert q\right\Vert _{L_{2}\left( S_{T}^{\pm }\right)
}^{2}+\left\Vert f\right\Vert _{L_{2}\left( Q_{T}^{\pm }\right) }^{2}\right]
.
\end{equation*}%
Combining this with (\ref{4.14}), we obtain%
\begin{equation*}
\left\Vert u\right\Vert _{H^{1}\left( G_{\gamma +2\delta }\cup G_{\gamma
+2\delta }\left( x_{0}\right) \right) }^{2}\leq K\exp \left( -2\lambda
\delta \right) \left\Vert u\right\Vert _{H^{1}\left( Q_{T}^{\pm }\right)
}^{2}+Ke^{2\lambda R^{2}}\left[ \left\Vert p\right\Vert _{H^{1}\left(
S_{T}^{\pm }\right) }^{2}+\left\Vert q\right\Vert _{L_{2}\left( S_{T}^{\pm
}\right) }^{2}+\left\Vert f\right\Vert _{L_{2}\left( Q_{T}^{\pm }\right)
}^{2}\right] .
\end{equation*}%
This, (\ref{4.17}) and the mean value theorem imply that there exists a
number $t_{0}\in \left[ 0,\sigma \right] $ such that
\begin{eqnarray}
&&\left\Vert u\left( x,t_{0}\right) \right\Vert _{H^{1}\left( \Omega \right)
}^{2}+\left\Vert u_{t}\left( x,t_{0}\right) \right\Vert _{L_{2}\left( \Omega
\right) }^{2}  \label{4.18} \\
&\leq &K\exp \left( -2\lambda \delta \right) \left\Vert u\right\Vert
_{H^{1}\left( Q_{T}^{\pm }\right) }^{2}+Ke^{2\lambda R^{2}}\left[ \left\Vert
p\right\Vert _{H^{1}\left( S_{T}^{\pm }\right) }^{2}+\left\Vert q\right\Vert
_{L_{2}\left( S_{T}^{\pm }\right) }^{2}+\left\Vert f\right\Vert
_{L_{2}\left( Q_{T}^{\pm }\right) }^{2}\right] .  \notag
\end{eqnarray}

Let $u_{tt}-c^{2}\left( x\right) u:=Z\left( x,t\right) .$ Then by (\ref{4.5}%
)
\begin{equation}
\left\vert Z\right\vert \leq A\left( \left\vert \nabla u\right\vert
+\left\vert u_{t}\right\vert +\left\vert u\right\vert +\left\vert
f\right\vert \right) ,\forall \left( x,t\right) \in Q_{T}^{\pm }.
\label{4.300}
\end{equation}%
Consider the initial boundary value problem with reversed time%
\begin{eqnarray}
u_{tt}-c^{2}\left( x\right) u &=&Z\left( x,t\right) ,\forall \left(
x,t\right) \in \left\{ x\in \Omega ,t\in \left( -T,t_{0}\right) \right\} ,
\label{4.301} \\
u\left( x,t_{0}\right) &=&u_{0}\left( x\right) ,u_{t}\left( x,t_{0}\right)
=u_{1}\left( x\right) ,  \label{4.302} \\
u &\mid &_{\left( x,t\right) \in \partial \Omega \times \left(
-T,t_{0}\right) }=p\left( x,t\right) .  \label{4.303}
\end{eqnarray}%
Next, consider the same initial boundary value problem but in the time
cylinder $\left( x,t\right) \in \left\{ x\in \Omega ,t\in \left(
t_{0},T\right) \right\} .$ Recall that the hyperbolic equation can be solved
in both positive and negative directions of time. Hence, the standard method
of energy estimates being applied to two latter problems combined with
inequalities (\ref{4.18}) and (\ref{4.300}) leads to
\begin{equation}
\left\Vert u\right\Vert _{H^{1}\left( Q_{T}^{\pm }\right) }^{2}\leq K\exp
\left( -2\lambda \delta \right) \left\Vert u\right\Vert _{H^{1}\left(
Q_{T}^{\pm }\right) }^{2}+Ke^{2\lambda R^{2}}\left[ \left\Vert p\right\Vert
_{H^{1}\left( S_{T}^{\pm }\right) }^{2}+\left\Vert q\right\Vert
_{L_{2}\left( S_{T}^{\pm }\right) }^{2}+\left\Vert f\right\Vert
_{L_{2}\left( Q_{T}^{\pm }\right) }^{2}\right] .  \label{4.19}
\end{equation}%
Choosing $\lambda =\lambda \left( K,\delta \right) $ so large that $K\exp
\left( -2\lambda \delta \right) \leq 1/2,$ we obtain the target estimate (%
\ref{4.8}) from (\ref{4.19}). $\square $

To apply QRM (Section 5.4.2), we need Theorem 5.2. The proof of this theorem
is almost identical with the proof of Theorem 5.1.

\textbf{Theorem 5.2}. \emph{Let the function }$u\in C^{2}\left( \overline{Q}%
_{T}^{\pm }\right) $\emph{\ satisfies the Dirichlet and Neumann boundary
conditions (\ref{4.4}) as well as the following integral inequality}%
\begin{equation*}
\int\limits_{Q_{T}^{\pm }}\left( u_{tt}-c^{2}\left( x\right) \Delta
u-\sum\limits_{j=1}^{n+1}b_{j}\left( x,t\right) u_{x_{j}}-b_{0}\left(
x,t\right) u\right) ^{2}dxdt\leq S^{2}
\end{equation*}%
\emph{and the rest of conditions of Theorem 5.1 is in place. Here }$%
S=const.>0.$\emph{\ Then there exists a constant }$\eta _{0}=\eta _{0}\left(
R,d,\alpha \right) \in \left( 0,1\right] $\emph{\ such that if }$T>R/\sqrt{%
\eta _{0}},$\emph{\ then with a constant }$K=K\left( A,R,T,d,\alpha \right)
=const.>0$\emph{\ the following Lipschitz stability estimate holds for the
function }$u,$%
\begin{equation*}
\left\Vert u\right\Vert _{H^{1}\left( Q_{T}^{\pm }\right) }\leq B\left[
\left\Vert p\right\Vert _{H^{1}\left( S_{T}^{\pm }\right) }+\left\Vert
q\right\Vert _{L_{2}\left( S_{T}^{\pm }\right) }+S\right] .
\end{equation*}%
\emph{In particular, if }$c\left( x\right) \equiv 1,$\emph{\ then one can
take }$\eta _{0}=1$\emph{\ and }$T>R$\emph{.}

\subsection{Thermoacoustic tomography}

\label{sec:5.2}

In thermoacoustic tomography (TAT) a short radio frequency pulse is sent in
a biological tissue, see papers of Agranovsky and Kuchment \cite{AK} and
Finch and Rakesh \cite{FR}. Some energy is absorbed. It is well known that
malignant legions absorb much more energy than healthy ones. Then the tissue
expands and radiates a pressure wave.\ The propagation of this wave can be
modeled as the solution of the following Cauchy problem
\begin{eqnarray}
u_{tt} &=&c^{2}\left( x\right) \Delta u,x\in \mathbb{R}^{3},  \label{4.20} \\
u\left( x,0\right) &=&f\left( x\right) ,u_{t}\left( x,0\right) =0.
\label{4.21}
\end{eqnarray}%
The function $u\left( x,t\right) $ is measured by transducers at certain
locations either at the boundary of the medium of interest or outside of
this medium. The function $f\left( x\right) $ characterizes the absorption
of the medium. Hence, if one would know the function $f\left( x\right) $,
then one would know locations of malignant spots. The inverse problem
consists in determining the initial condition $f\left( x\right) $ using
those measurements. Hence, we obtain the following inverse problem

\textbf{Inverse Problem 5.1 (IP5.1).} Let $\Omega \subset \mathbb{R}^{3}$ be
a bounded domain with $\partial \Omega \in C^{1}.$ Consider the Cauchy
problem (\ref{4.20}), (\ref{4.21}). Suppose that
\begin{equation}
f\left( x\right) =0,x\in \mathbb{R}^{3}\diagdown \Omega .  \label{4.22}
\end{equation}%
Determine the function $f\left( x\right) $ inside of the domain $\Omega $
assuming that the following function $p\left( x,t\right) $ is known%
\begin{equation}
u\mid _{S_{T}}=p\left( x,t\right) .  \label{4.23}
\end{equation}

We show below in this section how Theorem 5.1 implies the Lipschitz
stability estimate for IP5.1. Assume that $c\left( x\right) =1$ for $x\in
\mathbb{R}^{3}\diagdown \Omega .$ Then (\ref{4.20})-(\ref{4.23}) imply that
the function $u\left( x,t\right) $ is the solution of the following initial
boundary value problem in $\left( \mathbb{R}^{3}\diagdown \Omega \right)
\times \left( 0,T\right) $%
\begin{eqnarray}
u_{tt} &=&\Delta u,x\in \mathbb{R}^{3}\diagdown \Omega ,t\in \left(
0,T\right) ,  \notag \\
u\left( x,0\right) &=&u_{t}\left( x,0\right) =0,x\in \mathbb{R}^{3}\diagdown
\Omega ,  \label{4.24} \\
u &\mid &_{S_{T}}=p\left( x,t\right) .  \notag
\end{eqnarray}%
Under certain well known conditions the solution $u\in H^{2}\left( \left(
\mathbb{R}^{3}\diagdown \Omega \right) \times \left( 0,T\right) \right) $ of
this problem exists, is unique and depends continuously on the boundary data
$p\left( x,t\right) $. Just as above consider the case $\Omega =\left\{
\left\vert x\right\vert <R\right\} .$ For $\left\vert x\right\vert \geq R$
let $\overline{p}\left( x,t\right) =p\left( x,t\right) \left\vert
x\right\vert R^{-1}\rho \left( x\right) ,$ where the function $\rho \in
C^{2}\left( \left\vert x\right\vert \geq R\right) ,\rho \left( x\right) =0$
for $x\in \left\{ \left\vert x\right\vert \geq 3R\right\} $ and $\rho \left(
x\right) =1$ for $x\in \left\{ \left\vert x\right\vert \in \left[ R,2R\right]
\right\} .$ Let $v\left( x,t\right) =u\left( x,t\right) -\overline{p}\left(
x,t\right) .$ Substituting $v$ in equations (\ref{4.24}) and using the
standard method of energy estimates (see, e.g. the book of Ladyzhenskaya
\cite{Lad} for this method), we obtain the following stability estimate $%
\left\Vert u\right\Vert _{H^{2}\left( \left( \mathbb{R}^{3}\diagdown \Omega
\right) \times \left( 0,T\right) \right) }\leq C\left\Vert p\right\Vert
_{H^{3}\left( S_{T}\right) }.$ Let $q\left( x,t\right) =\partial _{n}u\mid
_{S_{T}}.$ Hence, trace theorem implies that
\begin{equation}
\left\Vert q\right\Vert _{L_{2}\left( S_{T}\right) }=\left\Vert \partial
_{n}u\right\Vert _{L_{2}\left( S_{T}\right) }\leq C\left\Vert p\right\Vert
_{H^{3}\left( S_{T}\right) }.  \label{4.25}
\end{equation}%
Here $C>0$ denotes different positive constants depending only on $R$ and
the function $\rho \left( x\right) .$

Consider now the even extension $\widetilde{u}\left( x,t\right) $ with
respect to $t$ of the function $u\left( x,t\right) .$ Assuming that $u\in
C^{2}\left( \overline{Q}_{T}\right) ,$ we obtain $\widetilde{u}\in
C^{2}\left( \overline{Q}_{T}^{\pm }\right) $ and also%
\begin{eqnarray}
\widetilde{u}_{tt} &=&c^{2}\left( x\right) \Delta \widetilde{u},\left(
x,t\right) \in Q_{T}^{\pm },  \label{4.26} \\
\widetilde{u} &\mid &_{S_{T}^{\pm }}=\widetilde{p}\left( x,t\right)
,\partial _{n}\widetilde{u}\mid _{S_{T}^{\pm }}=\widetilde{q}\left(
x,t\right) ,  \label{4.27}
\end{eqnarray}%
where functions $\widetilde{p}$ and $\widetilde{q}$ are even extensions of
functions $p$ and $q$ respectively. In addition, by (\ref{4.21})
\begin{equation}
\widetilde{u}\left( x,0\right) =f\left( x\right) .  \label{4.28}
\end{equation}%
Hence, Theorem 5.1, (\ref{4.25})-(\ref{4.28}) and trace theorem imply
Theorem 5.3. We need here $u\in H^{4}\left( \mathbb{R}^{3}\times \left(
0,T\right) \right) $ in order to make sure that the function $p\in
H^{3}\left( S_{T}\right) .$

\textbf{Theorem 5.3}. \emph{Let the domain }$\Omega =\left\{ \left\vert
x\right\vert <R\right\} .$\emph{\ Let the function }$c\in C^{1}\left(
\mathbb{R}^{3}\right) ,c=1$\emph{\ in }$\mathbb{R}^{3}\diagdown \Omega $%
\emph{\ and also }$c$ \emph{satisfies conditions (\ref{4.6_2}). In the case }%
$c\neq const.$\emph{\ we assume that condition (\ref{4.6}) is fulfilled.
Suppose that functions }$c\left( x\right) $\emph{\ and }$f\left( x\right) $%
\emph{\ are such that there exists the solution }$u\in C^{2}\left( \mathbb{R}%
^{3}\times \left[ 0,T\right] \right) \cap H^{4}\left( \mathbb{R}^{3}\times
\left( 0,T\right) \right) $\emph{\ of the Cauchy problem (\ref{4.20}), (\ref%
{4.21}).\ Also, let condition (\ref{4.22}) be satisfied. Then there exists a
constant }$\eta _{0}=\eta _{0}\left( R,d,\alpha \right) \in \left( 0,1\right]
$\emph{\ such that if }$T>R/\sqrt{\eta _{0}}$\emph{, then with a constant }$%
K=K\left( A,R,T,d,\alpha \right) >0$\emph{\ the following Lipschitz
stability estimate holds for IP5.1 }%
\begin{equation*}
\left\Vert f\right\Vert _{L_{2}\left( \Omega \right) }\leq K\left\Vert
p\right\Vert _{H^{3}\left( S_{T}\right) }.
\end{equation*}%
\emph{In particular, if }$c\left( x\right) \equiv 1,$\emph{\ then one can
take }$\eta _{0}=1$\emph{\ and }$T>R$\emph{.}

\subsection{Logarithmic stability in the case of a general hyperbolic
operator of the second order with $x-$dependent coefficients}

\label{sec:5.3}

Condition (\ref{4.6}) is used in Theorem 5.1 because it is linked with the
existence of the Carleman estimate for the hyperbolic case, see the end of
Section 4.1. Clearly (\ref{4.6}) is a restrictive condition.\ Therefore, the
next question is whether a stability estimate can be obtained for an analog
of IP5.1 in the case of an arbitrary hyperbolic operator of the second
order. For the first time, this question was positively addressed by the
author in \cite{Kltherm}. We formulate main results of \cite{Kltherm} in
this section without proofs.

The idea is to apply an analog of the above Reznickaya's transform (\ref%
{3.39}). This way the hyperbolic PDE is transformed in a similar parabolic
PDE. And the function $f\left( x\right) $ becomes the initial condition for
that parabolic PDE. On the other hand, logarithmic stability estimates for
the inverse problem of determination of the initial condition of a general
parabolic equation from lateral Cauchy data were obtained by Klibanov \cite%
{Kl1} in the case of a finite domain $\Omega \subset \mathbb{R}^{n}$ and by
Klibanov and Tikhonravov \cite{KT2} in the case of an infinite domain $%
\Omega \subseteq \mathbb{R}^{n}.$ Thus, modifications of these results can
be applied. Results of both publications \cite{Kl1,KT2} were obtained via
Carleman estimates. The difference between logarithmic stability estimates
for initial conditions in \cite{Kl1,KT2} and those in the book of Payne \cite%
{Payne} is that in \cite{Kl1,KT2} the case of a general elliptic operator
with coefficients depending on $\left( x,t\right) $ was considered. On the
other hand, the technique of \cite{Payne} works only for self-adjoint
elliptic operators with coefficients depending only on $x$. Since the method
of \cite{Kl1,KT2} works not only for parabolic PDEs but for integral
inequalities as well (see Theorem 2.2 in \cite{Kltherm} as well as Theorem
2.3 in Section 2.2 above), then it enables us to prove convergence of the
QRM, unlike the technique of \cite{Payne}.

We refer to Li, Yamamoto and Zou \cite{LYZ} for another logarithmic
stability estimate of the initial condition of a parabolic equation with the
self-adjoint operator $L$ in a finite domain. A Carleman estimate was also
used in this reference. An interesting feature of \cite{LYZ} is that
observations are performed on an internal subdomain for times $t\in \left(
\tau ,T\right) $ where $\tau >0.$ In addition, a numerical method was
developed in \cite{LYZ}.

\subsubsection{Statements of inverse problems}

\label{sec:5.3.1}

Let $\Omega \subset \left\{ x_{1}>0\right\} $ be a bounded domain with the
boundary $\partial \Omega \in C^{3}$. Denote $P=\left\{ x_{1}=0\right\}
,P_{T}=P\times \left( 0,T\right) ,\forall T>0.$Let $k\geq 0$ be an integer
and $\alpha \in \left( 0,1\right) $. Consider the elliptic operator $L$ of
the second order,
\begin{eqnarray}
Lu &=&\sum\limits_{i,j=1}^{n}a_{i,j}\left( x\right)
u_{x_{i}x_{j}}+\sum\limits_{j=1}^{n}b_{j}\left( x\right)
u_{x_{j}}+b_{0}\left( x\right) u,x\in \mathbb{R}^{n},  \label{1.2} \\
a_{i,j} &\in &C^{k+\alpha }\left( \mathbb{R}^{n}\right) \cap C^{1}\left(
\mathbb{R}^{n}\right) ,b_{j},b_{0}\in C^{k+\alpha }\left( \mathbb{R}%
^{n}\right) ,k\geq 2,\alpha \in \left( 0,1\right) ,  \label{1.3} \\
\mu _{1}\left\vert \eta \right\vert ^{2} &\leq
&\sum\limits_{i,j=1}^{n}a_{i,j}\left( x\right) \eta _{i}\eta _{j}\leq \mu
_{2}\left\vert \eta \right\vert ^{2},\forall x,\eta \in \mathbb{R}^{n};\mu
_{1},\mu _{2}=const.>0.  \label{1.4}
\end{eqnarray}%
Let the function $f\left( x\right) $ be such that%
\begin{equation}
f\in C^{p}\left( \mathbb{R}^{n}\right) ,p\geq 3,f\left( x\right) =0,x\in
\mathbb{R}^{n}\diagdown \Omega .  \label{1.5}
\end{equation}%
Consider the following Cauchy problem%
\begin{eqnarray}
u_{tt} &=&Lu,x\in \mathbb{R}^{n},t\in \left( 0,\infty \right) ,  \label{1.6}
\\
u\left( x,0\right) &=&f\left( x\right) ,u_{t}\left( x,0\right) =0.
\label{1.7}
\end{eqnarray}%
We use everywhere below in Section 5 the following assumption.

\textbf{Assumption 5.1}. We assume that in (\ref{1.3}), (\ref{1.5}) integers
$k\geq 2,p\geq 4$, coefficients of the operator $L$ and the initial
condition $f$ are such that there exists unique solution $u\in C^{4}\left(
\mathbb{R}^{n}\times \left[ 0,T\right] \right) ,\forall T>0$ of the problem (%
\ref{1.6}), (\ref{1.7}) satisfying
\begin{equation}
\left\Vert u\right\Vert _{C^{4}\left( \mathbb{R}^{n}\times \left[ 0,T\right]
\right) }\leq Be^{dT},\forall T>0,  \label{1.171}
\end{equation}%
where the constants $B=B\left( L,\overline{B}\right) >0,d=d\left( L,%
\overline{B}\right) >0$ depend only from the coefficients of the operator $L$
and the upper estimate $\overline{B}$ of the norm $\left\Vert f\right\Vert
_{C^{p}\left( \overline{\Omega }\right) }\leq \overline{B}.$

Note that (\ref{1.5}) as well as the finite speed of propagation of the
solution of problem (\ref{1.6}), (\ref{1.7}) guarantee that the function $%
u\left( x,t\right) $ has a finite support $\Psi \left( T\right) \subset
\mathbb{R}^{n},\forall t\in \left( 0,T\right) ,\forall T>0,$ see, e.g. the
book of Ladyzhenskaya \cite{Lad}. Hence, $C^{4}\left( \mathbb{R}^{n}\times %
\left[ 0,T\right] \right) $ in Assumption 5.1 is actually the space $%
C^{4}\left( \overline{\Psi \left( T\right) }\times \left[ 0,T\right] \right)
.$ Using the classical tool of energy estimates \cite{Lad}, one can easily
find non-restrictive sufficient conditions imposed on coefficients of the
operator $L$ and the function $f$ guaranteeing the smoothness $u\in
C^{4}\left( \mathbb{R}^{n}\times \left[ 0,T\right] \right) ,\forall T>0$ as
well as (\ref{1.171}). We are not doing this here for brevity. We consider
the following two Inverse Problems.

\textbf{Inverse Problem 5.2 (IP5.2).}\emph{\ Suppose that conditions (\ref%
{1.2})-(\ref{1.5}) and Assumption 5.1 hold. Let }

$u\in C^{4}\left( \mathbb{R}^{n}\times \left[ 0,T\right] \right) ,\forall
T>0 $\emph{\ be the solution of the problem (\ref{1.6}), (\ref{1.7}). Assume
that the function }$f\left( x\right) $\emph{\ is unknown. Determine this
function, assuming that the following function }$\varphi _{1}\left(
x,t\right) $\emph{\ is known}%
\begin{equation}
u\mid _{S_{\infty }}=\varphi _{1}\left( x,t\right) .  \label{1.8}
\end{equation}

\textbf{Inverse Problem 5.3 (IP5.3).} \emph{Suppose that conditions (\ref%
{1.2})-(\ref{1.5}) and Assumption 5.1 hold. Let }$u\in C^{4}\left( \mathbb{R}%
^{n}\times \left[ 0,T\right] \right) ,\forall T>0$\emph{\ be the solution of
the problem (\ref{1.6}), (\ref{1.7}). Assume that the function }$f\left(
x\right) $\emph{\ is unknown. Determine this function, assuming that the
following function }$\varphi _{2}\left( x,t\right) $\emph{\ is known}%
\begin{equation}
u\mid _{x\in P_{\infty }}=\varphi _{2}\left( x,t\right) .  \label{1.9}
\end{equation}

IP5.2 has complete data collection, since the function $\varphi _{1}$ is
known at the entire boundary of the domain of interest $\Omega .$ On the
other hand, IP5.3 is a special case of incomplete data collection, since $%
\Omega \subset \left\{ x_{1}>0\right\} .$

In stability estimates one is usually interested to see how the solution
varies for a small variation of the input data. Therefore, following (\ref%
{1.171}), (\ref{1.8}) and (\ref{1.9}), we assume that in the case of IP5.2%
\begin{equation}
\left\Vert \varphi _{1}\right\Vert _{C^{4}\left( \overline{S}_{T}\right)
}\leq \delta e^{dT},\forall T>0,  \label{1.142}
\end{equation}%
and in the case of IP5.3%
\begin{equation}
\left\Vert \varphi _{2}\right\Vert _{C^{4}\left( \overline{P}_{T}\right)
}\leq \delta e^{dT},\forall T>0,  \label{1.143}
\end{equation}%
where $\delta \in \left( 0,1\right) $ is a sufficiently small number. Note
that it is not necessary that $\delta =B,$ where $B$ is the number from (\ref%
{1.171}). Indeed, while the number $B$ is involved in the estimate of the
norm $\left\Vert u\right\Vert _{C^{4}\left( \mathbb{R}^{n}\times \left[ 0,T%
\right] \right) },\forall T>0$ in the entire space, the number $\delta $ is
a part of the estimate of the norm of the boundary data for both IP5.2 and
IP5.3.

\textbf{Remarks 5.1.}

\textbf{1}. The number $\delta $ can be viewed as an upper estimate of the
level of the error in the data $\varphi _{1},\varphi _{2}.$ Hence, Theorems
5.3 and 5.4 below address the question of estimating variations of the
solution $f$ of either IP5.2 or IP5.3 via the upper estimate of the level of
the error in the data.

\textbf{2}. Since the kernel of the transform $\mathcal{L}$ in (\ref{1.10})
decays rapidly with $\tau \rightarrow \infty ,$ then the condition $t\in
\left( 0,\infty \right) $ in (\ref{1.8}), (\ref{1.9}) is not a serious
restriction from the applied standpoint. In addition, if having the data in (%
\ref{1.8}), (\ref{1.9}) only on a finite time interval $t\in \left(
0,T\right) $ and knowing an upper estimate of a norm of the function $f$ in (%
\ref{1.7}), one can estimate the error in the integral (\ref{1.10}) when
integrating over $\tau \in \left( T,\infty \right) .$ Next, this error can
be incorporated in the stability estimates of theorems of this section.

\subsubsection{Transformation to the parabolic case}

\label{sec:5.3.2}

Consider the following analog of the Reznickaya's transform (\ref{3.39})
\begin{equation}
\mathcal{L}g=\overline{g}\left( t\right) =\frac{1}{\sqrt{\pi t}}%
\int\limits_{0}^{\infty }\exp \left( -\frac{\tau ^{2}}{4t}\right) g\left(
\tau \right) d\tau .  \label{1.10}
\end{equation}%
The transformation (\ref{1.10}) is valid for, e.g. all functions $g\in C%
\left[ 0,\infty \right) $ which satisfy $\left\vert g\left( t\right)
\right\vert \leq A_{g}e^{k_{g}t},$ where $A_{g}$ and $k_{g}$ are positive
constants depending on $g$. It follows from (\ref{1.171})\ that the solution
$u\left( x,t\right) $ of the problem (\ref{1.6}), (\ref{1.7}) satisfies this
condition together with its derivatives up to the fourth order. Obviously%
\begin{equation*}
\frac{\partial }{\partial t}\left[ \frac{1}{\sqrt{\pi t}}\exp \left( -\frac{%
\tau ^{2}}{4t}\right) \right] =\frac{\partial ^{2}}{\partial \tau ^{2}}\left[
\frac{1}{\sqrt{\pi t}}\exp \left( -\frac{\tau ^{2}}{4t}\right) \right] .
\end{equation*}%
Hence,%
\begin{equation}
\mathcal{L}\left( g^{\prime \prime }\right) =\overline{g}^{\prime }\left(
t\right) ,\forall g\in C^{2}\left[ 0,\infty \right) \text{ such that }%
g^{\prime }\left( 0\right) =0.  \label{1.101}
\end{equation}%
Changing variables in (\ref{1.10}) $\tau \Leftrightarrow z,z:=\tau /2\sqrt{t}%
,$ we obtain $\lim_{t\rightarrow 0^{+}}\overline{g}\left( t\right) =g\left(
0\right) .$ Denote
\begin{equation}
v:=\mathcal{L}u.  \label{1.100}
\end{equation}%
It follows from (\ref{1.171}) and (\ref{1.101}) that
\begin{equation}
v\in C^{2+\alpha ,1+\alpha /2}\left( \mathbb{R}^{n}\times \left[ 0,T\right]
\right) ,\forall \alpha \in \left( 0,1\right) ,\forall T>0.  \label{1.102}
\end{equation}%
By (\ref{1.6}), (\ref{1.7}) and (\ref{1.102}) the function $v\left(
x,t\right) $ is the solution of the following parabolic Cauchy problem%
\begin{eqnarray}
v_{t} &=&Lv,x\in \mathbb{R}^{n},t>0,  \label{1.11} \\
v\left( x,0\right) &=&f\left( x\right) .  \label{1.12}
\end{eqnarray}%
We refer here to the well known uniqueness result for the solution $v\in
C^{2+\alpha ,1+\alpha /2}\left( \mathbb{R}^{n}\times \left[ 0,T\right]
\right) ,\forall T>0$ of the problem (\ref{1.11}), (\ref{1.12}), see, e.g.
the book of Ladyzhenskaya, Solonnikov and Uralceva \cite{LSU}.

Below in Sections 5.3.2 and 5.3.3 we work only with the function $v$. Thus,
we set everywhere below $T:=1$ for the sake of definiteness. Denote $%
S_{1}=\partial \Omega \times \left( 0,1\right) ,P_{1}=P\times \left(
0,1\right) ,$%
\begin{equation}
\mathcal{L\varphi }_{1}:=\overline{\varphi }_{1}\left( x,t\right) =v\mid
_{S_{1}},\text{ }\mathcal{L\varphi }_{2}:=\overline{\varphi }_{2}\left(
x,t\right) =v\mid _{P_{1}}.  \label{1.13}
\end{equation}%
Then%
\begin{equation}
\overline{\varphi }_{1}\in C^{2+\alpha ,1+\alpha /2}\left( \overline{S}%
_{1}\right) ,\overline{\varphi }_{2}\in C^{2+\alpha ,1+\alpha /2}\left(
\overline{P}_{1}\right) .  \label{1.14}
\end{equation}%
Let
\begin{equation}
\overline{\psi }_{1}\left( x,t\right) =\partial _{\nu }v\mid _{S_{1}},%
\overline{\psi }_{2}\left( x,t\right) =\partial _{x_{1}}v\mid _{P_{1}}.
\label{1.140}
\end{equation}%
By Theorem 5.2 of Chapter IV of \cite{LSU}, (\ref{1.171}) and (\ref{1.13})-(%
\ref{1.140}) there exist numbers $C\left( \Omega ,L\right) ,C\left(
P,L\right) >0$ depending only on listed parameters such that%
\begin{eqnarray}
\left\Vert \overline{\psi }_{1}\right\Vert _{C^{1+\alpha ,\alpha /2}\left(
\overline{S}_{1}\right) } &\leq &C\left( \Omega ,L\right) \left\Vert
\overline{\varphi }_{1}\right\Vert _{C^{2+\alpha ,1+\alpha /2}\left(
\overline{S}_{1}\right) },  \label{1.14_1} \\
\left\Vert \overline{\psi }_{2}\right\Vert _{C^{1+\alpha ,\alpha /2}\left(
\overline{P}_{1}\right) } &\leq &C\left( P,L\right) \left\Vert \overline{%
\varphi }_{2}\right\Vert _{C^{2+\alpha ,1+\alpha /2}\left( \overline{P}%
_{1}\right) }.  \label{1.141}
\end{eqnarray}

We now describe an elementary and well known procedure of finding the normal
derivative of the function $v$ either at $S_{1}$ (in the case of IP5.2) or
at $P_{1}$ (in the case of IP5.3). In fact, an analog of this procedure was
described in Section 5.2 for the hyperbolic case, see (\ref{4.24}), (\ref%
{4.25}). In the case of IP5.2 we solve the initial boundary value problem
for equation (\ref{1.11}) for $\left( x,t\right) \in \left( \mathbb{R}%
^{n}\diagdown \Omega \right) \times \left( 0,1\right) $ with the zero
initial condition in $\mathbb{R}^{n}\diagdown \Omega $ (because of (\ref{1.5}%
)) and the Dirichlet boundary condition $v\mid _{S_{1}}=\overline{\varphi }%
_{1}.$ Then we uniquely find the normal derivative $\partial _{\nu }v\mid
_{S_{1}}=\overline{\psi }_{1}$. Similarly, in the case of IP5.3, we uniquely
find the Neumann boundary condition $\partial _{x_{1}}v\mid _{P_{1}}=%
\overline{\psi }_{2}$. Estimates (\ref{1.14_1}), (\ref{1.141}) ensure the
stability of this procedure.

Therefore, problems IP5.2 and IP5.3 are replaced with a corresponding
inverse problem for the parabolic PDE (\ref{1.11}) with the lateral Cauchy
data (\ref{1.13}), (\ref{1.140}). These data are given at $S_{1}$ for IP5.2
and at $P_{1}$ for IP5.3. Uniqueness of the solution of each of these
parabolic inverse problems follows from Theorem 2.2 and Remark 2.2.

Using (\ref{1.142}), (\ref{1.10}), (\ref{1.101}) and (\ref{1.13})-(\ref%
{1.141}), we obtain%
\begin{equation}
\left\Vert \overline{\varphi }_{1}\right\Vert _{C^{2+\alpha ,1+\alpha
/2}\left( \overline{S}_{1}\right) }+\left\Vert \overline{\psi }%
_{1}\right\Vert _{C^{1+\alpha ,\alpha /2}\left( \overline{S}_{1}\right)
}\leq C_{1}\left( \Omega ,L,d\right) \delta .  \label{1.144}
\end{equation}%
Next, using (\ref{1.143}) instead of (\ref{1.142}), we obtain%
\begin{equation}
\left\Vert \overline{\varphi }_{2}\right\Vert _{C^{2+\alpha ,1+\alpha
/2}\left( \overline{P}_{1}\right) }+\left\Vert \overline{\psi }%
_{2}\right\Vert _{C^{1+\alpha ,\alpha /2}\left( \overline{P}_{1}\right)
}\leq C_{2}\left( P,L,d\right) \delta ,  \label{1.145}
\end{equation}%
where constants $C_{1}\left( \Omega ,L,d\right) ,C_{2}\left( P,L,d\right) >0$
depend only on listed parameters. It follows from (\ref{1.144}) that with a
different constant $\overline{C}:=\overline{C}\left( \Omega ,L,d\right) >0$%
\begin{equation}
\left\Vert \overline{\varphi }_{1}\right\Vert _{H^{1}\left( S_{1}\right)
}+\left\Vert \overline{\psi }_{1}\right\Vert _{L_{2}\left( S_{1}\right)
}\leq \overline{C}\delta .  \label{1.146}
\end{equation}

\subsubsection{Logarithmic stability estimates for IP5.2 and IP5.3}

\label{sec:5.3.3}

\textbf{Theorem 5.4.} \emph{Consider IP5.2. Let Assumption 5.1 holds and
conditions (\ref{1.5}), (\ref{1.142}) be valid. Also, assume that the upper
bound }$F$\emph{\ of the norm }$\left\Vert \nabla f\right\Vert _{L_{2}\left(
\Omega \right) }$\emph{\ is given, }$\left\Vert \nabla f\right\Vert
_{L_{2}\left( \Omega \right) }\leq F.$ \emph{Then there exists a
sufficiently small number }$\delta _{0}=\delta _{0}\left( L,\Omega \right)
\in \left( 0,1\right) $\emph{\ and a constant }$M_{1}=M_{1}\left( L,\Omega
\right) >0,$\emph{\ both dependent only on listed parameters, such that if
the number }$\delta $\emph{\ in (\ref{1.142}) is so small that }$\overline{C}%
\delta \in \left( 0,\delta _{0}\right) $\emph{, then the following
logarithmic stability estimate holds}%
\begin{equation*}
\left\Vert f\right\Vert _{L_{2}\left( \Omega \right) }\leq \frac{M_{1}F}{%
\sqrt{\ln \left[ \left( \overline{C}\delta \right) ^{-1}\right] }}.\text{ }
\end{equation*}%
\emph{Here} $\overline{C}=\overline{C}\left( \Omega ,L,d\right) >0$ \emph{is
the number in (\ref{1.146}). }

Consider now IP5.3. Denote $\overline{x}=\left( x_{2},...,x_{n}\right) .$
Changing variables $\left( x^{\prime },t^{\prime }\right) =\left( \sqrt{b}%
x,dt\right) $ with an appropriate constant $b>0$ and keeping the same
notations for new variables for brevity, we obtain that without loss of
generality we can assume that
\begin{equation}
\Omega \subset \left\{ x_{1}+\left\vert \overline{x}\right\vert ^{2}<\frac{1%
}{4},x_{1}>0\right\} .  \label{1.23}
\end{equation}%
Denote%
\begin{eqnarray}
\Phi &=&\left\{ \left( x,t\right) :x_{1}\in \left( 0,1\right) ,\overline{x}%
=\left( x_{2},x_{3},...,x_{n}\right) \in \left( -1,1\right) ^{n-1},t\in
\left( 0,1\right) \right\} ,  \label{3.91} \\
\partial _{1}\Phi &=&\overline{\Phi }\cap P=\left\{ \left( x,t\right)
:x_{1}=0,\overline{x}\in \left( -1,1\right) ^{n-1},t\in \left( 0,1\right)
\right\} .  \label{3.92}
\end{eqnarray}%
Recall that (\ref{1.143}) implies (\ref{1.145}). Hence, assuming that (\ref%
{1.143}) holds and using (\ref{3.91}), (\ref{3.92}), we derive that there
exists a constant $\widetilde{C}=\widetilde{C}\left( L,\Phi ,d\right) >0$
such that
\begin{equation}
\left\Vert \overline{\varphi }_{2}\right\Vert _{H^{1}\left( \partial
_{1}\Phi \right) }+\left\Vert \overline{\psi }_{2}\right\Vert _{L_{2}\left(
\partial _{1}\Phi \right) }\leq \widetilde{C}\delta .  \label{3.94}
\end{equation}

\textbf{Theorem 5.5.} \emph{Consider IP5.3. Let Assumption 5.1 holds and (%
\ref{1.5}), (\ref{1.143}) be valid. Also, assume that for a certain }$\alpha
\in \left( 0,1\right) $\emph{\ the upper bound }$F$\emph{\ of the norm }$%
\left\Vert f\right\Vert _{C^{2+\alpha }\left( \overline{\Omega }\right) }$%
\emph{\ is given, i.e. }$\left\Vert f\right\Vert _{C^{2+\alpha }\left(
\overline{\Omega }\right) }\leq F$. \emph{Then there exists a sufficiently
small number }$\delta _{0}=\delta _{0}\left( L,\Phi \right) \in \left(
0,1\right) $\emph{\ and a constant }$M_{2}=M_{2}\left( L,\Phi \right) >0,$%
\emph{\ both dependent only on listed parameters, such that if the number }$%
\delta $\emph{\ in (\ref{1.143}) is so small that }$\widetilde{C}\delta \in
\left( 0,\delta _{0}\right) $\emph{, then the following logarithmic
stability estimate holds}
\begin{equation*}
\left\Vert f\right\Vert _{L_{2}\left( \Omega \right) }\leq \frac{M_{2}F}{%
\sqrt{\ln \left[ \left( \widetilde{C}\delta \right) ^{-1}\right] }}.\text{ }
\end{equation*}%
\emph{Here }$\widetilde{C}=\widetilde{C}\left( L,\Phi ,d\right) >0$\emph{\
is the number from (\ref{3.94}).}

\subsection{QRM in the hyperbolic case}

\label{sec:5.4}

Although the Quasi-Reversibility Method (QRM) was discussed in Section 2.5,
it is worth to discuss it here again for the specific hyperbolic case. The
reason is that we now have two types of stability estimates, which are
different from the H\"{o}lder stability estimate of Section 2.5: the
Lipschitz stability estimates of Theorems 5.1, 5.2 and the logarithmic
stability estimates of Theorems 5.3, 5.4. We start from IP5.3. The case of
IP 5.2 is not discussed here since it is similar.

\subsubsection{QRM for IP5.3}

\label{sec:5.4.1}

To use embedding theorem, we work in this section only in 3d. The 2d case is
similar. We now increase the required smoothness of the solution of the
problem (\ref{1.6}), (\ref{1.7}). To do this, we replace Assumption 5.1 with
Assumption 5.2, where we use the norm $\left\Vert u\right\Vert _{C^{8}\left(
\mathbb{R}^{n}\times \left[ 0,T\right] \right) }$ instead of the norm $%
\left\Vert u\right\Vert _{C^{4}\left( \mathbb{R}^{n}\times \left[ 0,T\right]
\right) }$ of Assumption 5.1.

\textbf{Assumption 5.2.} We assume that integers $k,p$ in (\ref{1.3}), (\ref%
{1.5}), coefficients of the operator $L$ and the initial condition $f$ are
such that there exists unique solution $u\in C^{12}\left( \mathbb{R}%
^{n}\times \left[ 0,T\right] \right) ,\forall T>0$ of the problem (\ref{1.6}%
), (\ref{1.7}) satisfying
\begin{equation}
\left\Vert u\right\Vert _{C^{12}\left( \mathbb{R}^{n}\times \left[ 0,T\right]
\right) }\leq Be^{dT},\forall T>0,  \label{4.100}
\end{equation}%
where the constants $B=B\left( L,\overline{B}\right) >0,d=d\left( L,%
\overline{B}\right) >0$ depend only from the coefficients of the operator $L$
and an upper estimate $\overline{B}$ of the norm $\left\Vert f\right\Vert
_{C^{p}\left( \overline{\Omega }\right) }\leq \overline{B}.$

We assume in Section 5.4.1 that Assumption 5.2 holds. Hence, Theorem 5.2 of
Chapter IV of the book \cite{LSU}, (\ref{4.100}) and (\ref{1.13})-(\ref%
{1.140}) imply that functions
\begin{equation}
\overline{\varphi }_{2}\in C^{10+\alpha ,5+\alpha /2}\left( \overline{P}%
_{1}\right) ,\overline{\psi }_{2}\in C^{8+\alpha ,4+\alpha /2}\left(
\overline{P}_{1}\right)  \label{4.101}
\end{equation}%
and there exists a number $C_{3}\left( P,L\right) >0$ depending only on
listed parameters such that
\begin{equation}
\left\Vert \overline{\psi }_{2}\right\Vert _{C^{8+\alpha ,4+\alpha /2}\left(
\overline{P}_{1}\right) }\leq C_{3}\left( P,L\right) \left\Vert \overline{%
\varphi }_{2}\right\Vert _{C^{10+\alpha ,5+\alpha /2}\left( \overline{P}%
_{1}\right) }.  \label{4.102}
\end{equation}%
Let sets $\Phi ,\partial _{1}\Phi $ be the ones introduced in (\ref{3.91}), (%
\ref{3.92}).\ Then (\ref{4.101}) and (\ref{4.102}) imply that
\begin{equation}
\overline{\varphi }_{2},\overline{\psi }_{2}\in H^{8,4}\left( \partial
_{1}\Phi \right) ,\left\Vert \overline{\varphi }_{2}\right\Vert
_{H^{8,4}\left( \partial _{1}\Phi \right) }+\left\Vert \overline{\psi }%
_{2}\right\Vert _{H^{8,4}\left( \partial _{1}\Phi \right) }\leq C_{4}\left(
P,\Phi ,L\right) \left\Vert \overline{\varphi }_{2}\right\Vert
_{C^{10+\alpha ,5+\alpha /2}\left( \overline{P}_{1}\right) },  \label{4.103}
\end{equation}%
where the number $C_{4}\left( P,\Phi ,L\right) =const.>0$ depends only on
listed parameters.

Let the function $v\left( x,t\right) $ be the one defined in (\ref{1.100}).
Then $v\in C^{10+\alpha ,5+\alpha /2}\left( \mathbb{R}^{n}\times \left[ 0,T%
\right] \right) ,\forall T>0$ is the solution of the problem (\ref{1.11}), (%
\ref{1.12}). Denote%
\begin{eqnarray}
r\left( x,t\right) &=&\overline{\varphi }_{2}\left( x,t\right) +x_{1}%
\overline{\psi }_{2}\left( x,t\right) =\overline{\varphi }_{2}\left(
\overline{x},t\right) +x_{1}\overline{\psi }_{2}\left( \overline{x},t\right)
,  \label{4.104} \\
\widehat{v}\left( x,t\right) &=&v\left( x,t\right) -r\left( x,t\right) ,
\label{4.105} \\
p\left( x,t\right) &=&-\left( r_{t}-Lr\right) \left( x,t\right) .
\label{4.106}
\end{eqnarray}%
Let $H_{0}^{4}\left( \Phi \right) :=\left\{ u\in H^{4}\left( \Phi \right)
:u\mid _{\partial _{1}\Phi }=u_{x_{1}}\mid _{\partial _{1}\Phi }=0\right\} .$
Then
\begin{eqnarray}
\widehat{v}_{t}-L\widehat{v} &=&p\left( x,t\right) ,\left( x,t\right) \in
\Phi ,\widehat{v}\in H_{0}^{4}\left( \Phi \right) ,  \label{4.108} \\
\widehat{v} &\mid &_{\partial _{1}\Phi }=0,\widehat{v}_{x_{1}}\mid
_{\partial _{1}\Phi }=0.  \label{4.109}
\end{eqnarray}

To solve IP5.3 via the QRM, we minimize the following Tikhonov functional
\begin{equation}
J_{\gamma }\left( \widehat{v}\right) =\left\Vert \widehat{v}_{t}-L\widehat{v}%
-p\right\Vert _{L_{2}\left( \Phi \right) }^{2}+\gamma \left\Vert \widehat{v}%
\right\Vert _{H^{4}\left( \Phi \right) }^{2},\widehat{v}\in H_{0}^{4}\left(
\Phi \right) ,  \label{4.60}
\end{equation}%
where $\gamma >0$ is the regularization parameter. The requirement $\widehat{%
v}\in H^{4}\left( \Phi \right) $ is an over-smoothness. This condition is
imposed to ensure that the function $\widehat{v}\in C^{1}\left( \overline{%
\Phi }\right) $: because of the embedding theorem.\ Indeed, we need the
smoothness $\widehat{v}\in C^{1}\left( \overline{\Phi }\right) $ to apply
theorems 2.2 and 3.1 of \cite{Kltherm}. We imposed in Assumption 4.2 $u\in
C^{12}\left( \mathbb{R}^{n}\times \left[ 0,T\right] \right) ,\forall T>0$
only to guarantee that $\widehat{v}\in H^{4}\left( \Phi \right) \subset
C^{1}\left( \overline{\Phi }\right) .$ However, the author's numerical
experience with QRM has consistently demonstrated that one can significantly
relax the required smoothness in practical computation, see \cite%
{KKKN,KPK,KBK}. This is likely because one is not using an overly small grid
step size in finite differences when minimizing functionals like the one in (%
\ref{4.60}). Hence, one effectively works with a finite dimensional space
with not too many dimensions. This means that one can rely in this case on
the equivalence of all norms in finite dimensional spaces. Thus, most likely
one can replace in real computations $\gamma \left\Vert v\right\Vert
_{H^{4}\left( \Phi \right) }^{2}$ with $\gamma \left\Vert v\right\Vert
_{H^{2,1}\left( \Phi \right) }^{2}$.

Let $\left( ,\right) $ and $\left[ ,\right] $ be scalar products in $%
L_{2}\left( \Phi \right) $ and $H^{4}\left( \Phi \right) $ respectively.\
Let the function $u_{\gamma }\in H_{0}^{4}\left( \Phi \right) $ be a
minimizer of the functional (\ref{4.60}). Then the variational principle
implies that
\begin{equation*}
\left( \partial _{t}u_{\gamma }-Lu,\partial _{t}w-Lw\right) +\gamma \left[
u,w\right] =\left( p,w_{t}-Lw\right) ,\forall w\in H_{0}^{4}\left( \Phi
\right) .
\end{equation*}

\textbf{Lemma 5.1}. \emph{For every function }$p\in L_{2}\left( \Phi \right)
$\emph{\ and every }$\gamma >0$ \emph{there exists unique minimizer }$%
u_{\gamma }=u_{\gamma }\left( p\right) \in H_{0}^{4}\left( \Phi \right) $%
\emph{\ of the functional (\ref{4.60}). Furthermore, there exists a constant
}$M=M\left( L,\Phi \right) $ \emph{such that the following estimate holds }$%
\left\Vert u_{\gamma }\right\Vert _{H^{4}\left( \Phi \right) }\leq
M_{1}\gamma ^{-1/2}\left\Vert p\right\Vert _{L_{2}\left( \Phi \right) }.$%
\emph{\ }

Lemma 5.1 is an obvious analog of Lemma 2.5. The idea now is that if $%
u_{\gamma }\left( x,t\right) \in H_{0}^{4}\left( \Phi \right) $ is the
minimizer mentioned in Lemma 5.1, then the approximate solution of IP5.3 is
\begin{equation}
f_{\gamma }\left( x\right) =u_{\gamma }\left( x,0\right) +r\left( x,0\right)
.  \label{4.80}
\end{equation}%
The question of convergence of minimizers of $J_{\gamma }$ to the exact
solution is more difficult than the existence question of Lemma 5.1. To
address the question of convergence, we need first to introduce the exact
solution as well as the error in the data, just as this is always done in
the regularization theory \cite{BKok,BK,EHN,Kab,T}, also, see Theorem 2.7.
We assume that there exists an \textquotedblleft ideal" noiseless data $%
\varphi _{2}^{\ast }\in C^{12}\left( P_{\infty }\right) $. Following (\ref%
{1.143}) and (\ref{4.100}), we assume that with a sufficiently small number $%
\delta \in \left( 0,1\right) $ the following estimate holds%
\begin{equation}
\left\Vert \varphi _{2}-\varphi _{2}^{\ast }\right\Vert _{C^{12}\left(
\overline{P}_{T}\right) }\leq \delta e^{dT},\forall T>0.  \label{4.110}
\end{equation}%
Let $\overline{\varphi }_{2}^{\ast }=\mathcal{L}\varphi _{2}^{\ast },$ the
function $f^{\ast }$ satisfying (\ref{1.5}) is the solution of IP5.3 for the
case of the noiseless data $\varphi _{2}^{\ast },$ the function $u^{\ast
}\in C^{12}\left( \mathbb{R}^{n}\times \left[ 0,T\right] \right) ,\forall
T>0 $ satisfying (\ref{4.100}) is the solution of the Cauchy problem (\ref%
{1.6}), (\ref{1.7}) with $f:=f^{\ast }.$ Following (\ref{1.100}), denote $%
v^{\ast }=\mathcal{L}u^{\ast }.$ Let functions $\overline{\varphi }%
_{2}^{\ast },...,p^{\ast }$ have the same meaning as corresponding functions
in (\ref{4.101})-(\ref{4.106}), except that they are generated by the
noiseless data $\varphi _{2}^{\ast }.$ Then (\ref{4.103}) and (\ref{4.110})
imply that%
\begin{equation}
\left\Vert p-p^{\ast }\right\Vert _{L_{2}\left( \Phi \right) }\leq
C_{5}\left( \Phi ,L,d\right) \delta ,  \label{4.113}
\end{equation}%
where the constant $C_{5}\left( P,\Phi ,L,d\right) >0$ depends only on
listed parameters.

Theorem 5.6 establishes the convergence rate of the QRM. The proof of this
theorem is using (\ref{4.113}) and is similar with the proof of Theorem 3.1
of \cite{Kltherm}.

\textbf{Theorem 5.6}. \emph{Let Assumption 5.2 and condition (\ref{4.110})
be valid. Suppose that the regularization parameter }$\gamma =\gamma \left(
\delta \right) :=\delta \in \left( 0,1\right) .$\emph{\ Let the function }$%
u_{\gamma }\in H_{0}^{4}\left( \Phi \right) $\emph{\ be the unique minimizer
of the functional (\ref{4.60}) (Lemma 5.1). Then there exists a number }$%
Y=Y\left( \Phi ,L,B,d\right) >0$ \emph{and} \emph{a sufficiently small
number }$\delta _{0}=\delta _{0}\left( L,\Phi \right) \in \left( 0,1\right) $%
\emph{\ such that if \ }$\delta \in \left( 0,\delta _{0}\right) ,$\emph{\
then the following logarithmic convergence rate takes place}%
\begin{equation}
\left\Vert f_{\gamma \left( \delta \right) }-f^{\ast }\right\Vert
_{L_{2}\left( \Omega \right) }\leq \frac{Y}{\sqrt{\ln \left( \delta
^{-1}\right) }},  \label{4.111}
\end{equation}%
\emph{where the function }$f_{\gamma \left( \delta \right) }\left( x\right) $%
\emph{\ is defined in (\ref{4.80})}$.$\emph{\ In addition, for every }$%
\omega \in \left( 0,\omega _{0}\right) $\emph{\ there exists a number }$\rho
=\rho \left( L,\Phi ,B,d\right) \in \left( 0,1/2\right) $\emph{\ such that
the following convergence rate takes place }%
\begin{equation*}
\left\Vert u_{\gamma \left( \eta \right) }-\widehat{v}^{\ast }\right\Vert
_{H^{1,0}\left( D_{1/2}\right) }\leq Y\delta ^{\rho },
\end{equation*}%
\emph{where the domain }$D_{1/2}=\left\{ \left( x,t\right)
:x_{1}>0,x_{1}+\left\vert \overline{x}\right\vert ^{2}+\left( t-1/2\right)
^{2}<1/4\right\} .$

\subsubsection{QRM for Problem 5.1 of Section 5.1}

\label{sec:5.4.2}

In this section we assume that condition (\ref{4.6}) is in place.\ This
enables us to use the Lipschitz stability estimate (\ref{4.8}) of Theorem
5.1. We consider the problem of the recovery of the function $u\left(
x,t\right) $ from conditions (\ref{4.3}), (\ref{4.4}). We rewrite these
conditions now as%
\begin{eqnarray}
Au &:&=u_{tt}-c^{2}\left( x\right) \Delta
u-\sum\limits_{j=1}^{n+1}b_{j}\left( x,t\right) u_{x_{j}}-b_{0}\left(
x,t\right) u=f\left( x,t\right) ,\left( x,t\right) \in Q_{T}^{\pm },
\label{4.112} \\
u &\mid &_{S_{T}^{\pm }}=p\left( x,t\right) ,\partial _{n}u\mid _{S_{T}^{\pm
}}=q\left( x,t\right) ,  \label{4.114}
\end{eqnarray}%
where $u_{n+1}:=u_{t}.$ Suppose that there exists a function $F\left(
x,t\right) $ such that%
\begin{equation*}
F\in H^{2}\left( Q_{T}^{\pm }\right) ,F\mid _{S_{T}^{\pm }}=p\left(
x,t\right) ,\partial _{n}F\mid _{S_{T}^{\pm }}=q\left( x,t\right) .
\end{equation*}%
Denote $w=u-F,G=f-AF.$ Let $H_{0}^{2}\left( Q_{T}^{\pm }\right) =\left\{
U\in H^{2}\left( Q_{T}^{\pm }\right) :U\mid _{S_{T}^{\pm }}=\partial
_{n}U\mid _{S_{T}^{\pm }}=0\right\} .$ Using (\ref{4.112}) and (\ref{4.114}%
), we obtain%
\begin{equation}
Aw=G,\left( x,t\right) \in Q_{T}^{\pm },w\in H_{0}^{2}\left( Q_{T}^{\pm
}\right) .  \label{4.115}
\end{equation}%
Thus, now we are concerned with finding the function $w$ satisfying (\ref%
{4.115}). QRM for this problem amounts to the minimization of the following
Tikhonov functional%
\begin{equation}
V_{\gamma }\left( w\right) =\left\Vert Aw-G\right\Vert _{L_{2}\left(
Q_{T}^{\pm }\right) }^{2}+\gamma \left\Vert w\right\Vert _{H^{2}\left(
Q_{T}^{\pm }\right) }^{2},w\in H_{0}^{2}\left( Q_{T}^{\pm }\right) .
\label{4.117}
\end{equation}%
As usual, $\gamma >0$ is the regularization parameter here. Let $w_{\gamma
}\in H_{0}^{2}\left( Q_{T}^{\pm }\right) $ be a minimizer of the functional (%
\ref{4.117}), $\left( ,\right) $ be the scalar product in $L_{2}\left(
Q_{T}^{\pm }\right) $ and $\left[ ,\right] $ be the scalar product in $%
H^{2}\left( Q_{T}^{\pm }\right) .$ Then the variational principle implies
that
\begin{equation}
\left( Aw_{\gamma },Av\right) +\gamma \left[ w_{\gamma },v\right] =\left(
G,Av\right) ,\forall v\in H_{0}^{2}\left( Q_{T}^{\pm }\right) .
\label{4.118}
\end{equation}%
Lemma 5.2 is again an obvious analog of Lemma 2.5.

\textbf{Lemma 5.2}. \emph{Let the function }$G\in L_{2}\left( Q_{T}^{\pm
}\right) .$\emph{\ Then for every }$\gamma >0$\emph{\ there exists unique
minimizer }$w_{\gamma }\in H_{0}^{2}\left( Q_{T}^{\pm }\right) $\emph{\ of
the functional (\ref{4.118}). Furthermore, with a constant }$%
C_{6}=C_{6}\left( A,Q_{T}^{\pm }\right) >0$\emph{\ depending only on listed
parameters the following estimate holds}%
\begin{equation*}
\left\Vert w_{\gamma }\right\Vert _{H^{2}\left( Q_{T}^{\pm }\right) }\leq
\frac{C}{\sqrt{\gamma }}\left\Vert G\right\Vert _{L_{2}\left( Q_{T}^{\pm
}\right) }.
\end{equation*}

\textbf{Theorem 5.7} (convergence). \emph{Let the domain }$\Omega =\left\{
\left\vert x\right\vert <R\right\} .$\emph{\ Assume that the coefficient }$%
c\in C^{1}\left( \overline{\Omega }\right) $\emph{\ of the operator }$A$%
\emph{\ in (\ref{4.112}) satisfies conditions (\ref{4.6}), (\ref{4.6_2}) and
other coefficients of the operator }$A$\emph{\ are such that }$b_{j}\in
C\left( \overline{Q}_{T}^{\pm }\right) ,j\in \left[ 0,n+1\right] .$\emph{\
Let }$T>R/\sqrt{\eta _{0}}$\emph{, where the constant }$\eta _{0}=\eta
_{0}\left( R,d,\alpha \right) \in \left( 0,1\right] $\emph{\ was defined in
Theorem 5.1. Assume that there exists exact solution }$w^{\ast }\in
H_{0}^{2}\left( Q_{T}^{\pm }\right) $\emph{\ of the problem (\ref{4.115})
with the exact data }$G^{\ast }.$\emph{\ Let }$w_{\gamma }\in
H_{0}^{2}\left( Q_{T}^{\pm }\right) $\emph{\ be the unique minimizer of the
functional (\ref{4.117}) (Lemma 5.2). Then there exists a constant }$%
K=K\left( A,R,T,d,\alpha \right) >0$\emph{\ such that}%
\begin{equation}
\left\Vert w_{\gamma }-w^{\ast }\right\Vert _{H^{1}\left( Q_{T}^{\pm
}\right) }\leq K\left( \left\Vert G-G^{\ast }\right\Vert _{L_{2}\left(
Q_{T}^{\pm }\right) }+\sqrt{\gamma }\left\Vert w^{\ast }\right\Vert
_{H^{2}\left( Q_{T}^{\pm }\right) }\right) .  \label{4.119}
\end{equation}%
\emph{In particular, if }$c\left( x\right) \equiv 1,$\emph{\ then it is
sufficient to have }$T>R.$\emph{\ Also, if }$\left\Vert G-G^{\ast
}\right\Vert _{L_{2}\left( Q_{T}^{\pm }\right) }\leq \delta ,$\emph{\ where }%
$\delta \in \left( 0,1\right) $\emph{\ is the level of the error in the
data, and if }$\gamma \in \left( 0,\delta ^{2}\right] ,$\emph{\ then }$%
\left\Vert w_{\gamma }-w^{\ast }\right\Vert _{H^{1}\left( Q_{T}^{\pm
}\right) }\leq K\delta .$

\textbf{Proof}. We have
\begin{equation*}
\left( Aw^{\ast },Av\right) +\gamma \left[ w^{\ast },v\right] =\left(
G^{\ast },Av\right) +\gamma \left[ w^{\ast },v\right] ,\forall v\in
H_{0}^{2}\left( Q_{T}^{\pm }\right) .
\end{equation*}%
Subtract this equality from (\ref{4.118}) and denote $\widetilde{w}%
=w_{\gamma }-w^{\ast },\widetilde{G}=G-G^{\ast }.$ Then $\widetilde{w}\in
H_{0}^{2}\left( Q_{T}^{\pm }\right) $ and
\begin{equation*}
\left( A\widetilde{w},Av\right) +\gamma \left[ \widetilde{w},v\right]
=\left( \widetilde{G},Av\right) -\gamma \left[ w^{\ast },v\right] ,\forall
v\in H_{0}^{2}\left( Q_{T}^{\pm }\right) .
\end{equation*}%
Hence, Cauchy-Bunyakovsky inequality implies that%
\begin{equation}
\left\Vert A\widetilde{w}\right\Vert _{L_{2}\left( Q_{T}^{\pm }\right)
}^{2}+\gamma \left\Vert \widetilde{w}\right\Vert _{H^{2}\left( Q_{T}^{\pm
}\right) }^{2}\leq \left\Vert \widetilde{G}\right\Vert _{L_{2}\left(
Q_{T}^{\pm }\right) }^{2}+\gamma \left\Vert w^{\ast }\right\Vert
_{H^{2}\left( Q_{T}^{\pm }\right) }^{2}.  \label{4.120}
\end{equation}%
Applying Theorem 5.2 to (\ref{4.120}), we obtain (\ref{4.119}). $\square $

\subsection{Published results}

\label{sec:5.5}

Stability estimates and convergent numerical methods for Problem 5.1,
Problem 5.2, IP5.1, IP5.2 and IP5.3 with an arbitrary time independent
principal part of the operator $L$ in (\ref{1.2}) were not obtained prior to
the work \cite{Kltherm}. The Lipschitz stability estimate (\ref{4.8}) is
important for the control theory, since it is used for proofs of exact
controllability theorems. For the first time, estimate (\ref{4.8}) was
proved in 1986 by Lop Fat Ho \cite{L} for the equation $u_{tt}-\Delta u=0$
with the aim of applying to the control theory. However, the method of
multipliers of \cite{L} cannot handle neither variable lower order terms of
the operator $L$ nor a variable coefficient $c\left( x\right) .$ On the
other hand, Carleman estimates are not sensitive to lower order terms of PDE
operators and also can handle the case of a variable coefficient $c\left(
x\right) .$

For the first time, the Carleman estimate was applied to this problem by
Klibanov and Malinsky \cite{KlibM}. In \cite{KlibM} Theorem 5.1 for equation
(\ref{4.3}) with the lateral Cauchy data (\ref{4.4}) and $c\left( x\right)
\equiv 1$ was proved. Using (\ref{4.8}), an analog of Theorem 5.7 was also
proved in \cite{KlibM}. Next, the result of \cite{KlibM} was extended by
Kazemi and Klibanov \cite{Kaz} and also by the author in \cite{K} to a more
general case of the hyperbolic inequality (\ref{4.5}) with $c\left( x\right)
\equiv 1$. Although in publications \cite{Kaz,KlibM,K} $c\equiv 1,$ it is
clear from them that the key idea is in applying the Carleman estimate,
while a specific form of the principal part of the hyperbolic operator is
less important. This thought is reflected in the proof of Theorem 3.4.8 of
the book of Isakov \cite{Is}. Thus, Theorem 5.1 for the variable coefficient
$c\left( x\right) $ satisfying an analog of (\ref{4.6_1}) was obtained in
section 2.4 of the book\ of Klibanov and Timonov \cite{KT} as well as in the
paper of Clason and Klibanov \cite{ClK}. The idea of \cite{Kaz} was used in
the control theory by Lasiecka, Triggiani and Yao \cite{LT1}, Lasiecka,
Triggiani and Zhang \cite{LT2,LT3,LT4} and by Triggiani and Yao \cite{Trig}.
Isakov and Yamamoto \cite{Y4} used that idea to prove a stronger version of
Theorem 5.1.

To prove Lipschitz stability without condition (\ref{4.6}), one can impose
some conditions of Riemannian geometry on the principal part of the operator
$L$ in (\ref{1.2}), see Bardos, Lebeau and Rauch \cite{Bardos}, Lasiecka,
Triggiani and Yao \cite{LT1,Trig}, Lasiecka, Triggiani and Zhang \cite%
{LT2,LT3,LT4}, Romanov \cite{Rom1,Rom2} and Stefanov and Uhlmann \cite{SU}.
In \cite{LT1,LT2,LT3,LT4,Rom1,Rom2,Trig} Carleman estimates were used.

Analogs of Theorem 5.7 about convergence of QRM were proved in Klibanov and
Rakesh \cite{KR}, Clason and Klibanov \cite{ClK} and Klibanov, Kuzhuget,
Kabanikhin and Nechaev \cite{KKKN}.\ Numerical testing of QRM was performed
in these references. This testing has consistently demonstrated a high
degree of robustness. For example, accurate results were obtained in \cite%
{KKKN} with up to 50\% noise in the data.

For explicit formulas for the reconstruction of the function $f\left(
x\right) $ for TAT (IP5.1) in the case when in (\ref{1.2}) $L\equiv \Delta $
we refer to Finch, Patch and Rakesh \cite{FPR}, Finch, Haltmeier and Rakesh
\cite{FHR}, the review paper of Kuchment and Kunyansky \cite{Kuch} and
Kunyansky \cite{KY}. These formulas lead to some stability estimates as well
as to numerical methods with good performances. Another numerical method for
TAT was proposed by Agranovsky and Kuchment \cite{AK}.

\section{Approximately Globally Convergent Numerical Method}

\label{sec:6}

The first step of the numerical method outlined in this section is the
elimination of the unknown coefficient from the underlying PDE, which is the
same step as in BK. In this section we briefly outline the recently
developed numerical method of Beilina and Klibanov referring for details to
\cite{AB,BK,BK1,BK2,BK3,BK4,BKK,BK20,JIIP12,KFBPS,KBB,KPK,KBK,KBKSNF,IEEE}.
Numerical tests are not presented here, since they are published in these
works.

Even though the field of Inverse Problems is an applied one and even though
MCIPs have been studied by many researchers since 1960-ies, the topic of
reliable numerical methods for them is still in its infancy. This is because
of \emph{enormous challenges}\textbf{\ }one inevitably faces when trying to
study this topic\textbf{.} Those challenges are caused by two factors
combined: nonlinearity and ill-posedness of MCIPs. In the case of single
measurement the third complicating factor is the minimal amount of available
information. Conventional least squares functionals for MCIPs suffer from
the phenomenon of multiple local minima and ravines. This leads to locally
convergent numerical methods, which require a good first guess about the
solution. However, the latter is impractical.

In the above cited series of recent publications some properties of
underlying PDE operators instead of least squares functionals were used. A
\emph{very important} feature of this numerical method is that it does not
require any knowledge of neither the medium inside of the domain of interest
nor of any point in a small neighborhood of the true solution. For the first
time, the following two goals were \emph{simultaneously} achieved for MCIPs
for a hyperbolic PDE with single measurement data:

\textbf{Goal 1}. The development of such a numerical method, which would
have a rigorous guarantee of obtaining at least one point in a small
neighborhood of the exact solution without any advanced knowledge of that
neighborhood.

\textbf{Goal 2}. This numerical method should have a good performance on
computationally simulated data. In addition, if experimental data are
available, then this method should demonstrate a good performance on these
data.

A \emph{crucial requirement} is to achieve both these goals \emph{%
simultaneously} rather than just only one of them. Because of the above
mentioned substantial challenges, it is natural to have the rigorous
guarantee of Goal 1 within the framework of a reasonable approximate
mathematical model. Since convergence is guaranteed in the framework of that
model, then we call our numerical method \emph{approximately globally
convergent}. In principle, any mathematical model can be called
\textquotedblleft approximate". Therefore, the validity of our model is
verified via a six-step procedure, see \cite{BK,JIIP12}. Basically this
procedure includes the proof of a convergence theorem and computational
results for both synthetic and experimental data. If convergence theorem
satisfying Goal 1 is proved and computational results are good ones (Goal
2), especially ones for experimental data, then that approximate
mathematical model is proclaimed as a valid one.

On the other hand, nothing works without such an approximate model. Indeed,
the author is unaware about such numerical methods for MCIPs with single
measurement data, which would: (1) simultaneously achieve Goals 1 and 2 and,
at the same time, (2) would not rely on some reasonable approximations,
which cannot be rigorously justified.

\subsection{Outline of the method}

\label{sec:6.1}

Let $\Omega \subset \mathbb{R}^{3}$ be a convex bounded domain with the
boundary $\partial \Omega \in C^{3}.$ Let $d=const.>2.$ We assume that the
coefficient $c\left( x\right) $ satisfies the following conditions
\begin{equation}
c\left( x\right) \in \lbrack 1,d],~~c\left( x\right) =1\text{ for }x\in
\mathbb{R}^{3}\diagdown \Omega ,c\in C^{\alpha }\left( \mathbb{R}^{3}\right)
.  \label{5.0}
\end{equation}%
We assume \emph{a priori} knowledge of the constant $d,$ which amounts to
the knowledge of the correctness set in the theory of Ill-Posed problems
\cite{BKok,BK,EHN,Kab,T}. Consider the Cauchy problem for the hyperbolic
equation
\begin{eqnarray}
c\left( x\right) u_{tt} &=&\Delta u\text{ in }\mathbb{R}^{3}\times \left(
0,\infty \right) ,  \label{5.1} \\
u\left( x,0\right) &=&0,\text{ }u_{t}\left( x,0\right) =\delta \left(
x-x_{0}\right) .  \label{5.2}
\end{eqnarray}%
A similar technique can be developed for MCIPs for the parabolic equation $%
c\left( x\right) v_{t}=\Delta v+a\left( x\right) v,$ where either of
coefficients $c\left( x\right) $ or $a\left( x\right) $ is unknown \cite{BK}%
. Equation (\ref{5.1}) governs, e.g. propagation of acoustic and
electromagnetic waves. In the acoustical case $c(x)=b^{-2}(x),$ where $%
b\left( x\right) $ is the sound speed. In the 2-D case of EM waves
propagation, the dimensionless coefficient is $c(x)=\varepsilon _{r}(x),$
where $\varepsilon _{r}(x)$ is the spatially distributed dielectric constant
of the medium. In \ the latter case the assumption $c\left( x\right) =1$ for
$x\in \mathbb{R}^{3}\diagdown \Omega $ in (\ref{5.0}) means that we have air
outside the medium of interest $\Omega .$ And the assumption $c\left(
x\right) \geq 1$ reflects the fact that the dielectric constants of almost
all materials exceed the one of the air. Equation (\ref{5.1}) was
successfully used in \cite{BK,BK4,KFBPS} to work with experimental data,
which are obviously in 3d. The latter was recently explained by Beilina in
\cite{BM}. It was shown in Test 4 of \cite{BM} that the component of the
electric field $E\left( x,t\right) =\left( E_{1},E_{2},E_{3}\right) \left(
x,t\right) ,$ which was originally initialized, strongly dominates two other
components.

\textbf{Multidimensional Coefficient Inverse Problem (MCIP).} \emph{Assume
that the coefficient }$c\left( x\right) $\emph{\ of equation (\ref{5.1})
satisfies condition (\ref{5.0}) and is unknown in the domain }$\Omega $\emph{%
. Determine the function }$c\left( x\right) $\emph{\ for }$x\in \Omega ,$%
\emph{\ assuming that the following function }$g\left( x,t\right) $\emph{\
is known for a single source position }$x_{0}\notin \overline{\Omega }$
\emph{in (\ref{5.2})}
\begin{equation}
u\left( x,t\right) =g\left( x,t\right) ,\forall \left( x,t\right) \in
\partial \Omega \times \left( 0,\infty \right) .  \label{5.4}
\end{equation}

The function $g\left( x,t\right) $ models time dependent measurements of the
wave field at the boundary of the domain of interest. The assumption of the
infinite time interval in (\ref{5.4}) is not a restrictive one, because we
work with the Laplace transform of the function $u\left( x,t\right) ,$ and
the kernel of this transform decays rapidly as $t\rightarrow \infty .$ In
this MCIP the data $g\left( x,t\right) $ are assumed to be known at the
entire boundary $\partial \Omega .$ The case of backscattering data can be
treated similarly, see Chapter 6 in the book \cite{BK} as well as \cite%
{JIIP12,KBKSNF,IEEE}.

Consider the Laplace transform of the function $u$,
\begin{equation}
w(x,s)=\int\limits_{0}^{\infty }u(x,t)e^{-st}dt,\text{ for }s>\underline{s}%
=const.>0.  \label{5.5}
\end{equation}%
We assume that the number $\underline{s}$ is sufficiently large, so that the
integral (\ref{5.5}) converges absolutely and that the same is valid for the
derivatives $D^{k}u,k=0,1,2$. We call the parameter $s$ \emph{pseudo
frequency}. It can be proven that
\begin{eqnarray}
\Delta w-s^{2}c\left( x\right) w &=&-\delta \left( x-x_{0}\right) ,\text{ }%
x\in \mathbb{R}^{3},  \label{5.6} \\
\lim_{\left\vert x\right\vert \rightarrow \infty }w\left( x,s\right) &=&0.
\label{5.7}
\end{eqnarray}%
Furthermore, for each value of $s>0$ the problem (\ref{5.6}), (\ref{5.7})
has unique solution of the form \cite{BK,JIIP12}%
\begin{equation}
w\left( x,s\right) =\frac{\exp \left( -s\left\vert x-x_{0}\right\vert
\right) }{4\pi \left\vert x-x_{0}\right\vert }+\overline{w}\left( x,s\right)
,\overline{w}\in C^{2+\alpha }\left( \mathbb{R}^{3}\right) ,\lim_{\left\vert
x\right\vert \rightarrow \infty }\overline{w}\left( x,s\right) =0.
\label{5.71}
\end{equation}%
Suppose that geodesic lines generated by the function $c\left( x\right) $
are regular and $c\left( x\right) $ is sufficiently smooth. Let $\tau \left(
x,x_{0}\right) $ be the length of the geodesic line connecting points $x$
and $x_{0}.$ Then
\begin{equation}
\left\vert D_{s}^{k}w(x,s)\right\vert _{2+\alpha }=\left\vert
D_{s}^{k}\left\{ \frac{\exp \left[ -s\tau \left( x,x_{0}\right) \right] }{%
f\left( x,x_{0}\right) }\right\} \right\vert _{2+\alpha }\left[ 1+O\left(
\frac{1}{s}\right) \right] ,s\rightarrow \infty ,k=0,1,  \label{5.8}
\end{equation}%
where $\left\vert \cdot \right\vert _{2+\alpha }=\left\Vert \cdot
\right\Vert _{C^{2+\alpha }\left( \overline{\Omega }\right) }$, $f\left(
x,x_{0}\right) $ is a certain function and $f\left( x,x_{0}\right) \neq 0$
for $x\in \overline{\Omega }.$ It is unclear how to effectively verify the
regularity of geodesic lines for generic functions $c\left( x\right) $.
Therefore, we assume below the asymptotic behavior (\ref{5.8}) without
linking it to the regularity of geodesic lines.

We have $w(x,s)>0$ \cite{BK,JIIP12}. Denote
\begin{equation*}
v\left( x,s\right) :=\frac{\ln w\left( x,s\right) }{s^{2}}.
\end{equation*}%
Since the source $x_{0}\notin \overline{\Omega }$, then
\begin{equation}
\Delta v+s^{2}\left( \nabla v\right) ^{2}=c(x),x\in \Omega .  \label{5.10}
\end{equation}%
Now we make the same step as the first step of BK. Differentiate both sides
of (\ref{5.10}) with respect to $s.$ Let $q\left( x,s\right) =\partial
_{s}v\left( x,s\right) .$ Then
\begin{eqnarray}
v\left( x,s\right) &=&-\int\limits_{s}^{\overline{s}}q\left( x,\tau \right)
d\tau +V\left( x,\overline{s}\right) ,  \label{5.11} \\
V\left( x,\overline{s}\right) &=&v\left( x,\overline{s}\right) =\frac{\ln
w\left( x,\overline{s}\right) }{\overline{s}^{2}}.  \label{5.12}
\end{eqnarray}%
Here the truncation pseudo frequency $\overline{s}>\underline{s}$ is a large
number. We call $V\left( x,\overline{s}\right) $ the \emph{tail function}.\
The tail function is unknown. By (\ref{5.8})
\begin{equation}
\left\vert V\left( x,\overline{s}\right) \right\vert _{2+\alpha }=O\left(
\overline{s}^{-1}\right) ,\text{ }\left\vert \partial _{\overline{s}}V\left(
x,\overline{s}\right) \right\vert _{2+\alpha }=O\left( \overline{s}%
^{-2}\right) ,\overline{s}\rightarrow \infty .  \label{5.13}
\end{equation}%
The number $\overline{s}$ is the main regularization parameter of our
numerical method. In the computational practice $\overline{s}$ is chosen in
numerical experiments\textbf{. }

Thus, we obtain from (\ref{5.10}), (\ref{5.11}) the following nonlinear
integral differential equation
\begin{equation}
\begin{split}
& \Delta q-2s^{2}\nabla q\int\limits_{s}^{\overline{s}}\nabla q\left( x,\tau
\right) d\tau +2s\left[ \int\limits_{s}^{\overline{s}}\nabla q\left( x,\tau
\right) d\tau \right] ^{2} \\
& +2s^{2}\nabla q\nabla V-4s\nabla V\int\limits_{s}^{\overline{s}}\nabla
q\left( x,\tau \right) d\tau +2s\left( \nabla V\right) ^{2}=0,x\in \Omega .
\end{split}
\label{5.14}
\end{equation}%
It follows from (\ref{5.4}) that
\begin{equation}
q\left( x,s\right) =\psi \left( x,s\right) ,\text{ }\forall \left(
x,s\right) \in \partial \Omega \times \left[ \underline{s},\overline{s}%
\right] ,  \label{5.15}
\end{equation}%
where $\psi \left( x,s\right) =s^{-2}\partial _{s}\ln \varphi -2s^{-3}\ln
\varphi $ and $\varphi \left( x,s\right) $ is the Laplace transform (\ref%
{5.5}) of the function $g\left( x,t\right) $ in (\ref{5.4}). We have two
unknown functions $q$ and $V$ in equation (\ref{5.14}). Therefore, to
approximate both of them, we approximate the function $q$ via
\textquotedblleft inner" iterations and the function $V$ is approximated via
\textquotedblleft outer" iterations. Suppose for a moment that functions $q$
and $V$ are approximated in $\Omega $ together with their derivatives $%
D_{x}^{\beta }q,D_{x}^{\beta }V,\left\vert \beta \right\vert \leq 2.$ Then
the corresponding approximation for the target coefficient can be found via (%
\ref{5.10}) as
\begin{equation}
c\left( x\right) =\Delta v+\underline{s}^{2}\left( \nabla v\right) ^{2},x\in
\Omega ,  \label{5.16}
\end{equation}%
where the function $v$ is approximated via (\ref{5.11}). We have found in
our numerical experiments that the optimal value of $s$ to use in (\ref{5.16}%
) is $s:=\underline{s}.$

To solve the problem (\ref{5.14}), (\ref{5.15}), we assume that $q\left(
x,s\right) $ is a piecewise constant function with respect $s.$ Hence, we
assume that there exists a partition $\underline{s}%
=s_{N}<s_{N-1}<...<s_{1}<s_{0}=\overline{s},s_{i-1}-s_{i}=h$ of the interval
$\left[ \underline{s},\overline{s}\right] $ with a sufficiently small grid
step size $h$ such that
\begin{equation*}
q\left( x,s\right) =q_{n}\left( x\right) \text{ for }s\in
(s_{n},s_{n-1}],q_{0}\equiv 0.
\end{equation*}%
We approximate the boundary condition (\ref{5.15}) as a piecewise constant
function, $q_{n}\left( x\right) :=\overline{\psi }_{n}\left( x\right) ,x\in
\partial \Omega ,$ where $\overline{\psi }_{n}\left( x\right) $ is the
average of the function $\psi \left( x,s\right) $ over the interval $\left(
s_{n},s_{n-1}\right) .$ Next, a certain system of elliptic equations for
functions $q_{n}\left( x\right) $ is derived from (\ref{5.14}) using the $s-$%
dependent so-called \textquotedblleft Carleman Weight Function" $\exp \left[
\lambda \left( s-s_{n-1}\right) \right] ,s\in \left( s_{n},s_{n-1}\right) ,$
where $\lambda >>1,\lambda h>1.$ Usually we use $\lambda =50$ in our
computations.

We solve elliptic Dirichlet boundary value problems for functions $%
q_{n}\left( x\right) $ sequentially, starting from $q_{1}\left( x\right) .$
An important new element of both Section 2.9 of the book \cite{BK} and the
paper \cite{JIIP12} is the choice of the first tail function $V_{1,1}\left(
x\right) ,$ see this section below. Having $V_{1,1}\left( x\right) ,$ we
solve the Dirichlet boundary value problem for the function $q_{1,1}\left(
x\right) .$\ Next, using (\ref{5.11}) with $V:=V_{1,1}$ and substituting it
in (\ref{5.10}), we find the first approximation $c_{1,1}\left( x\right) $
for our target coefficient $c\left( x\right) .$ This is the inner iteration,
i.e. when we work only inside of the domain $\Omega $. To perform the outer
iteration in the entire space $\mathbb{R}^{3}$, we solve the forward problem
(\ref{5.1}), (\ref{5.2}) with $c:=c_{1,1},$ calculate the Laplace transform (%
\ref{5.5}) at $s:=\overline{s}$ and then obtain an update $V_{1,2}\left(
x\right) $ for the tail function, using (\ref{5.12}). We repeat this
procedure $m$ times until stabilization occurs, thus getting functions $%
q_{1,i},c_{1,i},V_{1,i},i\in \left[ 1,m\right] .$ The number $m$ is found in
numerical experiments. Next, we set $%
q_{1}:=q_{1,m},c_{1}:=c_{1,m},V_{2,1}:=V_{1,m}$ \ and repeat the same for
functions $q_{2,i},c_{2,i},V_{2,i},i\in \left[ 1,m\right] .$ Similarly for
functions $q_{n,i},c_{n,i},V_{n,i},i\in \left[ 1,m\right] ,n\in \left[ 1,N%
\right] .$

We now describe how do we find the first tail function $V_{1,1}\left(
x\right) .$ Again following the Tikhonov concept \cite{BKok,BK,EHN,Kab,T},
we assume that there exists unique exact solution $c^{\ast }\left( x\right) $
of our MCIP with the noiseless data $g^{\ast }\left( x,t\right) $ (\ref{5.4}%
). We assume that the function $c^{\ast }\left( x\right) $ satisfies
condition (\ref{5.0}). Let $w^{\ast }\left( x,s\right) $ be the solution of
the problem (\ref{5.6})-(\ref{5.7}) satisfying (\ref{5.71}). Using (\ref%
{5.12}), we define the exact tail $V^{\ast }\left( x,s\right) $ for $s\geq
\overline{s}$ as%
\begin{equation*}
V^{\ast }\left( x,s\right) =\frac{\ln w^{\ast }\left( x,s\right) }{s^{2}}%
,\forall s\geq \overline{s}.
\end{equation*}%
Assuming that the asymptotic behavior (\ref{5.8}) holds and using (\ref{5.13}%
), we obtain%
\begin{equation}
V^{\ast }\left( x,s\right) =\frac{p^{\ast }\left( x\right) }{s}+O\left(
\frac{1}{s^{2}}\right) ,s\rightarrow \infty ,x\in \overline{\Omega }.
\label{5.300}
\end{equation}%
for a certain function $p^{\ast }\left( x\right) .$ We truncate the second
term of this asymptotic behavior. Thus, our\textbf{\ }approximate
mathematical model consists of the following assumption.

\textbf{Assumption 6.1. }There exists a function $p^{\ast }\left( x\right)
\in C^{2+\alpha }\left( \overline{\Omega }\right) $ such that the exact tail
function $V^{\ast }\left( x,s\right) $ has the form
\begin{equation}
V^{\ast }\left( x,s\right) :=\frac{p^{\ast }\left( x\right) }{s},\text{ }%
\forall s\geq \overline{s}.\text{ }  \label{5.35}
\end{equation}%
In addition,
\begin{equation}
\frac{p^{\ast }\left( x\right) }{s}=\frac{\ln w^{\ast }\left( x,s\right) }{%
s^{2}},\text{ }\forall s\geq \overline{s}.  \label{5.351}
\end{equation}

Since $q^{\ast }\left( x,s\right) =\partial _{s}V^{\ast }\left( x,s\right) $
for $s\geq \overline{s},$ we derive from (\ref{5.35}) that
\begin{equation}
q^{\ast }\left( x,\overline{s}\right) =-\frac{p^{\ast }\left( x\right) }{%
\overline{s}^{2}}\text{.}  \label{5.36}
\end{equation}%
Set in (\ref{5.14}) $q:=q^{\ast },V:=V^{\ast },s=\overline{s}$ and use (\ref%
{5.35}) and (\ref{5.36}). Then we obtain the following Dirichlet boundary
value problem for the function $p^{\ast }\left( x\right) $
\begin{eqnarray}
\Delta p^{\ast } &=&0\text{ in }\Omega ,\text{ }p^{\ast }\in C^{2+\alpha
}\left( \overline{\Omega }\right) ,  \label{5.37} \\
p^{\ast }|_{\partial \Omega } &=&-\overline{s}^{2}\psi ^{\ast }\left( x,%
\overline{s}\right) ,  \label{5.38}
\end{eqnarray}%
where $\psi ^{\ast }\left( x,s\right) $ is the exact function $\psi \left(
x,s\right) ,$ which corresponds to the function $g^{\ast }\left( x,t\right)
. $ The approximate equation (\ref{5.37}) is valid only within the framework
of Assumption 6.1. Although this equation is linear, formula (\ref{5.16})
for the reconstruction of the target coefficient $c^{\ast }$ is nonlinear.

Assuming that the function $\psi \left( x,\overline{s}\right) \in
C^{2+\alpha }\left( \partial \Omega \right) ,$ consider the solution $%
p\left( x\right) $ of the following boundary value problem
\begin{eqnarray}
\Delta p &=&0\text{ in }\Omega ,\text{ }p\in C^{2+\alpha }\left( \overline{%
\Omega }\right) ,  \label{5.40} \\
p|_{\partial \Omega } &=&-\overline{s}^{2}\psi \left( x,\overline{s}\right) .
\label{5.41}
\end{eqnarray}%
We choose the first tail function as
\begin{equation}
V_{1,1}\left( x\right) :=\frac{p\left( x\right) }{\overline{s}}.
\label{5.43}
\end{equation}%
By the Schauder theorem there exists unique solution $p$ of the problem (\ref%
{5.40}), (\ref{5.41}), see the book of Ladyzhenskaya and Uralceva \cite{LU}
for Schauder theorem.\ Furthermore, it follows from Schauder theorem as well
as from (\ref{5.37})-(\ref{5.43}) that with a number $M=M\left( \Omega
\right) >0$ the following estimate holds
\begin{equation}
\left\vert \nabla V_{1,1}-\nabla V^{\ast }\right\vert _{1+\alpha }\leq M%
\overline{s}\left\Vert \psi \left( x,\overline{s}\right) -\psi ^{\ast
}\left( x,\overline{s}\right) \right\Vert _{C^{2+\alpha }\left( \partial
\Omega \right) }.  \label{5.42}
\end{equation}

Therefore, our main approximations are (\ref{5.35}) and (\ref{5.351}).\emph{%
\ }These approximations mean the truncation of the term $O\left(
s^{-2}\right) $ in (\ref{5.300}). The goal of (\ref{5.35}) and (\ref{5.36})
is to obtain the accuracy estimate (\ref{5.42}) for the first tail. We point
out that these approximations are done only on the first iteration of our
method. It follows from (\ref{5.42}) that we obtain an approximation located
in a small neighborhood of the exact solution already on the first iteration
of our method, as long as the error in the boundary data $\psi ^{\ast
}\left( x,\overline{s}\right) $ is small.\ Convergence theorem 2.9.4 of \cite%
{BK} and theorem 5.1 of \cite{JIIP12} guarantee that all other solutions
obtained in the iterative process also provide good approximations, as long
as the number of iterations is not too large. This means that we should
develop a stopping criterion numerically. The latter was done in above cited
publications about this method.

Recall that the main point\emph{\ }of any locally convergent numerical
method is to get a good first guess about the solution. And this is exactly
what our approximately globally convergent method delivers. Hence, we can
apply the second stage of our two-stage numerical procedure to refine the
solution \cite{AB,BK,BK2,BK3,BK4,BKK,BK20}. More precisely, a locally
convergent Adaptive Finite Element Method (adaptivity) was applied. A good
numerical performance of this two-stage numerical procedure was demonstrated
in \cite{AB,BK,BK2,BK3,BK4,BKK,BK20}, including the most challenging case of
experimental data \cite{BK,BK4,BK20}. The adaptivity takes the solution
obtained on the globally convergent stage as the first guess for further
iterations. Suppose now that iterations of the first stage are stopped prior
that stopping criterion is in place. Then the adaptivity still refines the
solution quite well. For example, in Tests 2 and 3 on page 278 of \cite{BK}
adaptivity has started from functions $c_{n,i}\left( x\right) $ for those $n$
which were smaller than the number $\overline{N}$ where the stopping
criterion was achieved.\ Nevertheless, images of these tests were quite
accurate ones. Furthermore, this two-stage numerical procedure led to
accurate images from experimental data, see \cite{BK4}\ and Chapter 5 of
\cite{BK}. As to some other works on the adaptivity technique, see, e.g.
Beilina \cite{Beil1,Beil2} and Li, Xie and Zou \cite{Li}.

\subsection{Published non-local numerical methods for MCIPs}

\label{sec:6.2}

In parallel with the above approximately globally convergent method, another
one was developed by the group of researchers from University of Texas at
Arlington in collaboration with the author, see Klibanov, Su, Pantong, Shan
and Liu \cite{bib18}, Pantong, Su, Shan, Klibanov and Liu \cite{bib22},
Shan, Klibanov, Su, Pantong and Liu \cite{bib23}, Su, Shan, Liu and Klibanov
\cite{bib24} and Su, Klibanov, Liu, Lin, Pantong and Liu \cite{KSu2}. A
globally accelerated numerical method for optical tomography with continuous
wave source In this case the 2d MCIP for the elliptic equation%
\begin{equation}
\Delta u-a\left( x\right) u=-\delta \left( x-x_{0}\right) ,x\in \mathbb{R}%
^{2},\lim_{\left\vert x\right\vert \rightarrow \infty }u\left(
x,x_{0}\right) =0  \label{5.50}
\end{equation}%
was considered. Let $\Gamma \subset \left( \mathbb{R}^{2}\diagdown \Omega
\right) $ be a straight line. The MCIP consists in the determination of the
unknown coefficient $a\left( x\right) $ for $x\in \Omega $ in (\ref{5.50})
from the function $\varphi \left( x,x_{0}\right) =u\left( x,x_{0}\right)
\mid _{x\in \partial \Omega ,x_{0}\in \Gamma }.$ This is the problem of the
so-called Optical Diffusion Tomography with a direct application in optical
imaging of strokes in brains. In this case $x_{0}$ is the position of the
light source, $u\left( x,x_{0}\right) $ is the light intensity and the
function $a\left( x\right) $ is proportional to the absorption coefficient
of light. The most important difference between this problem and the one
discussed in Section 6.1 is that the asymptotic behavior of tails like the
one in (\ref{5.13}) is not the case here. Thus, tails are treated quite
differently in \cite{bib18,bib22,bib23,bib24,KSu2}. In \cite{KSu2} accurate
images from experimental data for a phantom medium were obtained.

In works of Bikowski, Knudsen and Mueller \cite{Bik}, DeAngelo and Mueller
\cite{Dean}, Hamilton, Herrera, Mueller and Von Herrmann \cite{Hamilton} and
Siltanen, Mueller and Isaacson \cite{Silt} non-local reconstruction
techniques are considered for the 2D problem of Electrical\ Impedance
Tomography. In particular, images from experimental data were obtained in
\cite{Dean}. In papers of Alexeenko, Burov and Rumyantseva \cite{Alex},
Burov, Morozov and Rumyantseva \cite{BMR} and Burov, Alekseenko and
Rumyantsevat \cite{BAR} the non-local reconstruction method of Novikov \cite%
{Nov1,Nov2,Nov3,Nov4} for the MCIP for an elliptic PDE with the data given
in the form of the scattering amplitude is implemented. A non-local
numerical method for a 2d MCIP for a hyperbolic PDE was developed by
Kabanikhin and Shishlenin in \cite{KabS1,KabS2}. It is based on a 2d analog
of the Gel'fand-Levitan-Krein equation. Just as the technique of \cite{BK},
algorithms of \cite{Alex,BMR,KabS1,KabS2} also use some reasonable
approximate mathematical models, which are not rigorously justified.

\begin{center}
\textbf{Acknowledgments}
\end{center}

This work was supported by the U.S. Army Research Laboratory and U.S. Army
Research Office under the grant number W911NF-11-1-0399 and by the National
Institutes of Health grant number 1R21NS052850-01A1.


\begin{thebibliography}{999}
\bibitem{} 

\bibitem{AK} M. Agranovsky and P. Kuchment, Uniqueness of reconstruction and
an inversion procedure for thermoacoustic and phtoacoustic tomography with
variable sound speed, \emph{Inverse Problems}, 23, 2089-2102, 2007.

\bibitem{Alex} N.V. Alexeenko, V.A. Burov and O.D. Rumyantseva, Solution of
a three-dimensional acoustical inverse scattering problem: II. Modified
Novikov algorithm, \emph{Acoust. Phys}., 54, 407-419, 2008.

\bibitem{AB} M.~Asadzadeh and L.~Beilina, \emph{A posteriori} error analysis
in a globally convergent numerical method for a hyperbolic coefficient
inverse problem, \emph{Inverse Problems}, 26, 115007, 2010.

\bibitem{BKok} A.B. Bakushinskii and M.Yu.\ Kokurin, \emph{Iterative Methods
for Approximate Solutions of Inverse Problems}, Springer, New York, 2004.

\bibitem{Bardos} C. Bardos, G. Lebeau and J. Rauch, Sharp sufficient
conditions for observation control and stabilization of waves from the
boundary, \emph{SIAM J. Contr. Opt}., 30, 1024-1065, 1992.

\bibitem{Baud1} L. Baudouin and J.-P. Puel, Uniqueness and stability in an
inverse problem for the Schr\"{o}dinger equation, \emph{Inverse Problems},
18, 1537-1554, 2002.

\bibitem{BMO} L. Baudouin, A.\ Mercado and A. Osses, A global Carleman
estimate in a transmission wave equation and application to a
one-measurement inverse problem, \emph{Inverse Problems}, 23, 257-258, 2007.

\bibitem{Baud3} L. Baudouin and A. Mercado, An inverse problem for Schr\"{o}%
dinger equations with discontinuous main coefficient, \emph{Applicable
Analysis}, 87, 1145-1165, 2008.

\bibitem{Baud2} L. Baudouin, E. Cr\'{e}peau and J. Valein, Global Carleman
estimate on a network for the wave equation and application to an inverse
problem, \emph{Mathematical Control and Related Fields}, 1, 1-24, 2011.

\bibitem{Baud4} L. Baudouin, E. Cerpa, E. Cr\'{e}peau and A. Mercado,
Lipschitz stabilty in an inverse problem for the Kuramoto-Sivashinsky
equation, \emph{Applicable Analysis}, to appear; preprint is available
online at \emph{arXiv}:1008.3279v2 [math.AP].

\bibitem{Beil1} L. Beilina, Adaptive finite element/difference method for
inverse elastic scattering waves, \emph{Applied and Computational Mathematics%
}, 2, 119-134, 2003.

\bibitem{Beil2} L. Beilina, Adaptive finite element method for a coefficient
inverse problem for the Maxwell's system, \emph{Applicable Analysis}, 90,
1461-1479, 2010.

\bibitem{BK} L. Beilina and M.V. Klibanov, \emph{Approximate Global
Convergence and Adaptivity for Coefficient Inverse Problems}, Springer, New
York, 2012.

\bibitem{BK1} L. Beilina and M.V. Klibanov, A globally convergent numerical
method for a coefficient inverse problem, \emph{SIAM J. Sci. Comp.,} 31%
\textbf{,} 478-509, 2008.

\bibitem{BK2} L. Beilina and M.V. Klibanov, Synthesis of global convergence
and adaptivity for a hyperbolic coefficient inverse problem in 3D, \emph{J.
Inverse and Ill-posed Problems,} 18\textbf{,} 85-132, 2010.

\bibitem{BK3} L. Beilina and M.V. Klibanov, \emph{A posteriori} error
estimates for the adaptivity technique for the Tikhonov functional and
global convergence for a coefficient inverse problem, \emph{Inverse Problems,%
} 26\textbf{,} 045012, 2010.

\bibitem{BK4} L. Beilina and M.V.Klibanov, Reconstruction of dielectrics
from experimental data via a hybrid globally convergent/adaptive inverse
algorithm, \emph{Inverse Problems}, 26, 125009, 2010.

\bibitem{BKK} L. Beilina, M.V. Klibanov and M.Yu Kokurin, Adaptivity with
relaxation for ill-posed problems and global convergence for a coefficient
inverse problem, \emph{Journal of Mathematical Sciences,} 167, 279-325, 2010.

\bibitem{BK20} L. Beilina, M.V. Klibanov, The philosophy of the approximate
global convergence for multidimensional coefficient inverse problems, \emph{%
Complex Variables and Elliptic Equations}, 57, 277-299, 2012.

\bibitem{JIIP12} L. Beilina and M.V. Klibanov, A new approximate
mathematical model for global convergence for a coefficient inverse problem
with backscattering data, \emph{J. Inverse and Ill-Posed Problems}, 20,
2012, to appear.

\bibitem{BM} L. Beilina, Energy estimates and numerical verification of the
stabilized domain decomposition finite element/finite difference approach
for the Maxwell's system in time domain, \emph{Central European Journal of
Mathematics}, accepted for publication; preprint is available online at
http://publications.lib.chalmers.se/publication/142368.

\bibitem{Bell1} M. Bellassoued, Uniqueness and stability in determining the
speed of propagation of second-order hyperbolic equation with variable
coefficients, \emph{Applicable Analysis}, 83, 983-1014, 2004.

\bibitem{Y6} M. Bellassoued and M.\ Yamamoto, Logarithmic stability in
determination of a coefficient in an acoustic equation by arbitrary boundary
observation, \emph{Journal de Math\'{e}matiques Pures et Appliqu\'{e}es},
85, 193-224, 2006.

\bibitem{Y20} M. Bellassoued and M.\ Yamamoto, Inverse source problem for a
transmission problem for a parabolic equation, \emph{J. Inverse and
Ill-Posed Problems}, 14, 47-56, 2006.

\bibitem{Y11} M. Bellassoued, O.Yu. Imanuvilov and M.\ Yamamoto, Inverse
problem of determining the density and two Lam\'{e} coefficients by boundary
data, \emph{SIAM J. Math. Anal}., 40, 238--265, 2008.

\bibitem{Bell2} M. Bellassoued, M. Cristofol and E. Soccorsi, Inverse
boundary value problem for the dynamical heterogeneous Maxwell's system,
\emph{Inverse Problems}, 28, 095009, 2012.

\bibitem{Ben2} A. Benabdallah, Y. Dermenjian and J. Le Rousseau, Carleman
estimates for the one-dimensional heat equation with a discontinous
coefficient and applications to controllability and an inverse problem,
\emph{J.\ Math. Analysis and Applications}, 336, 865-887, 2007.

\bibitem{Ben3} A. Benabdallah, P. Gaitan and J. Le Rousseau, Stability of
discontinous diffusion coefficients and initial conditions in an inverse
problem for the heat equation, \emph{SIAM J. Contr. Optim}., 46, 1849-1881,
2007.

\bibitem{Bik} J. Bikowski, K. Knudsen and J. L. Mueller, Direct numerical
reconstruction of conductivities in three dimensions using scattering
transforms, \emph{Inverse Problems}, 27, 015002, 2011.

\bibitem{Boul} M. Boulakia, C. Grandmont and A.\ Osses, Some inverse
stability results for the bi-stable reaction-diffusion equation using
Carleman estimates, \emph{C.R. Acad. Sci. Paris}, 347, 619-622, 2009.

\bibitem{Bourg1} L. Bourgeois, Convergence rates for the quasi-reversibility
method to solve the Cauchy problem for Laplace's equation, \emph{Inverse
Problems}, 22, 413-430, 2006.

\bibitem{Bourg2} L. Bourgeois and J. Darde, About stability and
regularization of ill-posed elliptic Cauchy problems: the case of Lipschitz
domains, \emph{Applicable Analysis}, 89, 1745-1768, 2010.

\bibitem{Buhan} M. de Buhan and A. Osses, Logarithmic stability in
determination of a 3D viscoelastic coefficient and a numerical example,
\emph{Inverse Problems}, 26, 095006, 2010.

\bibitem{BukhKlib} A.L. Bukhgeim and M.V. Klibanov, Uniqueness in the large
of a class of multidimensional inverse problems, \emph{Soviet Math. Doklady}%
, 17, 244-247, 1981.

\bibitem{Bukh1} A.L. Bukhgeim, Carleman estimates for Volterra operators and
uniqueness of inverse problems, in \emph{Non-Classical Problems of
Mathematical Physics}, pages 54-64, published by Computing Center of the
Siberian Branch of USSR Academy of Science, Novosibirsk, 1981 (in Russian).

\bibitem{Bukh2} A.L. Bukhgeim, \emph{Introduction In The Theory of Inverse
Problems}, VSP, Utrecht, The Netherlands, 2000.

\bibitem{BMR} V.A. Burov, S.A. Morozov and O.D. Rumyantseva, Reconstruction
of fine-scale structure of acoustical scatterers on large-scale contrast
background, \emph{\ Acoust. Imaging,} 26, 231-238, 2002.

\bibitem{BAR} V.A. Burov, N.V. Alekseenko and O.D. Rumyantseva,
Multifrequency generalization of the Novikov algorithm for the
two-dimensional inverse scattering problem \emph{Acoustical Physics,} 55,
843-856, 2009.

\bibitem{Cald} A.P. Calderon, Uniqueness in the Cauchy problem for partial
differential equations, \emph{American J. Math}., 80, 16-36, 1958.

\bibitem{Cao} H. Cao, M.V. Klibanov and S.V. Pereverzev, A Carleman estimate
and the balancing principle in the quasi-reversibility method for solving
the Cauchy problem for the Laplace equation, \emph{Inverse Problems}, 25,
35005, 2009.

\bibitem{Carl} T.~Carleman, Sur un probl\'{e}me d'unicit\'{e} pur les syste%
\'{e}mes d'\'{e}quations aux d\'{e}rive\'{e}s partielles \`{a} deux
varibales ind\'{e}pendantes, \emph{Ark.~Mat.~Astr.~Fys}, 26B, No. 17, 1--9,
1939.

\bibitem{Lorenzi} C. Cavaterra, A. Lorenzi and M. Yamamoto, A stability
result via Carleman estimates for an inverse source problem related to a
hyperbolic integro-differential equation, \emph{Computational and Applied
Mathematics}, 25, 229--250, 2006.

\bibitem{ClK} C. Clason and M.V. Klibanov, The quasi-reversibility method
for thermoacoustic tomography in a heterogeneous medim, \emph{SIAM J. Sci.
Comp}., 30, 1-23, 2007.

\bibitem{Cr1} M. Cristofol, P. Gaitan and H. Ramoul, Inverse problems for a $%
2\times 2$ reaction-diffusion system using a Carleman estimate with one
observation, \emph{Inverse Problems}, 22, 1561-1573, 2006.

\bibitem{Dean} M. DeAngelo and J.L. Mueller, 2D d-bar reconstructions of
human chest and tank using an improved approximation to the scattering
transform, \emph{Physiological Measurement}, 31, 221-232, 2010.

\bibitem{DO} A. Doubova and A. Osses, Rotated weights in global Carleman
estimates applied to an inverse problem for the wave equation, \emph{Inverse
Problems}, 22, 265-296, 2006.

\bibitem{EEK} H. Egger, H.W. Engl and M.V. Klibanov, Global uniqueness and H%
\"{o}lder stability for recovering a nonlinear source term in a parabolic
equation, \emph{Inverse Problems}, 21, 271-290, 2005.

\bibitem{EHN} H.W. Engl, M. Hanke and A. Neubauer, \emph{Regularization of
Inverse Problems}, Kluwer Academic Publishers, Boston, 2000.

\bibitem{Fan} J. Fan, M. Di Cristo, Y. Jiang, \ and G. Nakamura, Inverse
viscosity problem for the Navier--Stokes equation, \emph{J.\ Mathematical
Analysis and Applications}, 365, 750--757, 2010.

\bibitem{FPR} D. Finch, S.K Patch and Rakesh, Determining a function from
its mean values over a family of spheres, \emph{SIAM J. Math. Anal}., 35,
1213-1240, 2004.

\bibitem{FHR} D. Finch, M. Haltmeier and Rakesh, Inversion of spherical
means and the wave equation in even dimensions, \emph{SIAM J. Appl. Math}.,
68, 392-412, 2007.

\bibitem{FR} D. Finch and Rakesh, Recovering a function from its spherical
mean values in two and three dimensions, \emph{Photoacoustic Imaging and
Spectroscopy}, CRC Press, Boca Raton, Florida, 2009.

\bibitem{F} A. Friedman, \emph{Partial Differential Equations of Parabolic
Type}, Prentice Hall, Inc., Englewood Cliffs, N.J., 1964.

\bibitem{Furs} A.V. Fursikov and O. Yu. Imanuvilov, \emph{Controllability of
Evolution Equations}, Lecture Notes Series, 34, , Seoul National University,
Korea, 1996.

\bibitem{Hamilton} S.J. Hamilton, C.N.L. Herrera, J.L. Mueller and A. Von
Herrmann, A direct D-bar reconstruction algorithm for recovering a complex
conductivity in 2D, \emph{Inverse Problems}, 28, 095005, 2012.

\bibitem{Horm} L. H\"{o}rmander, \emph{Linear Partial Differential Operators}%
, Springer, Berlin, 1963.

\bibitem{Im} O. Yu. Imanuvilov, Controllability of parabolic equations,
\emph{Sbornik Math}., 186, 879-900, 1995.

\bibitem{Y8} O. Yu. Imanuvilov and M. Yamamoto, Lipschitz stability in
inverse parabolic problems by the Carleman estimate, \emph{Inverse Problems}%
, 14, 1229-1245, 1998.

\bibitem{Y2} O. Yu. Imanuvilov and M. Yamamoto, Global uniqueness and
stability in determining coefficients of wave equations, \emph{Commun. in
Partial Differential Equations}, 26, 1409-1425, 2001.

\bibitem{Y3} O. Yu. Imanuvilov and M. Yamamoto, Global Lipschitz stability
in an inverse hyperbolic problem by interior observations, \emph{Inverse
Problems}, 17, 717-728, 2001.

\bibitem{Y5} O. Yu. Imanuvilov and M. Yamamoto, Determination of a
coefficient in an acoustic equation with a single measuerement, \emph{%
Inverse Problems}, 19, 157-171, 2003.

\bibitem{Y10} O. Yu. Imanuvilov, V. Isakov and M. Yamamoto, An inverse
problem for the dynamical Lam\'{e} system with two sets of boundary data,
\emph{Comm. Pure and Applied Math}., 56, 1-17, 2003.

\bibitem{Y4} V. Isakov and M. Yamamoto, Carleman estimate with the Neumann
boundary condition and its applications to the observability inequality and
inverse hyperbolic problems, \emph{Contemporary Mathematics}, 268, 191-225,
2000.

\bibitem{Is0} V.\ Isakov, \emph{Inverse Source Problems}, American
Mathematical Society, Providence, RI, 1990.

\bibitem{IsMilan} V.\ Isakov, Carleman estimates and applications to inverse
problems, \emph{Milan J. of Mathematics}, 72, 249-271, 2004.

\bibitem{Is} V. Isakov, \emph{Inverse Problems for Partial Differential
Equations}, Second Edition, Springer, New York, 2006.

\bibitem{KabS1} S.I. Kabanikhin, A.D. Satybaev and M.A. Shishlenin, \emph{%
Direct Methods for Solving Multidimensional Inverse Hyperbolic Problems},
VSP, Utrecht, The Netherlands, 2004.

\bibitem{KabS2} S.I. Kabanikhin and M.A. Shishlenin, Numerical algorithm for
two-dimensional inverse acoustic problem based on Gel'fand-Levitan-Krein
equation, \emph{J.\ Inverse and Ill-Posed Problems}, 18, 979-995, 2011.

\bibitem{Kab} S.I. Kabanikhin, \emph{Inverse and Ill-Posed Problems. Theory
and Applications}, De Gruyter, Berlin, 2012.

\bibitem{Kalt} B. Kaltenbacher and M. V. Klibanov, An inverse problem for a
nonlinear parabolic equation with applications in population dynamics and
magnetics, \emph{SIAM J. Math. Anal}., 39, 1863-1889, 2008.

\bibitem{Kaz} M. Kazemi and M.V. Klibanov, Stability estimates for ill-posed
Cauchy problem involving hyperbolic equation and inequalities, \emph{%
Applicable Analysis}, 50, 93-102, 1993.

\bibitem{Khai} A. Kha\u{\i}darov, On stability estimates amd inverse
problems for second order hyperbolic equations, \emph{Math. USSR Sbornik},
58, 267-277, 1987.

\bibitem{Klib1} M. V. Klibanov, Uniqueness of solutions in the `large' of
some multidimensional inverse problems, in \emph{Non-Classical Problems of
Mathematical Physics}, pages 101-114, 1981, published by Computing Center of
the Siberian Branch of the USSR Academy of Science, Novosibirsk (in Russian).

\bibitem{Klib2} M. V. Klibanov, On a class of inverse problems, \emph{Soviet
Math. Doklady}, 26, 248-250, 1982.

\bibitem{Klib3} M. V. Klibanov, Inverse problems in the `large' and Carleman
bounds, \emph{Differential Equations}, 20, 755-760, 1984.

\bibitem{Klib4} M. V. Klibanov, Uniqueness of inverse problems in the large
for one class of differential equations, \emph{Differential Equations}, 20,
1667-1671, 1984.

\bibitem{KlibMaxw} M.~V.~Klibanov, Uniqueness of solutions of two inverse
problems for the Maxwellian system, \emph{USSR J. Computational Mathematics
and Mathematical Physics}, 26, 1063-1071, 1986

\bibitem{KlibPar1} M.V. Klibanov, A class of inverse problems for nonlinear
parabolic equations, \emph{Siberian Math. J}., 27, 698-707, 1987.

\bibitem{KlibSant} M.V. Klibanov and F. Santosa, A computational
quasi-reversibility method for Cauchy problems for Laplace's equation, \emph{%
SIAM J. Appl. Math.}, 51, 1653-1675, 1991.

\bibitem{KlibM} M.V. Klibanov and J. Malinsky, Newton-Kantorovich method for
3-dimensional potential inverse scattering problem and stability for the
hyperbolic Cauchy problem with time dependent data, \emph{Inverse Problems},
7, 577-596, 1991.

\bibitem{Klib5} M.~V.~Klibanov, Inverse problems and {C}arleman estimates,
\emph{Inverse Problems}, 8, 575--596, 1992.

\bibitem{KR} M.V. Klibanov and Rakesh, Numerical solution of a timelike
Cauchy problem for the wave equation, \emph{Math.\ Meth. in Appl. Sci}., 15,
559-570, 1992.

\bibitem{KlibSurvey} M.V. Klibanov, Carleman estimates and inverse problems
in the last two decades, \emph{Surveys on Solutions Methods for Inverse
Problems}, Wien, Springer, 119-146, 2000.

\bibitem{KlibPar2} M.V. Klibanov, Global uniqueness of a multidimensional
inverse problem for a nonlinear parabolic equation, \emph{Inverse Problems},
22, 495-514, 2004.

\bibitem{KT} M.V. Klibanov and A. Timonov, \emph{Carleman Estimates for
Coefficient Inverse Problems and Numerical Applications}, VSP, Utrecht, The
Netherlands, 2004.

\bibitem{K} M.V. Klibanov, Lipschitz stability for hyperbolic inequalities
in octants with the lateral Cauchy data and refocusing in time reversal,
\emph{J. Inverse and Ill-Posed Problems}, 13, 353-363, 2005.

\bibitem{KYam} M.V. Klibanov and M. Yamamoto, Lipschitz stability for an
inverse problem for an \ acoustic equation, \emph{Applicable Analysis}, 85,
515-538, 2006.

\bibitem{Kl1} M.V. Klibanov, Estimates of initial conditions of parabolic
equations and inequalities via lateral Cauchy data, \emph{Inverse Problems},
22, 495-514, 2006.

\bibitem{Kl2} M.V. Klibanov and A.V. Tikhonravov, Estimates of initial
conditions of parabolic equations and inequalities in infinite domains via
lateral Cauchy data, \emph{Journal of Differential Equations}, 237, 198-224,
2007.

\bibitem{KKKN} M.V. Klibanov, A.V. Kuzhuget, S.I. Kabanikhin and D.V.\
Nechaev, A new version of the quasi-reversibility method for the
thermoacoustic tomography and a coefficient inverse problem, \emph{%
Applicable Analysis}, 87, 1227-1254, 2008.

\bibitem{KFBPS} M.~V.~Klibanov, M.~A.~Fiddy, L.~Beilina, N.~Pantong and
J.~Schenk, Picosecond scale experimental verification of a globally
convergent numerical method for a coefficient inverse problem, \emph{Inverse
Problems}, 26, 045003, 2010.

\bibitem{KBB} M.V. Klibanov, A.B. Bakushinskii and L. Beilina, Why a
minimizer of the Tikhonov functional is closer to the exact solution than
the first guess,\emph{\ J.\ Inverse and Ill-Posed Problems}, 19, 83-105,
2011.

\bibitem{Klib6} M.V. Klibanov, Uniqueness of an inverse problem with single
measurement data generated by a plane wave in partial finite differences,
\emph{Inverse Problems}, 27, 115005, 2011.

\bibitem{Kltherm} M.V. Klibanov, Thermoacoustic tomography with an arbitrary
elliptic operator, \emph{Arxiv}: 1208.5187v1 [math-ph].

\bibitem{bib18} M.V. Klibanov, J. Su, N. Pantong, H. Shan and H. Liu, A
globally convergent numerical method for an inverse elliptic problem of
optical tomography, \emph{Applicable Analysis}, 89, 861-891, 2010.

\bibitem{Kuch} P. Kuchment and L. Kunyansky, Mathematics of thermoacoustic
tomography, \emph{European J. Applied Mathematics}, 19, 191-224, 2008.

\bibitem{KY} L. Kunyansky, Thermoacoustic tomography with detectors on an
open curve: an efficient reconstruction algorithm, \emph{Inverse Problems},
24, 055021, 2008.

\bibitem{KuzhKl} A.V. Kuzhuget and M.V. Klibanov, Global convergence for a
1-D inverse problem with application to imaging of land mines, \emph{%
Applicable Analysis,} 89, 125-157, 2010.

\bibitem{KPK} A.V. Kuzhuget, N. Pantong and M.V. Klibanov, A globally
convergent numerical method for a coefficient inverse problem with
backscattering data, \emph{Methods and Applications of Analysis,} 18, 47-68,
2011.

\bibitem{KBK} A.V. Kuzhuget, L.\ Beilina and M.V. Klibanov, Approximate
global convergence and quasireversibility of a coefficient inverse problem
with backscattering data, \emph{Journal of Mathematical Sciences}, 181,
19-49, 2012.

\bibitem{KBKSNF} A.V. Kuzhuget, L. Beilina, M.V. Klibanov, A. Sullivan,\ L.
Nguyen and M.A. Fiddy, Blind experimental data collected in the field and an
approximately globally convergent inverse algorithm, \emph{Inverse Problems}%
, 28, 095007, 2012.

\bibitem{IEEE} A.V. Kuzhuget, L. Beilina, M.V. Klibanov, A. Sullivan,\ L.
Nguyen and M.A. Fiddy, Quantitative image recovery from measured blind
backscattered data using a globally convergent inverse method, \emph{IEEE
Transactions of Geoscience and Remote Sensing}, accepted for publication, to
be published in 2012.

\bibitem{LSU} O.A. Ladyzhenskaya, V.A. Solonnikov and N.N. Uralceva, \emph{%
Linear and Quasilinear Equations of Parabolic Type}, AMS, Providence, R.I.,
1968.

\bibitem{LU} O.~A.~ Ladyzhenskaya and N.~N.~Uralceva, \emph{Linear and
Quasilinear Elliptic Equations}, Academic Press, New York, 1969.

\bibitem{Lad} O.A. Ladyzhenskaya, \emph{Boundary Value Problems of
Mathematical Physics}, Springer, New York, 1985.

\bibitem{LT1} I. Lasiecka, R. Triggiani and P.F. Yao, Inverse/observability
estimates for second order hyperbolic equations with variable coefficients,
\emph{J. Math. Anal. Appl.}, 235, 13-57, 1999.

\bibitem{LT2} I. Lasiecka, R. Triggiani and X. Zhang, Nonconservative wave
equations with unobserved Neumann BC: global uniqueness and observability,
\emph{AMS Contemp. Math}., 268, 227-326, 2000.

\bibitem{LT3} I. Lasiecka, R. Triggiani and X. Zhang, Global uniqueness,
observability and stabilization of non-conservative Schr\"{o}dinger
equations via pointwise Carleman estimates. Part I: $H^{1}\left( \Omega
\right) -$estimates, \emph{J.\ of Inverse and Ill-Posed Problems}, 12, 1-81,
2004.

\bibitem{LT4} I. Lasiecka, R. Triggiani and X. Zhang, Global uniqueness,
observability and stabilization of non-conservative Schr\"{o}dinger
equations via pointwise Carleman estimates. Part II: $L_{2}\left( \Omega
\right) -$estimates, \emph{J.\ of Inverse and Ill-Posed Problems}, 12,
182-231, 2004.

\bibitem{LL} R. Lattes and J.-L. Lions, \emph{The Method of
Quasireversibility: Applications to Partial Differential Equations,}
Elsevier, New York, 1969.

\bibitem{LRS} M.M. Lavrentiev, V.G. Romanov and S.P. Shishatskii, \emph{%
Ill-Posed Problems of Mathematical Physics and Analysis}, AMS, Providence,
R.I., 1986.

\bibitem{LiSIAM} S.\ Li, An inverse problem for Maxwell's equations in
bi-isotropic media, \emph{SIAM J. Math. Anal}., 37, 1027-1043, 2005.

\bibitem{LiYam} S.\ Li and M. Yamamoto, An inverse source problem for
Maxwell's equations in anisotropic media, \emph{Applicable Analysis}, 84,
1051-1067, 2005.

\bibitem{LYZ} J.\ Li, M. Yamamoto and J. Zou, Conditional stability and
numerical reconstruction of initial temperature, \emph{Communications on
Pure and Applied Analysis}, 8, 361-382, 2009.

\bibitem{Li} J. Li, J. Xie and J. Zou, An adaptive finite element
reconstruction of distributed fluxes, \emph{Inverse Problems}, 27, 075009,
2011.

\bibitem{Trig2} S. Liu and R. Triggiani, Global uniqueness in determining
electric potentials for a system of strongly coupled Schr\"{o}\"{o}dinger
equations with magnetic potential terms, \emph{J. of Inverse and Ill-Posed
Problems}, 19, 223-254, 2011.

\bibitem{Trig3} S. Liu and R. Triggiani, Global uniqueness and stability in
determining the damping coefficient of an inverse hyperbolic problem with
non homogeneous Neumann B.C. through an additional Dirichlet boundary trace,
\emph{SIAM J. Math. Anal}., 43, 1631-1666, 2011.

\bibitem{Trig4} S. Liu and R. Triggiani, Global uniqueness and stability in
determining the damping and potential coefficients of an inverse hyperbolic
problem, \emph{Nonlinear Analysis: Real World Applications, }12, 1562-1590,
2011.

\bibitem{Trig5} S. Liu and R. Triggiani, Global uniqueness and stability in
determining in determining the damping coefficient of an inverse hyperbolic
problem with non-homogeneous Dirichlet BC through an additional localized
Neumann boundary condition, \emph{Applicable Analysis}, 91, 1551-1581, 2012.

\bibitem{L} Lop Fat Ho, Observabilit\'{e} frontier\`{e} de l'equation des
ondes, \emph{C.R. Acad. Sc. Paris}, t. 302, Ser. I, No. 12, 443-446, 1986.

\bibitem{Lu} Q. L\"{u}, Carleman estimate for stochastic parabolic equations
and inverse stochastic parabolic problems, \emph{Inverse Problems}, 28,
045008, 2012.

\bibitem{Mercado} A. Mercado, A. Osses and L. Rosier, Inverse problems for
the Schr\"{o}dinger equation via Carleman inequalities with degenerate
weights, \emph{Inverse Problems}, 24, 015017, 2008.

\bibitem{Nov1} R.G. Novikov, Multidimensional inverse spectral problem for
the equation $-\triangle \psi +(v(x)-Eu(x))\psi =0,$ \emph{Functional
Analysis and Its Applications,} 22\textbf{,} 11-22, 1988.

\bibitem{Nov2} R.G. Novikov, The inverse scattering problem on a fixed
energy level for the two-dimensional Schr\"{o}dinger~~operator, \emph{J.
Func. Anal. and Its Applications, }103\textbf{,} 409-463, 1992.

\bibitem{Nov3} R.G. Novikov, The $\partial -$bar approach to approximate
inverse scattering at fixed energy in three dimensions, \emph{Int.} \emph{%
Math.\ Res. Reports}, 6, 287-349, 2005.

\bibitem{Nov4} R.G. Novikov and M. Santacesaria, Monochromatic
reconstruction algorithms for two-dimensional multi-channel inverse
problems, \emph{International Mathematics Research Notices}, to appear.

\bibitem{bib22} N. Pantong, J. Su, H. Shan, M.V. Klibanov and H. Liu, A
globally accelerated reconstruction algorithm for diffusion tomography with
continuous-wave source in arbitrary convex shape domain, \emph{Journal of
the Optical Society of America,} \emph{A}, 26, 456-472, 2009.

\bibitem{Payne} L.E. Payne, \emph{Improperly Posed Problems in Partial
Differential Equations}, SIAM, Philadeplhia, 1975.

\bibitem{Poisson} O.\ Poisson, Carleman estimates for the heat equation with
discontinuous diffusion coefficients, \emph{Applicable Analysis}, 87,
1129-1144, 2008.

\bibitem{Y1} J.-P. Puel and M. Yamamoto, On a global estimate in a linear
inverse hyperbolic problem, \emph{Inverse Problems}, 12, 995-1002, 1996.

\bibitem{Rezn} K.G. Reznitckaya, Connection between solutions of different
types of Cauchy problems and inverse problems, in \emph{Mathematical
Problems of Geophysics}, issue 5, part 1, 55-62, 1974, published by
Computing Center of the Siberian Branch of the USSR Academy of Science,
Novosibirsk (in Russian).

\bibitem{Rob1} L. Robbiano, Th\'{e}oreme d'unicite adapt\'{e} au contr$%
\widehat{\text{o}}$le des solutions des probl\`{e}mes hyperboliques, Comm.
Partial Differential Equations, 16, 789-800, 1991.

\bibitem{Rob2} L. Robbiano, Fonction de co\^{u}t et contr$\widehat{\text{o}}$%
le des solutions des \'{e}quations hypeboliques, \emph{Asymptotic Analysis},
10, 95-105, 1995.

\bibitem{Rom} V.G. Romanov, \emph{Inverse Problems of Mathematical Physics},
VNU Press, Utrecht, The Netherlands, 1986.

\bibitem{RomKab} V.G. Romanov and S.I. Kabanikhin, \emph{Inverse Problems
for Maxwell's Equations}, de Gruyter, Berlin, 1994.

\bibitem{Rom1} V.G. Romanov, Estimates of a solution to a differential
inequality related to a second order hyperbolic operator an Cauchy data on a
timelike surface, \emph{Doklady Mathematics}, 73, 51-53, 2006.

\bibitem{Rom2} V.G. Romanov, Stability estimates in inverse problems for
hyperbolic equations, \emph{Milan J. Math}., 74, 357-385, 2006.

\bibitem{Rom3} V.G. Romanov and M. Yamamoto, Recovering a Lam\'{e} kernel in
a viscoelastic equation by a single boundary measurement, \emph{Applicable
Analysis}, 89, 377-390, 2010.

\bibitem{bib23} H. Shan, M.V. Klibanov, J. Su, N. Pantong and H. Liu, A
globally accelerated numerical method for optical tomography with continuous
wave source, \emph{J. Inverse and Ill-Posed Problems, }16, 765-792, 2008.

\bibitem{Silt} S. Siltanen, J. Mueller and D. Isaacson, An implementation of
the reconstruction algorithm of A. Nachman for the 2-D inverse conductivity
problem, \emph{Inverse Problems}, 16, 681-699, 2000.

\bibitem{SU} P. Stefanov and G. Uhlmann, Thermoacoustic tomography with
variable sound speed, \emph{Inverse Problems}, 25, 075011, 2009.

\bibitem{bib24} J. Su, H. Shan, H. Liu and M.V. Klibanov, Reconstruction
method from a multiplesite continuous-wave source for three-dimensional
optical tomography, \emph{J. Optical Society of America A, }23, 2388-2395,
2006.

\bibitem{KSu2} J. Su, M. V. Klibanov, Y. Liu, Z. Lin, N. Pantong and H. Liu,
{Optical imaging of phantoms from real data by an approximately globally
convergent inverse algorithm, \emph{Arxiv} 1208.5175v1 [math-ph].}

\bibitem{T} A.N. Tikhonov, A.V. Goncharsky, V.V. Stepanov and A.G. Yagola,
\emph{Numerical Methods for the Solution of Ill-Posed Problems}, Kluwer,
London, 1995.

\bibitem{Trig} R. Triggiani and P.F. Yao, Carleman estimates with no lower
order terms for general Riemannian wave equations. Global uniqueness and
observability in one shot, \emph{Appl. Math. and Optimization}, 46, 331-375,
2002.

\bibitem{Wu1} B. Wu and J. Liu, Conditional stability and uniqueness for
determining two coefficients in a hyperbolic-parabolic system, \emph{Inverse
Problems}, 27, 075013, 2011.

\bibitem{Wu2} B. Wu and J. Liu, Determination of an unknown source for a
thermoelastic system with a memory effect, \emph{Inverse Problems}, 28,
095012, 2012.

\bibitem{Y40} M. Yamamoto, On an inverse problem of determining source terms
in Maxwell's equations with a single measurement, in \emph{Inverse Problems,
Tomography, and Image Processing}, edited by A. Ramm, pages 241-256, Plenum
Press,\ New York, 1998.

\bibitem{Y9} M. Yamamoto and J. Zou, Simultaneous reconstruction of the
initial temperature and heat radiative coefficient, \emph{Inverse Problems},
17, 1181-1202, 2001.

\bibitem{Y7} M. Yamamoto, Carleman estimates for parabolic equations.
Topical Review. \emph{Inverse Problems}, 25, 123013, 2009.

\bibitem{Y21} G. Yuan and M.\ Yamamoto, Lipschitz stability in inverse
problems for a Kirchhoff plate equation, \emph{Asymptotic Analysis}, 53,
29-60, 2007.

\bibitem{Y22} G. Yuan and M.\ Yamamoto, Carleman estimates for the Schr\"{o}%
dinger equation and applications to an inverse problem and an observability
inequality, \emph{Chinese Annals of Mathematics, Series B}, 31, 555-578,
2010.
\end{thebibliography}
\end{document}